\documentclass[11pt]{article}
\usepackage{amsfonts}
\usepackage[latin9]{inputenc}
\usepackage{geometry}
\usepackage{setspace}
\usepackage{amsmath}
\usepackage{amssymb}
\usepackage{dcolumn}
\usepackage{graphicx}
\usepackage[position=bottom]{subfig}
\usepackage[authoryear]{natbib}
\usepackage[colorlinks,linkcolor=blue,citecolor=blue]{hyperref}

\newcolumntype{d}{D{.}{.}{-1}}
\newtheorem{theorem}{Theorem}
\newtheorem{lemma}{Lemma}
\newtheorem{corollary}{Corollary}
\newtheorem{definition}{Definition}
\newtheorem{example}{Example}
\newtheorem{assumption}{Assumption}
\newenvironment{proof}[1][Proof]{\noindent \textbf{#1.} }{\  \rule{0.5em}{0.5em}}

\begin{document}

\title{Local polynomial estimation of time-varying parameters in nonlinear
models\thanks{%
The authors would like to thank Sokbae Lee and participants at various
seminars and conferences for valuable comments and suggestions.}}
\author{Dennis Kristensen\thanks{%
Department of Economics, University College London; e-mail: \href{mailto:}{%
d.kristensen@ucl.ac.uk}} \and Young Jun Lee\thanks{%
Korea Institute for International Economic Policy; e-mail: \href{mailto:}{%
y.lee@kiep.go.kr}}}
\maketitle

\begin{abstract}
We develop a novel asymptotic theory for local polynomial extremum
estimators of time-varying parameters in a broad class of nonlinear time
series models. We show the proposed estimators are consistent and follow
normal distributions in large samples under weak conditions. We also provide
a precise characterisation of the leading bias term due to smoothing, which
has not been done before. We demonstrate the usefulness of our general
results by establishing primitive conditions for local
(quasi-)maximum-likelihood estimators of time-varying models threshold
autoregressions, ARCH models and Poisson autogressions with exogenous
co--variates, to be normally distributed in large samples and characterise
their leading biases. An empirical study of US corporate default counts
demonstrates the applicability of the proposed local linear estimator for
Poisson autoregression, shedding new light on the dynamic properties of US
corporate defaults.

\bigskip{}

\end{abstract}

\section{Introduction\label{sec: intro}}

There is ample empirical evidence of time-varying parameters in many
econometric models; see, e.g., \cite{Inoue2011}, \citet{giacomini2016}, \cite%
{Caldara2012}, \cite{Ghysels1990} and \cite{Christoffersen2012}. Most
studies aiming at accommodating this feature assume a fully parametric model
for this time variation; one example of this is structural break models.
This has the advantage that the time-varying version of a given model stays
parametric and can be estimated using existing methods. The disadvantage is
that the researcher runs the risk of choosing a misspecified model for the
time variation.

To reduce this risk, methods that treat the problem of time--varying
parameters as a structured nonparametric one have been developed: They
assume that the sequence of time-varying parameters arise as values of an
underlying function which is then estimated nonparametrically; one popular
class of estimators that falls in this category are local estimators which
includes the so-called rolling-window estimator. However, the existing
literature has mostly focused on local constant (Nadaraya-Watson) kernel
estimators of the time-varying parameters; see, e.g., \citet{dahlhaus2017}
and \citet{bardet2022}.

We here propose to estimate the time--varying parameters using local
polynomial estimators since these are known to have a number of attractive
properties compared to the local constant one; see, e.g., \cite{fan1995JASA}%
. Under very weak restrictions on the model being estimated and the time
series data being used, we develop an asymptotic theory for the estimators.
Specifically, we show that they are normally distributed in large samples
and provide a complete charactersation of the leading variance and bias
components.

The class of estimators includes as special cases the local constant
estimator and the local linear estimator. We find that the local constant
estimator requires stronger regularity conditions to be well-behaved in
large samples and will generally suffer from additional biases in the
interior of the domain compared to the local linear estimator. These
additional biases involve the so-called derivative process of the stationary
approximation to data which is not present in the biases of the local linear
one. Moreover, the local linear estimator enjoys the well-known automatic
boundary adjustment property: At the beginning and end of the sample it will
perform better than the local constant one. This feature is important since
often the main interest is on the values that the time-varying parameters
take at the end of the sample.

The two most closely related papers to ours are \citet{dahlhaus2017} and %
\citet{bardet2022} who develop a general theory for local constant
estimators of time--varying parameters in Markov models and infinite memory
models, respectively. Our framework encompasses theirs as special cases and
we consider a broader class of estimators than they do. Moreover, our proof
techniques are different and, when specialising to local constant
estimators, allow us to arrive at the same results as they do under weaker
conditions: First, our theory imposes minimum requirements on the
data--generating process with the main assumption being that it is local
stationary. Second, it imposes weaker restrictions on the bandwidth used in
the estimation; in particular, we can allow for the bandwidth being chosen
using standard bandwidth selection rules, such as cross--validation, which
is not the case in \citet{dahlhaus2017} and \citet{bardet2022}. Second, we
characterise the leading bias term of the estimators, which is in contrast
to \citet{dahlhaus2017} and \citet{bardet2022}. To demonstrate these
attractive features of our general theory, we apply it to a class of Markov
models with exogenous co--variates.

Another important feature of our theory is that it also applies to
discrete-valued time series models, such as Poisson autoregressions. Such
models are not covered by the theories of \citet{dahlhaus2017} and %
\citet{bardet2022} since these require the model of interest to be smooth;
this condition is not satisfied when the time series is discrete--valued. In
contrast, our theory for local linear estimators impose very weak smoothness
conditions on the model due to our new proof techniques and so applies to
discrete--valued time series models without any modifications. For the local
constant estimator, we combine the ideas of \citet{Truquet2019}, who analyse
time--varying discrete-valued time series models, with our main result to
obtain a complete analysis of this estimator.

We also contribute to the literature on asymptotic analysis of local
polynomial estimators of varying-coefficient models by extending existing
results \citep[see, e.g.,][]{fan1995JASA,loader2006} to cover situations
where the objective functions are non-concave. This proves to be a
non--trivial extension, but at the same time an important one since the
log--likelihood functions of many non-linear models are non-concave.

As an empirical application, we revisit the empirical study of \cite%
{agosto2016} where a Poisson autoregressive model with additional
co--variates was used to model and analyze US defaults. Using the proposed
methodology, we find substantial time--variation in the model parameters
that the original study was unable to capture. In particular, we find that
the ability of macroeconomic and financial variables to predict defaults
have varied substantially over time. A battery of informal tests of the
time--varying model against the time--invariant version finds strong support
for the former.

The remainder of the paper is organized as follows: Framework and estimators
are introduced in Section \ref{sec: model}. Section \ref{sec: asymptotics}
presents the asymptotic theory of the estimators. Section \ref{sec: ex}
provides examples of the theory when applied to particular models. We
present the results of two simulation studies and the empirical application
in Sections \ref{sec: simul} and \ref{sec: empirical}, respectively. All
proofs have been relegated to the Appendix.

\section{Framework\label{sec: model}}

We are given $n$ observations, $Z_{n,t}\in \left( \mathcal{Z},\left\Vert
\cdot \right\Vert \right) $, $t=1,\ldots ,n$, where $\left( \mathcal{Z}%
,\left\Vert \cdot \right\Vert \right) $ is a Banach space, from a
time-series model characterised by a finite--dimensional vector of unknown
parameters $\theta \in \Theta \subset \mathbb{R}^{d_{\theta }}$ to be
estimated. In most applications, $Z_{n,t}$ will also be finite--dimensional
but our theory allows for, e.g., functional data as well. We take as given
an objective function $\ell _{n,t}\left( \theta \right) =\ell \left( 
\mathcal{Z}_{n,t};\theta \right) \in \mathbb{R}$, where $\mathcal{Z}%
_{n,t}=\left( Z_{n,t},,....Z_{n,0},Z_{n,-1},....\right) $. Since we do not
observe the process before $t=1$, we here initialise the process at
deterministic values chosen by us, $Z_{n,-t}=z_{-t}$, $t\geq 1$. Under
regularity conditions stated below, the effect of the initial values will
vanish asymptotically.

The objective function is assumed to identify the data--generating parameter
as its maximiser, $\theta =\arg \max_{\theta ^{\prime }\in \Theta }\mathbb{E}%
\left[ \ell _{n,t}\left( \theta ^{\prime }\right) \right] $, if $\theta $
indeed was time--invariant and $\ell _{n,t}\left( \theta ^{\prime }\right) $
was stationary and ergodic. In this case, the natural estimator is to
replace population expectations by sample ones and estimate $\theta $ by $%
\hat{\theta}=\arg \max_{\theta ^{\prime }\in \Theta }\frac{1}{n}%
\sum_{t=1}^{n}\ell _{n,t}\left( \theta ^{\prime }\right) $.

Suppose now that in fact $\theta $ is varying over time so that $\mathcal{Z}%
_{n,t}$ is generated by $\theta _{n,t}=\theta (t/n)$, $t=1,...,n$, where $%
\theta :\left[ 0,1\right] \mapsto \Theta $ is an unknown function that
characterizes the time--variation in the parameters.\footnote{%
Data points now depends on sample size $n$ through $\theta \left( t/n\right) 
$, which is why we write $Z_{n,t}$ instead of of $Z_{t}$.} At the same time,
the objective function is still assumed to identify the parameter in the
sense that $\theta \left( t/n\right) =\arg \max_{\theta ^{\prime }\in \Theta
}\mathbb{E}\left[ \ell _{n,t}\left( \theta ^{\prime }\right) \right] $. We
then propose to estimate $\theta \left( u\right) $ at any given value $u\in %
\left[ 0,1\right] $ using local polynomial estimators: First, for $t/n$ in a
neighbourhood of $u$, we approximate $\theta \left( t/n\right) $ by the
following polynomial of order $m\geq 0$, 
\begin{equation}
\theta _{u,\beta }^{\ast }\left( t/n\right) :=\beta _{1}+\beta _{2}\left(
t/n-u\right) +\cdots +\beta _{m+1}\left( t/n-u\right) ^{m}/m!=D_{m}\left(
t/n-u\right) \beta ,  \label{eq: theta* def}
\end{equation}%
where $\beta =\left( \beta _{1}^{\prime },...,\beta _{m+1}^{\prime }\right)
^{\prime }\in \mathbb{R}^{\left( m+1\right) d_{\theta }}$ with $\beta
_{i+1}=\theta ^{\left( i\right) }\left( u\right) =\partial ^{i}\theta \left(
u\right) /\partial u^{i}\in \mathbb{R}^{d_{\theta }}$ and 
\begin{equation*}
D_{m}\left( u\right) =\left( 1,u,u^{2}/2\ldots ,u^{m}/m!\right) \otimes
I_{d_{\theta }}\in \mathbb{R}^{d_{\theta }\times \left( m+1\right) d_{\theta
}}.
\end{equation*}%
Next, to control the approximation error, $\theta \left( t/n\right) -\theta
_{u,\beta }^{\ast }\left( t/n\right) $, we introduce a kernel weighted
version of the "global" objective function evaluated at the polynomial
approximation, 
\begin{equation*}
L_{n}\left( \beta |u\right) =\frac{1}{n}\sum_{t=1}^{n}K_{b}\left(
t/n-u\right) \ell _{n,t}\left( \theta _{u,\beta }^{\ast }\left( t/n\right)
\right) ,
\end{equation*}%
where $K_{b}\left( \cdot \right) =K\left( \cdot /b\right) /b$, $K:\mathbb{R}%
\mapsto \mathbb{R}$ is a kernel function, and $b=b_{n}>0$ a bandwidth. The
kernel weights ensure that when $t/n-u$ is "large", the corresponding
observations are down weighted in the estimation, thereby controlling for
the aforementioned approximation error. We then estimate the polynomial
coefficients by%
\begin{equation}
\hat{\beta}\left( u\right) =\arg \max_{\beta \in \mathcal{B}}L_{n}\left(
\beta |u\right) ,  \label{eq: beta-hat def}
\end{equation}%
where%
\begin{equation*}
\mathcal{B}=\left\{ \beta \in \mathbb{R}^{\left( m+1\right) d_{\theta
}}:\theta _{u,\beta }^{\ast }\left( v\right) \in \Theta \text{ for }v\in %
\left[ 0,1\right] \right\} .
\end{equation*}%
The estimated $\beta $ coefficients are used as estimates of $\theta \left(
u\right) $ and its first $m$ derivatives, $\hat{\theta}^{\left( i\right)
}\left( u\right) =\hat{\beta}_{i+1}\left( u\right) $, $i=0,...,m$. When $m=0$%
, we recover the standard local-constant estimator. The above class of
estimators is similar to the ones considered in \citet{fan1995JASA} for
so--called varying--coefficient models, except that we consider time series
models with the "regressor" that we smooth over being normalized time, $t/n$%
, and do not restrict $\theta \mapsto \ell _{n,t}\left( \theta \right) $ to
be convex.

The choice of the order of the polynomial, $m$, should reflect the degree of
smoothness that we are willing to assume $u\mapsto \theta \left( u\right) $
has. If $\theta \left( u\right) $ is $m$ times differentiable, then we
should use this $m$ in the estimation for optimal control of the bias in the
nonparametric estimation. On the other hand, increasing the order of the
polynomial tend to increase the variability of the resulting estimator,
since more local parameters are introduced in the estimation. For a further
dicussion of this issue, we refer the reader to Section 3.3 of %
\citet{Fan2018}.

Our framework includes Markov processes and stochastic processes with
infinite memory \citep[see, e.g., ][]{doukhan2008,bardet2022} as special
cases. In Section \ref{sec: ex}, we apply our general theory to the
following class of Markov models with exogenous co--variates, 
\begin{equation}
Y_{n,t}=G\left( Y_{n,t-1},...,Y_{n,t-q},X_{n,t},X_{n,t-1},\varepsilon
_{t};\theta \left( t/n\right) \right) ,  \label{eq: model inf memory}
\end{equation}%
where $G:\mathcal{Y}^{q}\times \mathcal{X}\times \mathcal{E}\times \Theta $
is a known function, $X_{n,t-1}$ is a vector of exogenous co--variates and $%
\varepsilon _{t}$ is a sequence of errors. For a given specification of $G$
and the distribution of $\varepsilon _{t}$, we can then derive the
corresponding log--likelihood for the model with time--invariant parameters, 
$\ell _{n,t}\left( \theta \right) =\ell \left( Z_{n,t},X_{n,t-1};\theta
\right) $, where $Z_{n,t}=\left( Y_{n,t},X_{n,t}\right) $. Below, we provide
three examples of models that our theory applies to:

\begin{example}
\label{exa: tv-TAR} Suppose $Y_{n,t}\in \mathbb{R}$ solves the following
time-varying threshold autoregressive with exogenous covariates (tv-TAR-X)
model with two regimes,%
\begin{equation}
Y_{n,t}=\omega +\sum_{i=1}^{q}\alpha _{1,i}\left( t/n\right)
Y_{n,t-i}^{+}+\sum_{i=1}^{q}\alpha _{2,i}\left( t/n\right)
Y_{n,t-1}^{-}+\gamma \left( t/n\right) ^{\prime }X_{n,t-1}+\varepsilon _{t},
\label{eq: tvTAR1}
\end{equation}%
where $y^{+}:=\max \left\{ y,0\right\} $ and $y^{-}:=\min \left\{
y,0\right\} $. Here, $X_{n,t-1}$ contains additional predictors and $%
\varepsilon _{t}$ is i.i.d. with $\mathbb{E}\left[ \varepsilon _{t}\right]
=0 $ and $\mathbb{E}\left[ \varepsilon _{t}^{2}\right] <\infty $. A natural
estimator of $\theta =\left( \omega ,\alpha _{1}^{\prime },\alpha
_{2}^{\prime },\gamma ^{\prime }\right) ^{\prime }$ is the least-squares one
so that $\ell _{n,t}\left( \theta \right) =-\left( Y_{n,t}-\omega
-\sum_{i=1}^{q}\alpha _{1,i}Y_{n,t-1}^{+}-\sum_{i=1}^{q}\alpha
_{2,i}Y_{n,t-1}^{-}-\gamma ^{\prime }X_{n,t-1}^{-}\right) ^{2}$.
\end{example}

\begin{example}
\label{exa: tv-ARCH}Suppose $Y_{n,t}\in \mathbb{R}_{+}$ solves the following
time--varying ARCH model with covariates (tv-ARCH-X),%
\begin{equation}
Y_{n,t}=\lambda _{n,t}\left( \theta \left( t/n\right) \right) \varepsilon
_{t}^{2},\text{ \ \ }\lambda _{n,t}\left( \theta \right) =\omega
+\sum_{i=1}^{q}\alpha _{i}Y_{n,t-1}+\gamma ^{\prime }X_{n,t-1},
\label{eq: tv-ARCH}
\end{equation}%
where $\theta =\left( \omega ,\alpha ^{\prime },\gamma ^{\prime }\right)
^{\prime }$, $\varepsilon _{t}$ is i.i.d. with $\mathbb{E}\left[ \varepsilon
_{t}^{2}\right] =1$ and $X_{n,t-1}$ contains additional predictors. The
Gaussian log-likelihood function takes the form $\ell _{n,t}\left( \theta
\right) =-Y_{n,t}/\lambda _{n,t}\left( \theta \right) -\log \left( \lambda
_{n,t}\left( \theta \right) \right) $, where $\theta =\left( \omega ,\alpha
^{\prime },\gamma ^{\prime }\right) ^{\prime }$.
\end{example}

\begin{example}
\label{exa: tv-PARX} Suppose $Y_{n,t}\in \mathbb{Z}_{+}$ solves the
following time-varying Poisson autoregression with exogenous covariates
(tv-PARX),%
\begin{equation}
Y_{n,t}|\mathcal{F}_{n,t-1}\sim \mathrm{Poisson}\left( \lambda _{n,,t}\left(
\theta \left( t/n\right) \right) \right) ,\text{ \ \ }\lambda _{n,t}\left(
\theta \right) =\omega +\sum_{i=1}^{q}\alpha _{i}Y_{n,t-1}+\gamma ^{\prime
}X_{n,t-1},  \label{eq: tv-PARX}
\end{equation}%
where $\mathrm{Poisson}\left( \lambda \right) $ denotes the Poisson
distribution with intensity parameter $\lambda $, $\theta =\left( \omega
,\alpha ^{\prime },\gamma ^{\prime }\right) ^{\prime }$ and $X_{n,t-1}$
contains additional predictors. The log-likelihood function takes the form $%
\ell _{n,t}\left( \theta \right) =Y_{n,t}\log \left( \lambda _{n,t}\left(
\theta \right) \right) -\lambda _{n,t}\left( \theta \right) $.
\end{example}

\section{Asymptotic theory\label{sec: asymptotics}}

We here provide an asymptotic theory for $\hat{\beta}$. One complication of
this analysis is that the components of $\hat{\beta}$ converge with
different rates. We follow the existing literature and handle this issue by
introducing a re--scaled version of $\hat{\beta}$; see, e.g., \citet{han2014}
for a similar approach: Define $\hat{\alpha}=U_{n}\hat{\beta}=(\hat{\theta}%
\left( u\right) ^{\prime },b\hat{\theta}^{\left( 1\right) }\left( u\right)
^{\prime },...,b^{m}\hat{\theta}^{\left( m\right) }\left( u\right) ^{\prime
})^{\prime }$, where 
\begin{equation*}
U_{n}=diag\left\{ 1,b,...,b^{m}\right\} \otimes I_{d_{\theta }}\in \mathbb{R}%
^{\left( m+1\right) d_{\theta }\times \left( m+1\right) d_{\theta }}
\end{equation*}%
is a weighting matrix containing their relative convergence rates, Given
that $U_{n}$ is non-singular, the estimation problem (\ref{eq: beta-hat def}%
) is equivalent to solving%
\begin{equation}
\hat{\alpha}=\arg \max_{\alpha \in \mathcal{A}}Q_{n}\left( \alpha |u\right)
,\quad Q_{n}\left( \alpha |u\right) =\frac{1}{n}\sum_{t=1}^{n}K_{b}\left(
t/n-u\right) \ell _{n,t}\left( D_{m,b}\left( t/n-u\right) \alpha \right) ,
\label{eq: alpha-hat def}
\end{equation}%
where $D_{m,b}\left( u\right) =D_{m}\left( u/b\right) $ and 
\begin{equation}
\mathcal{A}=\left\{ \alpha \in \mathbb{R}^{\left( m+1\right) d_{\theta
}}:D_{m}\left( v\right) \alpha \in \Theta ,\forall v\in \mathcal{K}\right\}
\label{eq: A def}
\end{equation}
with $\mathcal{K}$ denoting the support of $K$. We will then analyze the
properties of $\hat{\alpha}$.

Due to the time--varying parameters, $Z_{n,t}$ will generally be
non--stationary. To develop an asymptotic theory that allows for this
feature, we will rely on the concept of local stationarity as introduced by %
\citet{dahlhaus1997}; see also \citet{dahlhaus2006} and \citet{dahlhaus2017}%
. We first generalize this concept to sequences of random functions:

\begin{definition}
\label{def: LS}A triangular family of random sequences $W_{n,t}\left( \theta
\right) $, $\theta \in \Theta $, $t=1,2,...,n$ and $n\geq 1$, is uniformly
locally stationary on $\Theta $ (ULS$\left( p,q,\Theta \right) $) for some $%
p,q>0$ if there exists a family of processes $W_{t}^{\ast }\left( \theta
|u\right) $, $u\in \left[ 0,1\right] $, such that: (i) The process $\left\{
W_{t}^{\ast }\left( \theta |u\right) \right\} $ is stationary and ergodic
for all $\left( \theta ,u\right) \in \Theta \times \left[ 0,1\right] $; (ii)
for some $C<\infty $ and $\rho <1$,%
\begin{equation*}
\mathbb{E}\left[ \sup_{\theta \in \Theta }\left\Vert W_{n,t}\left( \theta
\right) -W_{t}^{\ast }\left( \theta |u\right) \right\Vert ^{p}\right]
^{1/p}\leq C\left( \left\vert \frac{t}{n}-u\right\vert ^{q}+\frac{1}{n^{q}}%
+\rho ^{t}\right) .
\end{equation*}
\end{definition}

If $W_{n,t}\left( \theta \right) =W_{n,t}$ does not depend on any
parameters, we write LS$\left( p,q\right) $. The above condition states that 
$W_{n,t}\left( \theta \right) $ may be non--stationary, but it is locally in
time well--approximated by a stationary version $W_{t}^{\ast }\left( \theta
|u\right) $. Compared to existing definitions of local stationarity, we
allow for an additional term $\rho ^{t}$ to appear in the approximation
error. This is needed in order to allow for the initial value of $%
W_{n,t}\left( \theta \right) $ to be chosen arbitrarily. In contrast, by not
including $\rho ^{t}$ in their defintions, most of the existing literature
implicitly assumes that $W_{n,t}\left( \theta \right) $ has been initialised
at $W_{n,0}\left( \theta \right) =W_{0}^{\ast }\left( \theta |u\right) $.
When used in the analysis of local estimators, this latter definition
implicitly requires that the data-generating process changes as the
researcher varies $u$ in the local log-likelihood which is a rather peculiar
assumption. In contrast, the above definition allows for $W_{n,0}\left(
\theta \right) $ to be initialized at a given fixed value -- as long as the
impact of this dies out with rate $\rho $.

For an example of how the additional error term appears in autoregressive
models, we refer the reader to the proof of Lemma \ref{lem: MarkovX LS} in
Section \ref{sec: ex} which allows for arbitrary initialisation of the
data-generating process. The additional error term due to different
initializations is here assumed to decay geometrically and so our definition
rules out long-memory type processes. This is mostly for simplicity and we
expect that most of our results can be generalized to allow for slower decay
rates.

We will then require that $\ell _{n,t}\left( \theta \right) $ is ULS$\left(
p,q,\Theta \right) $ with stationary approximation $\ell _{t}^{\ast }\left(
\theta |u\right) =\ell \left( \mathcal{Z}_{t}^{\ast }\left( u\right) ,\theta
\right) $ where $\mathcal{Z}_{t}^{\ast }\left( u\right) =\left( Z_{t}^{\ast
}\left( u\right) ,Z_{t-1}^{\ast }\left( u\right) ,....\right) $ is the
stationary solution to the model being estimated when $\theta _{n,t}=\theta
\left( u\right) $ is constant. To illustrate, consider again (\ref{eq: model
inf memory}). Under regularity conditions on $G$ and $\varepsilon _{t}$ (see
Section \ref{sec: ex} for details), the stationary solution will in this
case take the form 
\begin{equation*}
Y_{t}^{\ast }\left( u\right) =G\left( Y_{t-1}^{\ast }\left( u\right)
,...,Y_{t-q}^{\ast }\left( u\right) ,X_{t-1}^{\ast }\left( u\right)
,\varepsilon _{t};\theta \left( u\right) \right) ,
\end{equation*}%
where we impose the high--level condition that the exogenous co--variates
are locally stationary. If the data-generating process is locally
stationary, it follows under great generality that the likelihood and its
derivatives are also locally stationary, c.f. Section \ref{sec: ex}.

The next step in our proof is to establish a uniform law of large numbers
(ULLN) for the stationary approximation of $Q_{n}\left( \alpha |u\right) $, $%
Q_{n}^{\ast }\left( \alpha |u\right) =\frac{1}{n}\sum_{t=1}^{n}K_{b}\left(
t/n-u\right) \ell _{t}^{\ast }\left( D_{b}\left( t/n-u\right) \alpha
|u\right) $. A sufficient condition for a ULLN to hold is that $\theta
\mapsto \ell _{n,t}^{\ast }\left( \theta |u\right) $ is $L_{p}$\textit{%
-continuous}:

\begin{definition}
\label{Def: L_p-cont}A stationary process $W_{t}^{\ast }\left( \theta
|u\right) $ is said to be $L_{p}$\textit{-continuous} w.r.t. $\theta $ if $%
\mathbb{E}\left[ \left\Vert W_{t}^{\ast }\left( \theta |u\right) \right\Vert
^{p}\right] <\infty $ for all $\theta \in \Theta $ and 
\begin{equation*}
\forall \epsilon >0\exists \delta >0:\mathbb{E}\left[ \sup_{\theta ^{\prime
}:\left\Vert \theta -\theta ^{\prime }\right\Vert <\delta }\left\Vert
W_{t}^{\ast }\left( \theta |u\right) -W_{t}^{\ast }\left( \theta ^{\prime
}|u\right) \right\Vert ^{p}\right] ^{1/p}<\epsilon .
\end{equation*}
\end{definition}

Imposing $L_{p}$-continuity w.r.t. $\theta $ is weaker than almost surely
continuity: If $\theta \mapsto W_{t}^{\ast }\left( \theta |u\right) $ is
almost surely continuous with $\mathbb{E}\left[ \sup_{\theta \in \Theta
}\left\Vert W_{t}^{\ast }\left( \theta |u\right) \right\Vert ^{p}\right]
<\infty $ the process is also $L_{p}$-continuous since $DW_{t}(\delta
)=\sup_{\Vert \theta -\theta ^{\prime }\Vert \leq \delta }\left\Vert
W_{t}^{\ast }\left( \theta |u\right) -W_{t}^{\ast }\left( \theta ^{\prime
}|u\right) \right\Vert ^{p}$, $\delta >0$, will then satisfy $\lim_{\delta
\rightarrow 0}DW_{t}(\delta )=0$ almost surely and so, by dominated
convergence, $\lim_{\delta \rightarrow 0}\mathbb{E}[DW_{t}(\delta )]=0$. It
is easily verified that $L_{p}$-continuity w.r.t. $\theta $ implies
stochastic equicontinuity of $Q_{n}^{\ast }\left( \alpha |u\right) $ and so
a ULLN holds, c.f. Lemma \ref{thm: Aux result}(i) in Appendix \ref{subsec:
lemma}.

We are now ready to state the regularity conditions used to show consistency:

\begin{assumption}
\label{assu: kernel} (i) $K\left( \cdot \right) \geq 0$ has compact support $%
\mathcal{K}$ and $\int_{-\infty }^{+\infty }K\left( v\right) dv=1$; (ii) for
some $\Lambda <\infty $, $\left\vert K(v)-K(\tilde{v})\right\vert \leq
\Lambda \left\vert v-\tilde{v}\right\vert $, $v,\tilde{v}\in \mathbb{R}$;
(iii) $v\mapsto \theta \left( v\right) $ is continuous at $u$.
\end{assumption}

\begin{assumption}
\label{assu: compactness} (i) $\Theta $ is compact and the true value $%
\theta \left( u\right) \in \Theta $; (ii) $\theta \mapsto \ell _{t}^{\ast
}\left( \theta |u\right) $ is $L_{p}$\textit{-}continuous; (iii) $\theta
\mapsto \mathbb{E}\left[ \ell _{t}^{\ast }\left( \theta |u\right) \right] $
has a unique maximum at $\theta \left( u\right) \in \Theta $.
\end{assumption}

\begin{assumption}
\label{assu: LS}$\ell _{n,t}\left( \theta \right) $ is ULS$\left( p,q,\Theta
\right) $ for some $p\geq 1$ and $q>0$ with stationary approximation $\ell
_{t}^{\ast }\left( \theta |u\right) $;
\end{assumption}

Assumption \ref{assu: kernel}(i) imposes stronger than usual assumptions on $%
K$ and excludes, among others, the Gaussian kernel and higher-order kernels.
It includes, on the other hand, the Epanechnikov and the triangular kernel.
The restriction that $K\left( \cdot \right) \geq 0$ is used to ensure
identification of the parameters when $m>0$; without this, identification is
not necessarily guaranteed; see below for further discussion. For the
analysis of the local constant estimator ($m=0$), all subsequent results
will go through with $K$ having full support and taking negative values.

The compact support assumption greatly simplifies our analysis of local
polynomial estimation of non-concave models: In order to establish uniform
convergence of the likelihood we require $\Theta $ to be compact as is
standard in the literature. But under this restriction, it is easily checked
that $D_{m,b}\left( v\right) \alpha \notin \Theta $ as $b\rightarrow 0$ for
any given $\alpha =\left( \alpha _{1},...,\alpha _{m+1}\right) $ with $%
\alpha _{i}\neq 0$ for some $i\geq 1$ and any $v\neq 0$. Thus, to allow for
kernels with unbounded support, we would generally need the parameter space $%
\mathcal{A}$, as defined in (\ref{eq: A def}), to collapse at $\left\{
\left( \alpha _{1},0,...,0\right) :\alpha _{1}\in \Theta \right\} $ as $%
b\rightarrow 0$. Such shrinking behaviour in turn means that a formal Taylor
expansion of $\ell _{n,t}\left( D_{m,b}\left( v\right) \alpha \right) $
w.r.t. $\alpha $ is difficult to obtain and so standard arguments to
establish asymptotic normality of $\hat{\alpha}$ cannot be applied. On the
other hand, by restricting the support $\mathcal{K}$ to be compact, $%
K_{b}\left( v\right) \ell _{n,t}\left( D_{m,b}\left( v\right) \alpha \right) 
$ is well-defined for all $\alpha \in \mathcal{A}$ and $v\in \mathbb{R}$
(where we set $K_{b}\left( v\right) \ell _{n,t}\left( D_{m,b}\left( v\right)
\alpha \right) =0$ for $v/b\notin \mathcal{K}$). Moreover, $\left( \alpha
_{1},0,...,0\right) $ is an interior point of $\mathcal{A}$ and so in our
analysis of $\hat{\alpha}$ we can employ standard arguments involving a
Taylor expansion of the score function around this point.

Assumption \ref{assu: compactness} is standard in the analysis of
\textquotedblleft global\textquotedblright\ extremum estimators of
stationary models on the form $\tilde{\theta}\left( u\right) =\arg
\max_{\theta \in \Theta }\sum_{t=1}^{n}\ell _{t}^{\ast }\left( \theta
|u\right) /n$. In particular, for a given time series model, we can import
existing results for verification of Assumption \ref{assu: compactness}%
(ii)-(iii); see Section \ref{sec: ex} for more details. Assumption \ref%
{assu: compactness} in conjunction with $K\left( \cdot \right) \geq 0$
ensures that the local polynomial estimator identifies $\theta \left(
u\right) $. If we allow for kernels that take negative values, we have to
replace \ref{assu: compactness}(iii) with the following more abstract
identification condition: The function $Q^{\ast }\left( \alpha |u\right)
=\int K\left( v\right) \mathbb{E}\left[ \ell _{t}^{\ast }\left( D_{m}\left(
v\right) \alpha |u\right) \right] dv$ satisfies $Q^{\ast }\left( \alpha
|u\right) <Q^{\ast }\left( \left( \theta \left( u\right) ,0,...,0\right)
|u\right) $ for any $\alpha \neq \left( \theta \left( u\right)
,0,...,0\right) $. We have not been able to provide primitive conditions for
this to hold when $K$ can take negative values and so instead impose the
positivity constraint on $K$.

If the objective function $\theta \mapsto \ell _{n,t}\left( \theta \right) $
is concave and $\Theta $ is convex, we can replace Assumption \ref{assu: LS}
with the following pointwise versions: For any $\theta \in \Theta $, $\ell
_{n,t}\left( \theta \right) $ is locally stationary and $\mathbb{E}\left[
|\ell _{t}^{\ast }\left( \theta |u\right) |\right] <\infty $; see Theorem
2.7 in \citet{newey1994}. Under the above assumptions, the following
consistency result holds:

\begin{theorem}
\label{thm: Con} Let Assumptions \ref{assu: kernel}-\ref{assu: LS} hold.
Then, as $b\rightarrow 0$ and $nb\rightarrow \infty ,$ $\hat{\alpha}%
\rightarrow ^{p}\left( \theta \left( u\right) ,0,....,0\right) ^{\prime }$.
In particular, $\hat{\theta}\left( u\right) \rightarrow ^{p}\theta \left(
u\right) $.
\end{theorem}

Note that the above theorem only shows consistency of $\hat{\theta}\left(
u\right) $ and so at this stage we cannot make any statements regarding $%
\hat{\theta}^{\left( i\right) }\left( u\right) $, $i=1,...,m$. This is
similar to other results for nonlinear extremum estimators that converge
with different rates; see, e.g., Theorem 9 in \citet{han2014} where a global
consistency result is only provided for the component with the fastest rate.

However, under additional regularity conditions on the quasi-likelihood
function, we can provide a more precise analysis of the estimators,
including local consistency of $\hat{\theta}^{\left( k\right) }\left(
u\right) $, $1\leq k\leq m$. With $s_{n,t}\left( \theta \right) =\partial
\ell _{n,t}\left( \theta \right) /\left( \partial \theta \right) \in \mathbb{%
R}^{d_{\theta }}$ and $h_{n,t}\left( \theta \right) =\partial ^{2}\ell
_{n,t}\left( \theta \right) /(\partial \theta \partial \theta ^{^{\prime
}})\in \mathbb{R}^{d_{\theta }\times d_{\theta }}$, $D_{n,t}\left( u\right)
=D_{m,b}\left( t/n-u\right) $ and $K_{n,t}\left( u\right) =K_{b}(t/n-u)$,
the score and hessian of $Q_{n}\left( \alpha |u\right) $ are given by 
\begin{align*}
S_{n}\left( \alpha |u\right) & =\frac{\partial Q_{n}\left( \alpha |u\right) 
}{\partial \alpha }=\frac{1}{n}\sum_{t=1}^{n}K_{n,t}\left( u\right)
D_{n,t}\left( u\right) ^{\prime }s_{n,t}\left( D_{n,t}\left( u\right) \alpha
\right) , \\
H_{n}\left( \alpha |u\right) & =\frac{\partial ^{2}Q_{n}\left( \alpha
|u\right) }{\partial \alpha \partial \alpha ^{\prime }}=\frac{1}{n}%
\sum_{t=1}^{n}K_{n,t}\left( u\right) D_{n,t}\left( u\right) ^{\prime
}h_{n,t}\left( D_{n,t}\left( u\right) \alpha \right) D_{n,t}\left( u\right) .
\end{align*}%
It is easily checked that $\alpha _{0}:=U_{n}\beta _{0}$, where $\beta
_{0}=(\theta \left( u\right) ^{\prime },\theta ^{\left( 1\right) }\left(
u\right) ^{\prime },...,\theta ^{\left( m\right) }\left( u\right) ^{\prime
})^{\prime }$, belongs to the interior of $\mathcal{A}$ for all $n$ large
enough due to Assumption \ref{assu: smoothness}(ii) below in conjunction
with Assumption \ref{assu: compactness}. Due to the consistency result, so
will $\hat{\alpha}$ w.p.a. 1. Thus, $\hat{\alpha}$ will satisfy the
first-order condition of (\ref{eq: alpha-hat def}) which combined with the
mean-value theorem yields 
\begin{equation}
0=S_{n}\left( \hat{\alpha}|u\right) =S_{n}\left( \alpha _{0}|u\right)
+H_{n}\left( \bar{\alpha}|u\right) \left( \hat{\alpha}-\alpha _{0}\right)
=S_{n}\left( \alpha _{0}|u\right) +H_{n}\left( \bar{\alpha}|u\right) U_{n}(%
\hat{\beta}-\beta _{0}),  \label{eq: FOC}
\end{equation}%
where $\bar{\alpha}$ is situated on the line segment connecting $\hat{\alpha}
$ and $\alpha _{0}$. We then decompose the score function into a bias and
variance component, $S_{n}\left( \alpha _{0}|u\right) =B_{n}\left( u\right)
+S_{n}\left( u\right) $, where 
\begin{equation}
B_{n}\left( u\right) =\frac{1}{n}\sum_{t=1}^{n}K_{n,t}\left( u\right)
D_{n,t}\left( u\right) ^{\prime }b_{n,t},\;S_{n}\left( u\right) =\frac{1}{n}%
\sum_{t=1}^{n}K_{n,t}\left( u\right) D_{n,t}\left( u\right) ^{\prime
}s_{n,t},  \label{eq: Bn def}
\end{equation}%
$s_{n,t}=s_{n,t}\left( \theta \left( t/n\right) \right) $, and $%
b_{n,t}=s_{n,t}\left( \theta _{u}^{\ast }\left( t/n\right) \right)
-s_{n,t}\left( \theta \left( t/n\right) \right) $ with $\theta _{u}^{\ast
}\left( t/n\right) $ defined in eq. (\ref{eq: theta* def}). This
decomposition is different from the one usually employed in the analysis of
kernel estimators of time-varying coefficients where $s_{n,t}\left( \theta
\left( t/n\right) \right) $ is replaced by the stationary version of the
score function evaluated at $\theta \left( u\right) $, $s_{t}^{\ast }\left(
\theta \left( u\right) |u\right) $; see, e.g., \citet{dahlhaus2017} and %
\citet{dahlhaus2006}. This "usual" choice has as consequence that the
corresponding bias term in generally involves the time derivative process of
the score function and so the resulting analysis tends to impose stronger
regularity conditions on the model being estimated. By instead centering the
analysis around $s_{n,t}$, we can obtain the leading term of the bias $%
B_{n}\left( u\right) $ through a standard Taylor expansion w.r.t. $\theta $, 
\begin{equation}
b_{n,t}\cong h_{n,t}\left( \theta _{u}^{\ast }\left( t/n\right) \right)
\left\{ \theta _{u}^{\ast }\left( t/n\right) -\theta \left( t/n\right)
\right\} \cong -h_{n,t}\left( \theta \left( u\right) \right) \frac{\theta
^{\left( m+1\right) }\left( u\right) }{\left( m+1\right) !}\left\{
t/n-u\right\} ^{m+1}.  \label{eq: bias expansion}
\end{equation}%
Thus, our approach allows for a simpler derivation of the leading bias and
variance terms under the following weak regularity conditions, where here
and in the following $p$ and $q$ have to satisfy $p\geq 1$ and $q>0$, but
can otherwise vary depending on the particular application.

\begin{assumption}
\label{assu: smoothness}(i) $\theta \mapsto \ell _{n,t}\left( \theta \right) 
$ is twice continuously differentiable; (ii) $\theta \left( u\right) $ lies
in the interior of $\Theta $ and is $m+1$ times continuously differentiable;
(iii) $s_{n,t}$ is a martingale difference (MGD) array w.r.t. $\mathcal{F}%
_{n,t}=\mathcal{F}\left\{ Z_{n,t},Z_{n,t-1},\ldots \right\} $; (iv) $H\left(
u\right) \equiv \mathbb{E}\left[ h_{t}^{\ast }\left( \theta (u)|u\right) %
\right] $ is non-singular.
\end{assumption}

\begin{assumption}
\label{assu: score}(i) $\omega _{n,t}\left( \theta \right) =s_{n,t}\left(
\theta \right) s_{n,t}\left( \theta \right) ^{\prime }$ is ULS$\left( p,q,%
\mathcal{N}\left( u,\epsilon \right) \right) $, where, for some arbitrarily
small $\epsilon >0$, $\mathcal{N}\left( u,\epsilon \right) :=\left\{ \theta
\in \Theta :\left\Vert \theta -\theta \left( u\right) \right\Vert <\epsilon
\right\} $, and $\omega _{t}^{\ast }\left( \theta |u\right) $ is $L_{1}$%
-continuous at $\theta =\theta \left( u\right) $; (ii) $h_{n,t}\left( \theta
\right) $ is ULS$\left( p,q,\mathcal{N}\left( u,\epsilon \right) \right) $
with continuous stationary approximation \textup{$h_{t}^{\ast }\left( \theta
\left( u\right) \right) $} and
\end{assumption}

Similar to Assumption \ref{assu: compactness}, Assumption \ref{assu:
smoothness} contains standard regularity conditions used in the analysis of
regular parametric estimators on the form $\tilde{\theta}\left( u\right)
=\arg \max_{\theta \in \Theta }\sum_{t=1}^{n}\ell _{t}^{\ast }\left( \theta
|u\right) $. At the same time, Assumption \ref{assu: smoothness}(iii) is
non-standard compared to the existing literature (as discussed above) and,
together with \ref{assu: score}(i), allows us to apply a novel martingale
central limit theorem for locally stationary sequences to $S_{n}\left(
u\right) $, 
\begin{equation}
\sqrt{nb}S_{n}\left( u\right) \rightarrow ^{d}N\left( 0,\mathbb{K}%
_{2}\otimes \Omega \left( u\right) \right) ,\;\Omega \left( u\right) =%
\mathbb{E}\left[ \omega _{t}^{\ast }\left( \theta \left( u\right) |u\right) %
\right] ;  \label{eq: as norm 1}
\end{equation}%
see Lemma \ref{thm: Aux result}(iii) in Appendix \ref{subsec: lemma}. This
result can be seen as a generalisation of the standard CLT for stationary
and ergodic MGD's that allows for locally stationary processes. The MGD
assumption amounts to assuming that the time-varying model is correctly
specified and has to be verified on a case-by-case basis.

Finally, Assumption \ref{assu: score}(ii) together with the expansion in eq.
(\ref{eq: bias expansion}) is used to derive the limits of $B_{n}\left(
u\right) $ and $H_{n}\left( \bar{\alpha}|u\right) $,%
\begin{equation}
H_{n}\left( \bar{\alpha}|u\right) \rightarrow ^{p}\mathbb{K}_{1}\otimes
H\left( u\right) ,\text{ \ \ }B_{n}\left( u\right) =-b^{m+1}\mu _{1}\otimes
H\left( u\right) \frac{\theta ^{\left( m+1\right) }\left( u\right) }{\left(
m+1\right) !}+o_{P}\left( b^{m+1}\right) ,  \label{eq: as norm 2}
\end{equation}%
where $\mu _{i}=\int_{\mathbb{R}}K\left( v\right) v^{m+i}D_{m}\left(
v\right) dv$ and $\mathbb{K}_{i}=\int_{\mathbb{R}}K^{i}\left( v\right)
D_{m}\left( v\right) D_{m}\left( v\right) ^{\prime }dv$, $i\geq 1$.
Combining (\ref{eq: FOC}), (\ref{eq: as norm 1}) and (\ref{eq: as norm 2}),
we obtain:

\begin{theorem}
\label{thm: Norm1}Suppose that Assumptions \ref{assu: kernel}-\ref{assu:
score} hold. Then, as $b\rightarrow 0$ and $nb\rightarrow \infty $, 
\begin{equation*}
\sqrt{nb}U_{n}\left\{ \hat{\beta}-\beta _{0}-R_{n}Bias\left( u\right)
\right\} \rightarrow ^{d}N\left( 0,\mathbb{K}_{1}^{-1}\mathbb{K}_{2}\mathbb{K%
}_{1}^{-1}\otimes V(u)\right) ,
\end{equation*}%
where $R_{n}=diag\left\{ b^{m+1},b^{m},...,b\right\} \otimes I_{d_{\theta }}$%
, $V(u)=H\left( u\right) ^{-1}\Omega \left( u\right) H\left( u\right) ^{-1}$
and $Bias\left( u\right) =\mathbb{K}_{1}^{-1}\mu _{1}\otimes \frac{\theta
^{\left( m+1\right) }\left( u\right) }{\left( m+1\right) !}$.

In particular, for $i=0,1,...,m$, 
\begin{equation}
\sqrt{nb^{2i+1}}\left\{ \hat{\theta}^{\left( i\right) }\left( u\right)
-\theta ^{\left( i\right) }\left( u\right) -b^{m+1-i}Bias_{i}\left( u\right)
\right\} \rightarrow ^{d}N\left( 0,\kappa _{2,i}V\left( u\right) \right) ,
\label{eq: poly est norm}
\end{equation}%
where $Bias_{i}\left( u\right) =\kappa _{1,i}\frac{\theta ^{\left(
m+1\right) }\left( u\right) }{\left( m+1\right) !}$ and $\kappa _{1,i}$ and $%
\kappa _{2,i}$ denote the $i$th element of $\mathbb{K}_{1}^{-1}\mu _{1}$ and 
$\left( i,i\right) $th element of $\mathbb{K}_{1}^{-1}\mathbb{K}_{2}\mathbb{K%
}_{1}^{-1}$, respectively.
\end{theorem}

Similar to existing results for local polynomial estimators in a
cross-sectional setting, the leading bias term in (\ref{eq: poly est norm})
only depends on $\theta ^{\left( m+1\right) }\left( u\right) $ and so the
estimators adapt to the curvature of $\theta \left( u\right) $. The
asymptotic variance in Theorem \ref{thm: Norm1} can be estimated using
plug-in methods: It follows from the proof of Theorem \ref{thm: Norm1} that 
\begin{equation*}
\hat{W}\left( u\right) =\frac{1}{n}\sum_{t=1}^{n}K_{n,t}^{2}\left( u\right)
D_{n,t}\left( u\right) ^{\prime }s_{n,t}\left( D_{n,t}\left( u\right) \hat{%
\alpha}\right) s_{n,t}\left( D_{n,t}\left( u\right) \hat{\alpha}\right)
^{\prime }D_{n,t}\left( u\right)
\end{equation*}%
satisfies $\hat{W}\left( u\right) \rightarrow ^{p}\mathbb{K}_{2}\otimes
\Omega \left( u\right) $ while $H_{n}\left( \hat{\alpha}|u\right)
\rightarrow ^{p}\mathbb{K}_{1}\otimes H\left( u\right) $.

Compared to most existing asymptotic results for the local constant
estimator, such as \citet{dahlhaus2017}, above result with $m\geq 1$ impose
much weaker restrictions on the bandwidth. In particular, standard bandwidth
selection rules can be employed here but not under most of the existing
theories since their conditions require undersmoothing (that is, $%
b\rightarrow 0$ at a faster rate than the optimal one). This is due to the
fact that these theories do not provide a complete characterisation of the
leading bias term. The few papers that do characterise the leading bias
term, such as \cite{dahlhaus2006}, require the so--called time derivatives
of the stationary score function to exist and be well-behaved since these
enter their bias expressions. Our conditions and results, on the other hand,
do not require these and are analogous to the ones found in the literature
on local polynomial likelihood estimators; see, e.g., Theorem 1b of %
\citet{fan1995JASA}.

Equation (\ref{eq: poly est norm}) holds for any value of $m\geq 0$ and $%
i=0,...,m$. However, if $K$ is symmetric, then $\kappa _{1,i}=0$ when $m-i$
is even. In particular, for the local constant estimator ($m=i=0$), Theorem %
\ref{thm: Norm1} only informs us that the bias component of $\hat{\theta}%
\left( u\right) $ is $o_{p}\left( b\right) $ which is not a sharp rate. To
obtain the leading bias term in the cases where $m-i$ is even, a
higher-order expansion of $b_{n,t}$ in eq. (\ref{eq: Bn def}) is necessary.
This expansion requires additional assumptions involving aforementioned time
derivatives and standard derivatives w.r.t. $\theta $ of $h_{t}^{\ast
}\left( \theta \left( u\right) |u\right) $. To present these, we need the
following additional concept:

\begin{definition}
A stationary process $W_{t}^{\ast }\left( \theta |u\right) $ is said to be $%
L_{p}$\textit{-differentiable} w.r.t. $u$ if there exists a stationary and
ergodic process $\partial _{u}W_{t}^{\ast }\left( \theta |u\right) $ with $%
\mathbb{E}\left[ \left\Vert \partial _{u}W_{t}^{\ast }\left( \theta
|u\right) \right\Vert ^{p}\right] <\infty $ such that 
\begin{equation*}
\mathbb{E}\left[ \left\Vert W_{t}^{\ast }\left( \theta |u+\Delta \right)
-W_{t}^{\ast }\left( \theta |u\right) -\partial _{u}W_{t}^{\ast }\left(
\theta |u\right) \Delta \right\Vert ^{p}\right] ^{1/p}=o\left( \Delta
\right) ,\;\Delta \rightarrow 0.
\end{equation*}
\end{definition}

Our definition of time differentiability is slightly weaker compared to the
one found in \citet{dahlhaus2017} and other papers where differentiability
w.r.t. $u$ has to hold almost surely. With this definition in hand, we are
ready to introduce the following additional regularity conditions in order
to derive the leading bias term when $m-i$ is even:

\begin{assumption}
\label{assu: 3rdderiv} $\partial h_{n,t}\left( \theta \right) /\left(
\partial \theta _{i}\right) $ exists and is ULS$\left( 1,q,\mathcal{N}%
(u,\epsilon )\right) $ with $L_{1}$-continuous stationary approximation $%
\partial h_{t}^{\ast }\left( \theta |u\right) /\left( \partial \theta
_{i}\right) $, $i=1,...,d_{\theta }$.
\end{assumption}

\begin{assumption}
\label{assu: derivprocess}$h_{t}^{\ast }\left( \theta \left( u\right)
|v\right) $ is $L_{1}$-differentiable w.r.t. $v$ at $u$ with time-derivative 
$\partial _{u}h_{t}^{\ast }\left( \theta \left( u\right) |u\right) =\partial
h_{t}^{\ast }\left( \theta |u\right) /\left( \partial u\right) |_{\theta
=\theta \left( u\right) }\in \mathbb{R}^{d_{\theta }\times d_{\theta }}$.
\end{assumption}

\begin{assumption}
\label{assu: hessian shortdep}$\sum_{t=1}^{\infty }\left\vert \mathrm{Cov}%
\left( h_{ij,0}^{\ast }\left( \theta \left( u\right) |u\right)
,h_{ij,t}^{\ast }\left( \theta \left( u\right) |u\right) \right) \right\vert
<\infty $, $i,j=1,...,d_{\theta }.$
\end{assumption}

The time-derivative $\partial _{u}h_{t}^{\ast }\left( \theta |u\right) $
will generally involve time-derivatives of the underlying stationary
approximation of data: If $h_{t}^{\ast }\left( \theta |u\right) =h\left( 
\mathcal{Z}_{t}^{\ast }\left( u\right) ;\theta \right) $ for some function $%
h $ which is differentiable w.r.t. $\mathcal{Z}_{t}^{\ast }\left( u\right) $%
, then it takes the form $\partial _{u}h_{t}^{\ast }\left( \theta |u\right)
=\sum_{i=0}^{\infty }\partial h\left( z_{0},z_{1},z_{2},....;\theta \right)
/\left( \partial z_{i}\right) |_{z=\mathcal{Z}_{t}^{\ast }\left( u\right)
}\times \partial _{u}Z_{t-i}^{\ast }\left( u\right) $, where $\partial
_{u}Z_{i,t}^{\ast }\left( u\right) $ is the time derivative of $Z_{t}^{\ast
}\left( u\right) $. The short memory condition imposed in Assumption \ref%
{assu: hessian shortdep} is used to control the variance component of the
first-order bias term derived in Theorem \ref{thm: Norm1}. In Section \ref%
{sec: ex}, we use the concept of $\tau $--weak dependence \citep{doukhan2008}
to verify Assumption \ref{assu: hessian shortdep}.

Under the above additional assumptions, we obtain the following higher-order
expansion of the bias component:

\begin{theorem}
\label{thm: Norm2} Suppose Assumptions \ref{assu: kernel}-\ref{assu: hessian
shortdep} hold and $\theta \left( \cdot \right) $ is $m+2$ times
continuously differentiable. Then, as $b\rightarrow 0$ and $nb\rightarrow
\infty $, the bias $B_{n}\left( u\right) $ defined in (\ref{eq: Bn def})
satisfies, with $q$ given in Assumption \ref{assu: 3rdderiv},%
\begin{eqnarray}
B_{n}\left( u\right) &=&-b^{m+1}\left[ \mu _{1}\otimes H\left( u\right) 
\frac{\theta ^{\left( m+1\right) }\left( u\right) }{\left( m+1\right) !}%
+o_{P}\left( b\right) \right] -b^{m+2}\left[ \mu _{2}\otimes B_{1}\left(
u\right) +o_{p}\left( 1\right) \right]  \label{eq: Bn expansion} \\
&&-b^{2m+2}\left[ \mu _{2}\otimes B_{2}\left( u\right) +o_{p}\left( 1\right) %
\right] +O_{P}\left( 1/n^{\min \left\{ 1,q\right\} }\right) +o_{p}\left( 
\frac{1}{\sqrt{nb}}\right) ,  \notag
\end{eqnarray}%
where, with $\partial _{u}H\left( u\right) =\mathbb{E}\left[ \partial
_{u}h_{t}^{\ast }\left( \theta |u\right) \right] _{\theta =\theta \left(
u\right) }$ and $\partial _{\theta _{i}}H\left( u\right) =\mathbb{E}\left[
\partial h_{t}^{\ast }\left( \theta |u\right) /\left( \partial \theta
_{i}\right) \right] _{\theta =\theta \left( u\right) }$,%
\begin{align*}
B_{1}\left( u\right) & =H\left( u\right) \frac{\theta ^{\left( m+2\right)
}\left( u\right) }{\left( m+2\right) !}+\left( \partial _{u}H\left( u\right)
+\sum_{i=1}^{d_{\theta }}\theta _{i}^{\left( 1\right) }\left( u\right)
\partial _{\theta _{i}}H\left( u\right) \right) \frac{\theta ^{\left(
m+1\right) }\left( u\right) }{\left( m+1\right) !}, \\
B_{2}\left( u\right) & =\frac{-1}{2\left\{ \left( m+1\right) !\right\} ^{2}}%
\sum_{i=1}^{d_{\theta }}\theta _{i}^{\left( m+1\right) }\left( u\right)
\partial _{\theta _{i}}H\left( u\right) \theta ^{\left( m+1\right) }\left(
u\right) .
\end{align*}
\end{theorem}

As a special case, we obtain the following result for the local constant
estimator:

\begin{corollary}
\label{cor: LC norm}Under the assumptions of Theorem \ref{thm: Norm2} with $%
m=0$ together with $\int_{\mathbb{R}}K\left( v\right) vdv=0$, the local
constant estimator satisfies, as $b\rightarrow 0$, $nb^{3}\rightarrow \infty 
$ and $n^{\min \left\{ q,1\right\} }b\rightarrow \infty $, 
\begin{equation*}
\sqrt{nb}\left\{ \hat{\theta}\left( u\right) -\theta \left( u\right)
-b^{2}\mu _{2}Bias_{0}\left( u\right) \right\} \rightarrow ^{d}N\left(
0,\kappa _{2,0}V\left( u\right) \right) ,
\end{equation*}%
where $Bias_{0}\left( u\right) =H^{-1}\left( u\right) \left[
B_{1}(u)+B_{2}\left( u\right) \right] $ and $B_{1}(u)$ and $B_{2}(u)$ are
given in Theorem \ref{thm: Norm2}.

Two equivalent representations of $B_{1}(u)+B_{2}\left( u\right) $ are, with 
$s_{t}^{\ast }\left( \theta |u\right) $ denoting the stationary
approximation of $s_{n,t}\left( \theta \right) $,%
\begin{eqnarray*}
B_{1}(u)+B_{2}\left( u\right) &=&\frac{1}{2}\frac{\partial ^{2}}{\partial
v^{2}}\mathbb{E}\left[ s_{t}^{\ast }\left( \theta \left( v\right) |u\right) %
\right] _{v=u}+\frac{\partial ^{2}}{\partial u\partial v}\mathbb{E}\left[
s_{t}^{\ast }\left( \theta \left( v\right) |u\right) \right] _{v=u} \\
&=&-\frac{1}{2}\frac{\partial ^{2}}{\partial u^{2}}\mathbb{E}\left[
s_{t}^{\ast }\left( \theta \left( v\right) |u\right) \right] _{v=u},
\end{eqnarray*}%
where the second representation is only well-defined if $s_{t}^{\ast }\left(
\theta |u\right) $ is twice $L_{1}$-differentiable w.r.t. $u$.
\end{corollary}

To our knowledge this is the first complete characterization of the leading
bias term of local constant estimators in general time-varying parameter
models. The final characterisation of the bias, $B_{1}(u)+B_{2}\left(
u\right) =-\frac{1}{2}\partial _{u}^{2}\mathbb{E}\left[ s_{t}^{\ast }\left(
\theta \left( v\right) |u\right) \right] _{v=u}$, corresponds to the one
obtained in \citet{dahlhaus2006} for the time--varying ARCH model. This
characterisation, however, requires $s_{t}^{\ast }\left( \theta |u\right) $
to be twice differentiable w.r.t. $u$, while our characterisation only
requires $h_{t}^{\ast }\left( \theta |u\right) $ to be once differentiable
w.r.t $u$.

Comparing Theorems \ref{thm: Norm1} and \ref{thm: Norm2}, we see that the
local linear and local constant estimators share convergence rate and
asymptotic variance, but that the latter suffers from additional biases.
This is consistent with the theory found for local constant and local linear
estimators in a cross-sectional settingl; ; see, e.g., \cite{fan1995JASA}.

\subsection{Discrete--valued time series}

The above theory for the local constant estimator does not cover
discrete-valued time series models. Specifically, Theorem \ref{thm: Norm2}
requires $h_{t}^{\ast }\left( \theta |u\right) $ to be differentiable w.r.t. 
$u$, c.f. Assumption \ref{assu: derivprocess}. This property rarely holds
when $h_{t}^{\ast }\left( \theta |u\right) $ is a function of
discrete--valued random variables since these are generally not smooth
functions of the underlying parameters of the model; see \citet{Truquet2019}
for more details. Thus, Theorem \ref{thm: Norm2} does not apply to, for
example, the Poisson autoregressive model.

But Theorem \ref{thm: Norm1} still applies. We therefore combine the ideas
of \citet{Truquet2019,Truquet2020} with Theorem \ref{thm: Norm1} to obtain a
theory for the local constant estimator that also covers models with
discrete-valued outcomes. This is achived by replacing Assumptions \ref%
{assu: derivprocess} and \ref{assu: hessian shortdep} with the following
ones:

\begin{assumption}
\label{assu: derivprocess alt}$v\mapsto \mathbb{E}\left[ h_{t}^{\ast }\left(
\theta \left( u\right) |v\right) \right] $ is continuously differentiable at 
$u$.
\end{assumption}

\begin{assumption}
\label{assu: hessian shortdep alt}(i) $\bar{V}_{ijkl}\left(
v_{1},v_{2}\right) :=\sum_{t=1}^{\infty }\mathrm{Cov}\left( h_{ij,0}^{\ast
}\left( \theta \left( u\right) |v_{1}\right) ,h_{kl,t}^{\ast }\left( \theta
\left( u\right) |v_{2}\right) \right) $ exists for all $\left(
v_{1},v_{2}\right) $ in a neighbourhood of $\left( u,u\right) $ and all $%
\left( i,j,k,l\right) $ (ii) $\bar{V}_{ijkl}\left( v_{1},v_{2}\right) $ is
continuously differentiable at $\left( v_{1},v_{2}\right) =\left( u,u\right) 
$.
\end{assumption}

If $v\mapsto h_{t}^{\ast }\left( \theta \left( u\right) |v\right) $ is $%
L_{1} $-differentiable w.r.t. $v$ at $u$, then $\partial _{v}\mathbb{E}\left[
h_{t}^{\ast }\left( \theta \left( u\right) |v\right) \right] _{v=u}=\mathbb{E%
}\left[ \partial _{v}h_{t}^{\ast }\left( \theta \left( u\right) |v\right) %
\right] _{v=u}$. Thus, Assumption \ref{assu: derivprocess alt} is weaker
than Assumption \ref{assu: derivprocess} and is satisfied as long as the
cumulative distribution function of $h_{t}^{\ast }\left( \theta \left(
u\right) |v\right) $ is differentiable w.r.t. $v$, c.f. Lemma \ref{lemma:
time deriv conds} below. This property holds for\ many discrete-valued
models, including Poisson autoregressions and dynamic discrete choice
models, c.f. \citet{Truquet2019,Truquet2020}.

Assumption \ref{assu: hessian shortdep alt}, on the other hand, is stronger
than Assumption \ref{assu: hessian shortdep}. However, similar to Assumption %
\ref{assu: hessian shortdep}, part (i) is satisfied if $h_{t}^{\ast }\left(
\theta \left( u\right) |v\right) $ is $\tau $-weakly dependent for $v$ in a
neighbourhood of $u$ since this in turn implies that the joint process $%
\left( h_{t}^{\ast }\left( \theta \left( u\right) |v_{1}\right) ,h_{t}^{\ast
}\left( \theta \left( u\right) |v_{2}\right) \right) $ is weakly dependent
for $\left( v_{1},v_{2}\right) $ in a neighbourhood of $\left( u,u\right) $.
Moreover, part (ii) will hold under the same conditions that ensure
Assumption \ref{assu: derivprocess alt} holds, namely that the joint
distribution function of $\left( h_{0}^{\ast }\left( \theta \left( u\right)
|v_{1}\right) ,h_{t}^{\ast }\left( \theta \left( u\right) |v_{2}\right)
\right) $ is differentiable.

The following result shows that the results for the local constant estimator
remains essentially the same under Assumptions \ref{assu: derivprocess alt}--%
\ref{assu: hessian shortdep alt} in place of \ref{assu: derivprocess}--\ref%
{assu: hessian shortdep}:

\begin{theorem}
\label{thm: Norm3}Suppose Assumptions \ref{assu: kernel}-\ref{assu: 3rdderiv}
and \ref{assu: derivprocess alt}--\ref{assu: hessian shortdep alt} hold and $%
\theta \left( \cdot \right) $ is $m+2$ times continuously differentiable.
Then the conclusions of Theorem \ref{thm: Norm2} and Corollary \ref{cor: LC
norm} still holds, except that $\partial _{u}H\left( u\right) $ in the
expression of $B_{1}\left( u\right) $ is now defined as $\partial
_{u}H\left( u\right) =\partial _{v}E\left[ h_{t}^{\ast }\left( \theta \left(
u\right) |v\right) \right] _{v=u}$.
\end{theorem}

\subsection{Behaviour at boundary}

We have already seen that the local linear estimator has smaller biases than
the local constant one in the interior of its domain, $u\in \left(
0,1\right) $. Another well-known advantage of the local linear estimators in
a cross-sectional setting is that they exhibit automatic boundary
carpentering. This property also holds in our setting where the boundaries
are $u=0$ and $u=1$. Since the results for $u=1$ are similar, we here only
analyze the properties of the local constant ($m=0$) and the local linear ($%
m=1$) estimators at $u=cb$ for some constant $c>0$. Combining the
intermediate bias--variance analysis carried out in the proofs of Theorem %
\ref{thm: Norm1} and \ref{thm: Norm2} with the arguments of %
\citet{fan1995JASA}, we find that the two estimators remain asymptotically
normally distributed but their asymptotic biases and variances take
different forms:

\begin{corollary}
\label{cor: boundary}Let $\hat{\theta}_{0}\left( u\right) $ and $\hat{\theta}%
_{1}\left( u\right) $ be the local constant and local linear estimators,
respectively, of $\theta \left( u\right) $. Under the assumptions of Theorem %
\ref{thm: Norm2} with $m=1$ together with $\mu _{1}=\int_{\mathbb{R}}K\left(
v\right) vdv=0$, as $b\rightarrow 0$, $nb^{3}\rightarrow \infty $ and $%
n^{\min \left\{ q,1\right\} }b\rightarrow \infty $, and for $m\in \{0,1\}$, 
\begin{equation*}
\sqrt{nb}\left( \hat{\theta}_{m}\left( cb\right) -\theta \left( cb\right)
-b^{m+1}Bias_{m}\right) \overset{d}{\rightarrow }N\left( 0,a_{m}V\left(
0_{+}\right) \right) ,
\end{equation*}%
where $V\left( 0_{+}\right) =\lim_{u\downarrow 0}V\left( u\right) $ and 
\begin{gather*}
Bias_{0}=\left( \kappa _{1,0}^{c}\right) ^{-1}\kappa _{1,1}^{c}\theta
^{\left( 1\right) }\left( 0_{+}\right) ;\quad Bias_{1}=\frac{1}{2}\frac{%
\left( \kappa _{1,2}^{c}\right) ^{2}-\kappa _{1,1}^{c}\kappa _{1,3}^{c}}{%
\kappa _{1,0}^{c}\kappa _{1,2}^{c}-\left( \kappa _{1,1}^{c}\right) ^{2}}%
\theta ^{\left( 2\right) }\left( 0_{+}\right) ; \\
a_{0}=\kappa _{2,0}^{c}/\left( \kappa _{1,0}^{c}\right) ^{2};\quad a_{1}= 
\left[ \left( \kappa _{1,2}^{c}\right) ^{2}\kappa _{2,0}^{c}-2\kappa
_{1,1}^{c}\kappa _{1,2}^{c}\kappa _{2,1}^{c}+\left( \kappa _{1,1}\right)
^{2}\kappa _{2,2}^{c}\right] /\left[ \kappa _{1,0}^{c}\kappa
_{1,2}^{c}-\left( \kappa _{1,1}^{c}\right) ^{2}\right] ^{2}.
\end{gather*}
\end{corollary}

We refer to \citet{fan1995JASA} for the precise expressions of the constants 
$\kappa _{i,j}^{c}$, $i=1,2$ and $j=0,1,2$. At the boundary, both the biases
and variances of the local constant and local linear estimators are now
different. While the difference between two asymptotic variances is a
constant scale, compare $a_{0}$ and $a_{1}$ above, the biases are now of a
different order: The local linear estimator still enjoys a bias of order $%
O\left( b^{2}\right) $ while the bias of the local constant one blows up and
becomes of order $O\left( b\right) $. Thus, the local constant estimator
will generally suffer from significantly larger biases at the boundary
compared to the local linear one.

\section{Applications\label{sec: ex}}

To demonstrate the usefulness of our general results, we here apply our
theory to Markov models with exogenous co--variates and infinite memory
autoregressive models, respectively. In the process, we provide more
primitive conditions for the critical assumption of ULS.

By inspection of our Assumptions \ref{assu: kernel}--\ref{assu: score}, we
observe that, for a given parametric model and estimator, most of the
assumptions are easily verifiable using standard arguments known from the
literature on regular parametric estimators. The only ones that are
non--standard are Assumptions \ref{assu: LS} and \ref{assu: score}, which
require the reserarcher to show that $\ell _{n,t}\left( \theta \right) $ and
its first two derivatives are ULS. Similarly, Assumption \ref{assu: 3rdderiv}
is also a ULS requirement, while Assumption \ref{assu: hessian shortdep}
and/or \ref{assu: hessian shortdep alt} will hold if $h_{t}^{\ast }\left(
\theta \left( u\right) |u\right) $ is weakly dependent.

The following lemma provides sufficient conditions for a given
transformation of a time series to be ULS if the underlying time series is
ULS. It also shows that if the stationary version is $\tau $--weakly
dependent then so is the transformation.

\begin{lemma}
\label{lemma: LS transform}Let $\left( \mathcal{W},\left\Vert \cdot
\right\Vert \right) $ be a Banach space and $f:\left( \mathcal{W},\left\Vert
\cdot \right\Vert \right) ^{\infty }\times \Theta \mapsto \mathbb{R}^{d}$, $%
d<\infty $, be a given mapping. Then the following hold:

\begin{enumerate}
\item Suppose that: (i) some process $W_{n,t}\left( \theta \right) \in 
\mathcal{W}$ is ULS$\left( p,q,\Theta \right) $ with stationary
approximation $W_{t}^{\ast }\left( \theta |u\right) $ satisfying $\mathbb{E}%
\left[ \left\Vert W_{t}^{\ast }\left( \theta |u\right) \right\Vert ^{p}%
\right] <\infty $; (ii) for some $r,C\geq 0$ and $\left\{ a_{k,i}\right\}
_{i=1}^{\infty }$ with $\sum_{i=1}^{\infty }a_{k,i}<\infty $, $k=1,2$,%
\begin{equation}
\left\Vert f\left( w_{1},w_{2},....;\theta \right) -f\left( w_{1}^{\prime
},w_{2}^{\prime },....;\theta \right) \right\Vert \leq C\left(
1+\sum_{i=1}^{\infty }a_{1,i}\left\Vert w_{i}\right\Vert
^{r}+\sum_{i=1}^{\infty }a_{1,i}\left\Vert w^{\prime }\right\Vert
^{r}\right) \sum_{i=1}^{\infty }a_{2,i}\left\Vert w_{i}-w_{i}^{\prime
}\right\Vert ,  \label{eq: f conds}
\end{equation}%
for all $\theta \in \Theta $ and $w=\left( w_{1},w_{2},...\right) \in 
\mathcal{W}^{\infty }$ and $w^{\prime }=\left( w_{1}^{\prime },w_{2}^{\prime
},...\right) \in \mathcal{W}^{\infty }$. Then $f\left( \mathcal{W}%
_{n,t}\left( \theta \right) ;\theta \right) $ is ULS$\left( p/\left(
r+1\right) ,q,\Theta \right) $, where $\mathcal{W}_{n,t}\left( \theta
\right) =\left( W_{n,t}\left( \theta \right) ,W_{n,t-1}\left( \theta \right)
,....\right) $.

\item Suppose furthermore that $W_{t}^{\ast }\left( \theta |u\right) $ is $%
\tau $-weakly dependent with $\mathbb{E}\left[ \left\Vert W_{t}^{\ast
}\left( \theta |u\right) \right\Vert ^{p}\right] <\infty $. Then $f\left( 
\mathcal{W}_{t}^{\ast }\left( \theta |u\right) ;\theta \right) $ is also $%
\tau $-weakly dependent with $\mathbb{E}\left[ \left\Vert f\left( \mathcal{W}%
_{t}^{\ast }\left( \theta |u\right) ;\theta \right) \right\Vert ^{p/\left(
r+1\right) }\right] $. In particular, if $p/\left( r+1\right) >2$ then $%
\sum_{t=1}^{\infty }\left\vert \mathrm{Cov}\left( f\left( \mathcal{W}%
_{t}^{\ast }\left( \theta |u\right) ;\theta \right) ,f\left( \mathcal{W}%
_{t}^{\ast }\left( \theta |u\right) ;\theta \right) \right) \right\vert
<\infty $.
\end{enumerate}
\end{lemma}

In our applications below, we will then require $\ell _{n,t}\left( \theta
\right) $ and its first two derivatives to satisfy these conditions so that
Assumptions \ref{assu: LS}, \ref{assu: score}, \ref{assu: 3rdderiv} and \ref%
{assu: hessian shortdep} are satisfied.

Next, we here provide more primitive conditions for Assumption \ref{assu:
derivprocess alt} that can be be applied to discrete--valued random
variables:

\begin{lemma}
\label{lemma: time deriv conds}Suppose that the distribution function $%
F_{W}\left( w;\theta |u\right) $ of $W_{t}^{\ast }\left( \theta |u\right) $
satisfies $dF_{W}\left( w;\theta ,u\right) =p\left( w;\theta ,u\right) d\mu
\left( w\right) $ for some measure $\mu $, where $u\mapsto p\left( w;\theta
,u\right) $ is differentiable w.r.t. $\mu $-a.s.. Then for any function $%
f\left( w\right) $ with $E\left[ \left\Vert f\left( W_{t}^{\ast }\left(
\theta |u\right) ;u\right) \right\Vert \right] <\infty $, $\partial _{u}E%
\left[ f\left( W_{t}^{\ast }\left( \theta |u\right) \right) \right] =\int
f\left( w\right) \partial _{u}p\left( w;\theta ,u\right) d\mu \left(
w\right) $ assuming it is well defined.
\end{lemma}

What remains is to show that $u\mapsto p\left( w;\theta ,u\right) $ is
differentiable. This is fairly straightforward for Markov models, where we
can import results from, e.g., \cite{Truquet2020} and \cite{VazquezAbad1992}%
; see proof of Corollary \ref{cor: PARX} for an example of this. A
sufficient condition for $\int \left\vert f\left( w\right) \partial
_{u}p\left( w;\theta ,u\right) \right\vert d\mu \left( w\right) <\infty $ is 
$\int \left\vert f\left( w\right) \right\vert \frac{\left\vert \partial
_{u}p\left( w;\theta ,u\right) \right\vert }{p\left( w;\theta ,u\right) }%
p\left( w;\theta ,u\right) d\mu \left( w\right) <\infty $. For example, if $%
\left\vert \partial _{u}p\left( w;\theta ,u\right) \right\vert /p\left(
w;\theta ,u\right) \leq C\left( 1+\left\Vert w\right\Vert ^{r}\right) $, $%
r\geq 0$, then we need $E\left[ \left\Vert W_{t}^{\ast }\left( \theta
|u\right) \right\Vert ^{r}\left\Vert f\left( W_{t}^{\ast }\left( \theta
|u\right) ;u\right) \right\Vert \right] <\infty $.

We summarise our findings in the following corollary:

\begin{corollary}
\label{cor: Master}Suppose that (i) $K$ satisfies Assumption \ref{assu:
kernel}; (ii) the stationary version of the model, as summarised by $\theta
\mapsto \ell _{t}^{\ast }\left( \theta |u\right) $, satisfies Assumptions %
\ref{assu: compactness} and \ref{assu: smoothness}; (iii) $\ell _{n,t}\left(
\theta \right) $ and its first two derivatives satisfy (\ref{eq: f conds});
(iv) $Z_{n,t}$ is LS$\left( p,q\right) $ and $\mathbb{E}\left[ \left\Vert
Z_{t}^{\ast }\left( u\right) \right\Vert ^{p}\right] <\infty $. Then, the
conclusions of Theorem \ref{thm: Norm1} hold.

Suppose that, in addition, (v) the stationary distribution satisfies either
Assumption \ref{assu: derivprocess} or \ref{assu: derivprocess alt}; (vi) $%
Z_{t}^{\ast }\left( u\right) $ is $\tau $-weakly dependent; (vii) $%
h_{t}^{\ast }\left( \theta |u\right) $ satisfies (\ref{eq: f conds}) with $%
r\geq 0$ so that $p/\left( r+1\right) >2$. Then, the conclusions of Theorem %
\ref{thm: Norm3} hold.
\end{corollary}

If the model and estimator of interest is regular in the sense that its
stationary version satisfies above assumptions, then all that remains to be
shown is that $Z_{n,t}$ is locally stationary with $Z_{t}^{\ast }\left(
u\right) $ being weakly dependent. The next subsection provides primitive
conditions for this to hold for Markov models with exogenous co--variates,
while the second subsection discuss application to infinite memory models.
In the third subsection, we revisit the examples of Section \ref{sec: model}
and provide primitive conditions under which our main results apply to these.

\subsection{Markov models with exogenous co--variates}

We first consider $q$-Markov models without covariates on the form%
\begin{equation*}
Y_{n,t}=G\left( Y_{n,t-1},...,Y_{n,t-q},\varepsilon _{t},\theta \left(
t/n\right) \right) ,\quad t=1,\ldots ,n,
\end{equation*}%
where $G:\mathcal{Y}^{q}\times \mathcal{E}\times \Theta \mapsto \mathcal{Y}$
is some known mapping, $\varepsilon _{t}\in \mathcal{E}\subseteq \mathbb{R}%
^{d_{\varepsilon }}$ is a sequence of i.i.d. errors, and $\theta \left(
\cdot \right) \in \Theta $. Importantly, the initial value $Y_{n,0}$ can be
arbitrarily chosen which is in contrast to most of the existing literature.
Under regularity conditions, its corresponding stationary approximation $%
Y_{t}^{\ast }\left( u\right) $ will solve 
\begin{equation}
Y_{t}^{\ast }\left( u\right) =G\left( Y_{t-1}^{\ast }\left( u\right)
...,Y_{t-q}^{\ast }\left( u\right) ,\varepsilon _{t};\theta \left( u\right)
\right) ,\quad u\in \left[ 0,1\right] .  \label{eq: S-Markov}
\end{equation}%
We impose the following assumptions:

\begin{assumption}
\label{assu: LS-Markov}(i) $\sup_{\theta \in \Theta }\mathbb{E}\left[ \lVert
F\left( y_{0},\varepsilon _{t};\theta \right) \rVert ^{p}\right] <\infty $
for some $y_{0}\in \mathcal{Y}^{q}$ and $p>0$; (ii) there exists $\rho <1$
so that for all $y,y^{\prime }\in \mathcal{Y}^{q}$, $\mathbb{E}\left[ \lVert
F\left( y,\varepsilon _{t};\theta \right) -F\left( y^{\prime },\varepsilon
_{t};\theta \right) \rVert ^{p}\right] ^{1/p}\leq \rho \lVert y-y^{\prime
}\rVert $; (iii) there exist $\tilde{p}\geq 1$, $q>0$ and $r\geq 0$ so that
for all $\theta ,\theta ^{\prime }\in \Theta $, $\mathbb{E}\left[ \lVert
F\left( y,\varepsilon _{t};\theta \right) -F\left( y,\varepsilon _{t};\theta
^{\prime }\right) \rVert ^{\tilde{p}}\right] ^{1/\tilde{p}}\leq C\left(
1+\lVert y\rVert ^{r}\right) \lVert \theta -\theta ^{\prime }||^{q}$; (iv) $%
\mathbb{E}\left[ \lVert Y_{n,0}\rVert ^{\tilde{p}}\right] <\infty $ and $%
\mathbb{E}\left[ \lVert G\left( y,\varepsilon _{t};\theta \right) -G\left(
y,\varepsilon _{t};\theta ^{\prime }\right) \rVert ^{\tilde{p}}\right] ^{1/%
\tilde{p}}\leq C\left( 1+\lVert y\rVert ^{r}\right) \lVert \theta -\theta
^{\prime }||^{q};$
\end{assumption}

This assumption is similar to the one found in \citet{dahlhaus2017}, but we
here allow for $p\neq \tilde{p}.$ Assuming we can verify $\mathbb{E}\left[
\lVert Y_{t}^{\ast }\left( u\right) \rVert ^{\tilde{p}r}\right] <\infty $
this allows us to show higher-order local stationarity ($\tilde{p}>p$).
Furthermore, part (iv) only requires suitable moments of $Y_{n,0}$ but
otherwise the process can be initialized at any given value while most of
the existing literature implicitly assumes $Y_{n,0}=Y_{0}^{\ast }\left(
u\right) $.

\begin{lemma}
\label{thm: Markov LS} Under Assumptions \ref{assu: LS-Markov}(i)-(ii),
there exists a stationary and $\tau $-weakly dependent solution, $\left\{
Y_{t}^{\ast }\left( u\right) \right\} $ to (\ref{eq: S-Markov}) with $%
\sup_{u\in \left[ 0,1\right] }\mathbb{E}\left[ \lVert Y_{t}^{\ast }\left(
u\right) \rVert ^{p}\right] <\infty $. If furthermore Assumptions \ref{assu:
LS-Markov}(iii)-(iv) hold, $\sup_{u\in \left[ 0,1\right] }\mathbb{E}\left[
\lVert Y_{t}^{\ast }\left( u\right) \rVert ^{\tilde{p}r}\right] <\infty $
and $\theta \left( \cdot \right) \in \Theta $ is continuously
differentiable, then $Y_{n,t}$ is LS$\left( \tilde{p},q\right) $ with $%
\sup_{n,t}\mathbb{E}\left[ \lVert Y_{n,t}\rVert ^{\tilde{p}}\right] <\infty $
for any given initial value $Y_{n,0}$ with \thinspace $\mathbb{E}\left[
\lVert Y_{n,0}\rVert ^{p}\right] <\infty $.
\end{lemma}

Next, we extend the results to the following class of $q$-Markov models with
exogenous co-variates: 
\begin{equation}
Y_{n,t}=G\left( Y_{n,t-1},...,Y_{n,t-q},X_{n,t-1},\varepsilon _{t};\theta
\left( t/n\right) \right) ,  \label{eq: AR-X model}
\end{equation}%
\begin{equation}
X_{n,t}=H\left( X_{n,t-1},...,X_{n,t-q},\eta _{t},t/n\right) .
\label{eq: x model}
\end{equation}
We could allow for $X_{n,t}$ to exhibit richer dynamics, e.g., allow for $%
X_{n,t}$ to depend on lags of $Y_{n,t}$. However, this would lead to more
complicated assumptions and so we here maintain (\ref{eq: x model}) for
simplicity. We impose the following assumptions on (\ref{eq: AR-X model})--(%
\ref{eq: x model}):

\begin{assumption}
\label{assu: LS-MarkovX}(i) $\left( \varepsilon _{t},\eta _{t}\right) $ is
i.i.d. over time; (ii) for some $\rho _{x}<1$, $p\geq 1$ and $x_{0}\in 
\mathcal{X}^{q}$, $\mathbb{E}\left[ \left\Vert H\left( x;\eta _{t},u\right)
-H\left( \tilde{x};\eta _{t},u\right) \right\Vert ^{p}\right] \leq \rho
_{x}\left\Vert x-\tilde{x}\right\Vert ^{p}$ and $\mathbb{E}\left[ \left\Vert
H\left( x_{0};\eta _{t},u\right) \right\Vert ^{p}\right] <\infty $ for all $%
x,x^{\prime }\in \mathcal{X}^{q}$ and $u\in \left[ 0,1\right] $; (iii) for
some $\left( x_{0},y_{0}\right) \in \mathcal{X}^{q}\times \mathcal{Y}^{q}$,$%
L<\infty $ and $\rho _{y}<1$, $\sup_{\theta \in \Theta }\mathbb{E}\left[
\lVert G\left( y_{0},x_{0},\varepsilon _{t};\theta \right) \rVert ^{p}\right]
<\infty $ and 
\begin{equation*}
\mathbb{E}\left[ \lVert G\left( y,x,\varepsilon _{t};\theta \right) -G\left(
y^{\prime },x^{\prime },\varepsilon _{t};\theta \right) \rVert ^{p}\right]
^{1/p}\leq L\lVert x-x^{\prime }\rVert +\rho _{y}\lVert y-y^{\prime }\rVert
\end{equation*}%
for all $\left( x,y\right) ,\left( x^{\prime },y^{\prime }\right) \in 
\mathcal{X}^{q}\times \mathcal{Y}^{q}$; (iv) for some $\tilde{p}\geq 1$, $%
q>0 $ and $r\geq 0$ and all $\theta ,\theta ^{\prime }\in \Theta $, 
\begin{equation*}
\mathbb{E}\left[ \lVert G\left( y,x,\varepsilon _{t};\theta \right) -G\left(
y,x,\varepsilon _{t};\theta ^{\prime }\right) \rVert ^{\tilde{p}}\right] ^{1/%
\tilde{p}}\leq C\left( 1+\lVert x\rVert ^{r}+\lVert y\rVert ^{r}\right)
\lVert \theta -\theta ^{\prime }||^{q};
\end{equation*}%
\begin{equation*}
\mathbb{E}\left[ \lVert H\left( x,\eta _{t};u\right) -H\left( x,\eta
_{t};v\right) \rVert ^{\tilde{p}}\right] ^{1/\tilde{p}}\leq C\left( 1+\lVert
x\rVert ^{r}\right) \left\vert u-v\right\vert ^{q};
\end{equation*}%
(v) $\mathbb{E}\left[ \lVert Y_{n,0}\rVert ^{\tilde{p}}\right] <\infty $ and 
$\mathbb{E}\left[ \lVert X_{n,0}\rVert ^{\tilde{p}}\right] <\infty $.
\end{assumption}

\begin{lemma}
\label{lem: MarkovX LS} Under Assumptions \ref{assu: LS-MarkovX}(i)-(iii),
for any $u\in \left[ 0,1\right] $, there exists a stationary and $\tau $%
-weakly dependent solution $\left\{ Y_{t}^{\ast }\left( u\right)
,X_{t}^{\ast }\left( u\right) \right\} $ to 
\begin{eqnarray}
Y_{t}^{\ast }\left( u\right) &=&G\left( Y_{t-1}^{\ast }\left( u\right)
,X_{t-1}^{\ast }\left( u\right) ,\varepsilon _{t};\theta \left( u\right)
\right) ,  \label{eq: S-MarkovX} \\
X_{t}^{\ast }\left( u\right) &=&H\left( X_{t-1}^{\ast }\left( u\right) ,\eta
_{t},u\right)
\end{eqnarray}%
If furthermore \ref{assu: LS-MarkovX}(iv)-(v) hold, $\sup_{u\in \left[ 0,1%
\right] }\mathbb{E}\left[ \lVert X_{t}^{\ast }\left( u\right) \rVert ^{%
\tilde{p}r}\right] $, $\sup_{u\in \left[ 0,1\right] }\mathbb{E}\left[ \lVert
Y_{t}^{\ast }\left( u\right) \rVert ^{\tilde{p}r}\right] <\infty $ and $%
u\mapsto \theta \left( u\right) $ and $u\mapsto H\left( x,\eta _{t},u\right) 
$ is continuously differentiable, then $\left( Y_{n,t},X_{n,t}\right) $ is LS%
$\left( \tilde{p},q\right) $ with $\sup_{n,t}\mathbb{E}\left[ \lVert
Y_{n,t}\rVert ^{\tilde{p}}\right] <\infty $ and $\sup_{n,t}\mathbb{E}\left[
\lVert X_{n,t}\rVert ^{\tilde{p}}\right] <\infty $ for any given initial
values $\left( Y_{n,0},X_{n,0}\right) $ with $\mathbb{E}\left[ \lVert
Y_{n,0}\rVert ^{p}\right] <\infty $ and $\mathbb{E}\left[ \lVert
X_{n,0}\rVert ^{p}\right] <\infty $.
\end{lemma}

Combining Corollary \ref{cor: Master} with Lemma \ref{lem: MarkovX LS}, we
obtain a set of easily verifiable primitive conditions for local
M-estimators of time-varying parameters in Markov models to satisfy Theorems %
\ref{thm: Norm1} and \ref{thm: Norm3}; see Section \ref{sec: ex spec} for
examples of the verification procedure.

\subsection{Infinite memory models}

Next, we consider the following class of infinite memory models,%
\begin{equation*}
Y_{n,t}=G\left( Y_{n,t-1},Y_{n,t-2},....,\varepsilon _{t};\theta \left(
t/n\right) \right) .
\end{equation*}%
The local constant estimator of $\theta \left( u\right) $ was analysed in 
\cite{bardet2022} using techniques different from ours.\ Specifically, they
provide primitive conditions under which $Y_{n,t}$ satisfies%
\begin{equation*}
\sup_{\left[ n\left( u-cb\right) \right] \leq t\leq \left[ n\left(
u-cb\right) \right] +2cnb}E\left[ \left\Vert Y_{n,t}-Y_{t}^{\ast }\left(
u\right) \right\Vert ^{p}\right] ^{1/p}\leq C\left( b^{\rho }+\lambda _{%
\left[ n\left( u-cb\right) \right] +t}\right) ,
\end{equation*}%
where $Y_{t}^{\ast }\left( u\right) $ is the stationary solution to $%
Y_{t}^{\ast }\left( u\right) =G\left( Y_{t-1}^{\ast }\left( u\right)
,Y_{t-2}^{\ast }\left( u\right) ,....,\varepsilon _{t};\theta \left(
u\right) \right) $. But this result unfortunately does not suffice for our
theory since this requires $Y_{n,t}$ to satisfy our version of local
stationarity as given in Definition \ref{def: LS}. At the same time, \cite%
{bardet2022} do provide primitive conditions for the assumptions on the
stationary version of Corollary \ref{cor: Master} to hold. Thus, under the
high--level assumption that $Y_{n,t}$ satisfies our version of local
stationarity, we can combine the results of \cite{bardet2022} with our
Corollary \ref{cor: Master} to obtain a theory for local polynomial
estimators of infinite memory models. Under this high--level condition, we
are also able to provide a precise characterisation of the leading bias term
of their estimator, something which was missing in the analysis of \cite%
{bardet2022}. We leave it to future research to establish conditions under
which $Y_{n,t}$ is locally stationary a la Definition \ref{def: LS}.

\subsection{Specific examples\label{sec: ex spec}}

We here apply Corollary \ref{cor: Master} to Examples \ref{exa: tv-TAR}--\ref%
{exa: tv-PARX}. Note here that the first and third examples cannot be
analysed using existing theories since these rely on differentiability of
the model. Moreover, none of the existing theories allow for co--variates to
be included in the model. As such, the results below are new to the
literature.

Throughout this section the following assumption is maintained where $\theta
\left( u\right) $ and $\Theta $ are specified in each of the following
examples:

\begin{assumption}
\label{assu: Examples}(i) The kernel $K$ satisfies Assumption \ref{assu:
kernel}; (ii) $\theta \left( u\right) \in \text{Int}\left( \Theta \right) $
and $\theta \left( \cdot \right) $ is twice continuously differentiable;
(iii) the additional predictors $X_{n,t}$ of the model solves (\ref{eq: x
model}) where $H\left(x,\eta _{\tau };u \right) $ satisfies the conditions
in Assumption \ref{assu: LS-MarkovX} with $p=\tilde{p}=2$.
\end{assumption}

For the time--varying TAR-X model in Example \ref{exa: tv-TAR}, we obtain
the following result:

\begin{corollary}
\label{cor: tvTAR1}Let $\Theta =R^{2q+d_{X}+1}$ and suppose that%
\begin{equation}
\sup_{u\in \left[ 0,1\right] }\sum_{i=1}^{q}\max \left\{ \left\vert \alpha
_{1,i}\left( u\right) \right\vert ,\left\vert \alpha _{2,i}\left( u\right)
\right\vert \right\} <1,  \label{eq: TAR stat}
\end{equation}%
$\mathbb{E}\left[ \varepsilon _{t}^{2}\right] <\infty $ and Assumption \ref%
{assu: Examples} hold. Then, the results of Theorem \ref{thm: Norm1} apply
to the local linear estimators of the tv-TAR$(1)$ model (\ref{eq: tvTAR1})
with $H\left( u\right) =\mathbb{E}\left[ \tilde{X}_{t}^{\ast }\left(
u\right) \tilde{X}_{t}^{\ast }\left( u\right) ^{\prime }\right] $, $\Omega
\left( u\right) =\mathbb{E}\left[ \varepsilon _{t}^{2}\right] H\left(
u\right) $ and 
\begin{equation*}
\tilde{X}_{t}^{\ast }\left( u\right) =\left( Y_{t-1}^{\ast +}\left( u\right)
,,....,Y_{t-q}^{\ast +}\left( u\right) ,Y_{t-1}^{\ast -}\left( u\right)
,....,Y_{t-q}^{\ast -}\left( u\right) ,X_{t-1}^{\ast }\left( u\right)
\right) .
\end{equation*}%
If, in addition, $\mathbb{E}\left[ \varepsilon _{t}^{4}\right] <\infty $, $%
\varepsilon _{t}$ has a differentiable density and $H\left( x,\eta
_{t},u\right) $ satisfies Assumption 4.1(L3)--(L4) in \citet{dahlhaus2017},
then the results of Theorem \ref{thm: Norm2} hold for the local constant
estimators with $\partial _{\theta }H\left( u\right) =0$ and $\partial
_{u}H\left( u\right) =\partial H\left( \theta |u\right) /\left( \partial
u\right) |_{\theta =\theta \left( u\right) }$.
\end{corollary}

This appears to be the first result for the threshold AR--X\ model in the
literature. Note here that we do not need to restrict $\Theta $ to be
compact since here $\ell _{n,t}\left( \theta \right) $ is concave. The Eq. (%
\ref{eq: TAR stat}) ensures that the model indeed has a locally stationary
solution and is $\tau $-weakly dependent. The additional restrictions
imposed in the second part of the corollary are used to show that the
time--derivative of $\left( Y_{t}^{\ast }\left( u\right) ,X_{t}^{\ast
}\left( u\right) \right) $ exists so that Assumption \ref{assu: derivprocess}
holds.

Next, consider the local Gaussian QMLE of the tv--ARCH-X model given in
Example \ref{exa: tv-ARCH}. Under eq. (\ref{eq: ARCH stat}) below, there
exists a locally stationary solution to the model which takes the form 
\begin{equation*}
Y_{t}^{\ast }(u)=\lambda _{t}^{\ast }(u)\varepsilon _{t}^{2},\text{ \ \ }%
\lambda _{t}^{\ast }(u)=\theta (u)^{\prime }\tilde{X}_{t}^{\ast }(u),
\end{equation*}%
where $\tilde{X}_{t}^{\ast }(u)=\left( 1,Y_{t-1}^{\ast
}(u),...,Y_{t-q}^{\ast }(u),X_{t-1}^{\ast }\left( u\right) ^{\prime }\right)
^{\prime }$.

\begin{corollary}
\label{cor: ARCH}Assume that $\Theta \subseteq \mathbb{R}_{+}^{1+q+d_{X}}$
is compact with $\omega \geq \underline{\omega }>0$ for all $\theta \in
\Theta $, $X_{n,t-1}\in \mathbb{R}_{+}^{d_{X}}$, $\mathbb{E}\left[
\varepsilon _{t}^{4}\right] <\infty $,%
\begin{equation}
\sup_{u\in \left[ 0,1\right] }\sum_{i=1}^{q}\alpha _{i}\left( u\right) <1,
\label{eq: ARCH stat}
\end{equation}%
and Assumption \ref{assu: Examples}(i)--(iii) hold with $p=\tilde{p}=2$.
Then, the results of Theorem \ref{thm: Norm1} apply to the local linear
estimators of the tv-ARCH--X model (\ref{eq: tv-ARCH}) with 
\begin{equation*}
H\left( u\right) =-\mathbb{E}\left[ \frac{\partial _{\theta }\lambda
_{t}^{\ast }\left( u\right) \left( \partial _{\theta }\lambda _{t}^{\ast
}\left( u\right) \right) ^{\prime }}{\lambda _{t}^{\ast }(u)^{2}}\right]
,\;\ \ \Omega \left( u\right) =-\mathrm{Var}\left( \varepsilon
_{t}^{2}\right) H\left( u\right) .
\end{equation*}%
If, in addition, $\mathbb{E}\left[ \varepsilon _{t}^{4+\delta }\right]
<\infty $, for some $\delta >0$, and $H\left( x,\eta _{t},u\right) $
satisfies Assumption 4.1(L3)--(L4) in \citet{dahlhaus2017}, then the results
of Theorem \ref{thm: Norm2} apply to the local constant estimators of the
tv-ARCH--X model with $\partial _{u}H\left( u\right) =\partial H\left(
\theta |u\right) /\left( \partial u\right) |_{\theta =\theta \left( u\right)
}$ and%
\begin{equation*}
\partial _{\theta _{i}}H\left( u\right) =2\mathbb{E}\left[ \frac{\partial
_{\theta _{i}}\lambda _{t}^{\ast }\left( u\right) \partial _{\theta }\lambda
_{t}^{\ast }\left( u\right) \left( \partial _{\theta }\lambda _{t}^{\ast
}\left( u\right) \right) ^{\prime }}{\lambda _{t}^{\ast }(u)^{3}}\right] .
\end{equation*}
\end{corollary}

Our conditions on the model are more or less identical to \citet{bardet2022}
but allows for exogenous regressors to be included. Our conditions are
substantially weaker compared to \citet{dahlhaus2006} and \citet{inoue2019}
who impose much stronger moment conditions.

Finally, we apply our results to the local MLE of the tv--PARX model in
Example \ref{exa: tv-PARX}. Under (\ref{eq: PARX stat}) below, the tv--PARX\
process is locally stationary with stationary solution $Y_{t}^{\ast }(u)|%
\mathcal{F}_{t-1}^{\ast }(u) \sim \mathrm{Poisson}\left( \lambda _{t}^{\ast
}\left( u\right) \right) $, where $\lambda _{t}^{\ast }\left( u\right)
=\theta \left( u\right) ^{\prime }\tilde{X}_{t}^{\ast }\left( u\right) $ and 
$\tilde{X}_{t}^{\ast }(u)=\left( 1,Y_{t-1}^{\ast }(u),...,Y_{t-q}^{\ast
}(u),X_{t-1}^{\ast }\left( u\right) ^{\prime }\right) ^{\prime }$.

\begin{corollary}
\label{cor: PARX}Suppose that $\Theta \subseteq \mathbb{R}_{+}^{1+q+d_{X}}$
is compact with $\omega \geq \underline{\omega }>0$ for all $\theta \in
\Theta $, $X_{n,t-1}\in \mathbb{R}_{+}^{d_{X}}$, $\inf_{u}\omega \left(
u\right) >0$,%
\begin{equation}
\sup_{u\in \left[ 0,1\right] }\sum_{i=1}^{q}\alpha _{i}\left( u\right) <1,
\label{eq: PARX stat}
\end{equation}%
and Assumption \ref{assu: Examples}(i)--(iii) hold with $p=\tilde{p}=2$.
Then, the results of Theorems \ref{thm: Norm1} apply to the linear
estimators of the tv-PARX model (\ref{eq: tv-PARX}) with 
\begin{equation*}
\Omega \left( u\right) =\mathbb{E}\left[ \frac{\left( \partial _{\theta
}\lambda _{t}^{\ast }\left( u\right) \right) \left( \partial _{\theta
}\lambda _{t}^{\ast }\left( u\right) \right) ^{\prime }}{\lambda _{t}^{\ast
}\left( u\right) }\right] =-H\left( u\right) ,\text{ \ \ }\partial _{\theta
}\lambda _{t}^{\ast }\left( u\right) =\tilde{X}_{t}^{\ast }\left( u\right) .
\end{equation*}

If furthermore $H\left( x,\eta _{t},u\right) $ is bounded for all values of $%
\left( x,\eta _{t},u\right) $ and $\eta _{t}$ has full support, then the
local constant estimator satisfies Theorem \ref{thm: Norm2} with $\partial
_{u}H\left( u\right) =\partial H\left( \theta |u\right) /\left( \partial
u\right) |_{\theta =\theta \left( u\right) }$ and%
\begin{equation*}
\partial _{\theta _{i}}H\left( u\right) =2\mathbb{E}\left[ \frac{\partial
_{\theta _{i}}\lambda _{t}^{\ast }\left( u\right) \partial _{\theta }\lambda
_{t}^{\ast }\left( u\right) \left( \partial _{\theta }\lambda _{t}^{\ast
}\left( u\right) \right) ^{\prime }}{\lambda _{t}^{\ast }(u)^{3}}\right] .
\end{equation*}
\end{corollary}

\section{Simulation study\label{sec: simul}}

In this section, we examine the finite-sample performances of the local
constant and local linear estimators. All reported results are based on 1000
simulated data sets. The over--all performance of the estimators is
evaluated using the mean absolute deviation error (MADE), $MADE_{i}:=\frac{1%
}{n}\sum_{t=1}^{n}\mathbb{E}\left[ |\hat{\theta}_{i}\left( t/n\right)
-\theta _{i}\left( t/n\right) |\right] $, as well as their integrated bias,
variance, and mean squared error. All results are based on the Epanechnikov
kernel and with the bandwidth chosen using the cross-validation method
proposed in \citet{richter}.

\subsection{Time-varying ARCH}

We first consider the time-varying ARCH(1) in eq. (\ref{eq: tv-ARCH}) where $%
\varepsilon \sim i.i.d.N\left( 0,1\right) $, $\omega \left( u\right)
=0.7-0.5\sin \left( 4\pi u\right) $ and $\alpha \left( u\right)
=0.45+0.4\sin \left( 4\pi u\right) $. We estimate $\omega \left( u\right) $
and $\alpha \left( u\right) $ using both Gaussian log-likelihood and the WLS
method of \citet{fryzlewicz2008}. Table \ref{tab: tvarch1} reports the
performance of the estimators. For all sample sizes, the local MLE's perform
better than the local WLS estimators. For sample sizes of $n=250$ and $n=500$%
, the local constant MLE performs as well as the local linear MLE in terms
of the global measures, but the latter performs best in terms of IMSE and
MADE for $n=1000$. Thus, the over--all superiority of the local linear
estimator indicated by the theory appears to be a large--sample property.

\begin{table}[tbp]
\caption{Performance of the local constant (LC) and local linear (LL)
estimators for tvARCH model: Integrated squared bias (IBias2), integrated
variance (IVar), integrated mean squared errors (IMSE), and MADE.}
\label{tab: tvarch1}\centering
{%
\begin{tabular}{clcccccccc}
\hline
&  & \multicolumn{4}{c}{$\omega\left(u\right)$} & \multicolumn{4}{c}{$%
\alpha\left(u\right)$} \\ 
&  & \multicolumn{2}{c}{WLS} & \multicolumn{2}{c}{ML} & \multicolumn{2}{c}{
WLS} & \multicolumn{2}{c}{ML} \\ 
$n$ &  & LC & LL & LC & LL & LC & LL & LC & LL \\ \hline
250 & IBias2 & 0.0212 & 0.0222 & 0.0202 & 0.0204 & 0.0217 & 0.0279 & 0.0178
& 0.0213 \\ 
& IVar & 0.0647 & 0.0787 & 0.0549 & 0.0578 & 0.0734 & 0.0848 & 0.0771 & 
0.0720 \\ 
& IMSE & 0.0859 & 0.1009 & 0.0752 & 0.0782 & 0.0951 & 0.1126 & 0.0948 & 
0.0933 \\ 
& MADE & 0.2223 & 0.2386 & 0.2136 & 0.2141 & 0.2462 & 0.2643 & 0.2414 & 
0.2379 \\ 
& IBias2BD & 0.0397 & 0.0173 & 0.0310 & 0.0215 & 0.0226 & 0.0208 & 0.0181 & 
0.0144 \\ \hline
500 & IBias2 & 0.0116 & 0.0118 & 0.0091 & 0.0091 & 0.0107 & 0.0140 & 0.0076
& 0.0096 \\ 
& IVar & 0.0346 & 0.0390 & 0.0313 & 0.0322 & 0.0448 & 0.0484 & 0.0464 & 
0.0442 \\ 
& IMSE & 0.0461 & 0.0508 & 0.0405 & 0.0413 & 0.0555 & 0.0624 & 0.0541 & 
0.0537 \\ 
& MADE & 0.1651 & 0.1743 & 0.1555 & 0.1556 & 0.1878 & 0.1977 & 0.1813 & 
0.1794 \\ 
& IBias2BD & 0.0383 & 0.0071 & 0.0346 & 0.0128 & 0.0206 & 0.0092 & 0.0207 & 
0.0134 \\ \hline
1000 & IBias2 & 0.0067 & 0.0060 & 0.0045 & 0.0043 & 0.0060 & 0.0072 & 0.0038
& 0.0049 \\ 
& IVar & 0.0205 & 0.0225 & 0.0182 & 0.0183 & 0.0278 & 0.0297 & 0.0274 & 
0.0257 \\ 
& IMSE & 0.0272 & 0.0285 & 0.0228 & 0.0226 & 0.0337 & 0.0369 & 0.0312 & 
0.0306 \\ 
& MADE & 0.1261 & 0.1301 & 0.1152 & 0.1148 & 0.1463 & 0.1515 & 0.1374 & 
0.1353 \\ 
& IBias2BD & 0.0322 & 0.0017 & 0.0274 & 0.0049 & 0.0183 & 0.0039 & 0.0164 & 
0.0072 \\ \hline
\end{tabular}%
}
\end{table}

To compare the performance of the estimators near the end of the sample, we
also evaluate the bias of the estimators for the first and last 2.5\% of
time periods corresponding to $u\in \left[ 0,0.025\right] \cup \left[ 0.975,1%
\right] $. This is reported as IBias2BD in Table \ref{tab: tvarch1}. As
predicted by the theory, we find that relative to the local constant
versions the local linear WLS and ML estimators enjoy significantly smaller
biases near the boundaries of $\left[ 0,1\right] $ for all sample sizes.

\subsection{Time-varying Poisson autoregression with exogenous covariates
(PARX)}

We here report simulation results for the local linear MLE of the following
PARX(1) model with an additional exogenous regressor $X_{n,t}$, 
\begin{equation*}
\lambda_{n,t}=\omega\left(t/n\right)+\alpha\left(t/n\right)Y_{n,t-1}+\gamma%
\left(t/n\right)X_{n,t-1}^{2},
\end{equation*}
where $\omega\left(u\right)=0.6-0.3u+0.3\sin\left(2\pi u\right)$, $%
\alpha\left(u\right)=0.3+0.3u-0.3\sin\left(2\pi u\right)$ and $%
\gamma\left(u\right)=1-0.5\cos\left(\pi u\right)$. The dynamics of the
exogenous regressor was chosen as 
\begin{equation*}
X_{n,t}=\sqrt{\sigma\left(t/n\right)+\beta\left(t/n\right)X_{n,t-1}^{2}}%
\varepsilon_{t},\;\varepsilon_{t}\sim\text{i.i.d.}N\left(0,1\right).
\end{equation*}
where either

\begin{itemize}
\item DGP1: $\sigma\left(u\right)=0.8$ and $\beta\left(u\right)=0.4$ so that 
$X_{n,t}=X_{t}$ is strictly stationary.

\item DGP2: $\sigma\left(u\right)=0.8+0.4\cos\left(2\pi u\right)$ and $%
\beta\left(u\right)=0.4-0.2\sin\left(2\pi u\right)$ so $X_{n,t}$ is locally
stationary.
\end{itemize}

Table \ref{tab: tvparx1-1} reports the over-all performance of the
estimators in terms of integrated squared bias, variance, MSE and MADE. The
table shows that, in all sample sizes, the local linear estimator behaves
well for both DGP1 and DGP2. All bias, variance, and MADE decrease as the
sample size increases. Finally, similar to the case of the tvARCH model, the
local linear estimator performs well near the boundaries; we leave out these
results since they are similar to the ones reported for the tv-ARCH model
above.

\begin{table}[tbp]
\caption{Performance of the local linear estimators for tvPARX models:
Integrated squared bias (IBias2), integrated variance (IVar), integrated
mean squared errors (IMSE), and median of MADE}
\label{tab: tvparx1-1}\centering
{%
\begin{tabular}{clcccccc}
\hline
&  & \multicolumn{3}{c}{DGP1} & \multicolumn{3}{c}{DGP2} \\ 
$n$ &  & $\omega\left(u\right)$ & $\alpha\left(u\right)$ & $%
\gamma\left(u\right)$ & $\omega\left(u\right)$ & $\alpha\left(u\right)$ & $%
\gamma\left(u\right)$ \\ \hline
250 & IBias2 & 0.0330 & 0.0037 & 0.0004 & 0.0339 & 0.0035 & 0.0003 \\ 
& IVar & 0.1550 & 0.0078 & 0.0422 & 0.1409 & 0.0077 & 0.0383 \\ 
& IMSE & 0.1880 & 0.0115 & 0.0426 & 0.1748 & 0.0111 & 0.0386 \\ 
& MADE & 0.2407 & 0.0839 & 0.1463 & 0.2371 & 0.0832 & 0.1447 \\ \hline
500 & IBias2 & 0.0111 & 0.0016 & 0.0002 & 0.0137 & 0.0016 & 0.0002 \\ 
& IVar & 0.0528 & 0.0039 & 0.0216 & 0.0648 & 0.0042 & 0.0208 \\ 
& IMSE & 0.0639 & 0.0055 & 0.0218 & 0.0784 & 0.0057 & 0.0209 \\ 
& MADE & 0.1645 & 0.0586 & 0.1080 & 0.1717 & 0.0593 & 0.1079 \\ \hline
1000 & IBias2 & 0.0044 & 0.0007 & 0.0001 & 0.0044 & 0.0008 & 0.0001 \\ 
& IVar & 0.0266 & 0.0023 & 0.0115 & 0.0286 & 0.0022 & 0.0111 \\ 
& IMSE & 0.0310 & 0.0030 & 0.0116 & 0.0330 & 0.0030 & 0.0112 \\ 
& MADE & 0.1216 & 0.0426 & 0.0798 & 0.1199 & 0.0428 & 0.0798 \\ \hline
\end{tabular}%
}
\end{table}

\section{Empirical application\label{sec: empirical}}

\begin{figure}[tbp]
\includegraphics[width=\textwidth]{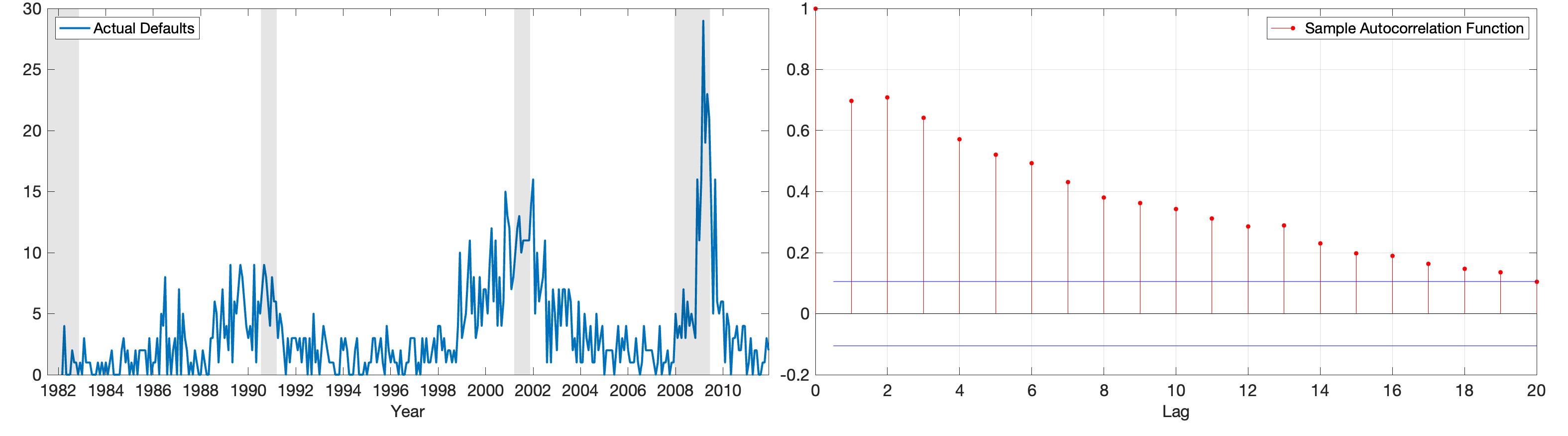}
\caption{Left: Number of defaults per month among Moody's rated US
industrial firms in the period 1982--2011; Right: autocorrelation function
of defaults.}
\label{fig: USdefaults}
\end{figure}

We here revisit the empirical analysis of US corporate defaults carried out
in \cite{agosto2016} with the aim of examining whether there is evidence of
structural instability in the time series. The data set consists of monthly
number of bankruptcies among Moody's rated industrial firms in the United
States for the period 1982--2011 ($n=360$ observations), collected from
Moody's Credit Risk Calculator (CRC). Figure \ref{fig: USdefaults} shows the
time series of default counts together with its sample autocorrelation
function, which reveals high temporal dependence in default counts and
existence of default clusters over time.

With $Y_{n,t}\in \left\{ 0,1,2,\ldots \right\} $, $t\geq 1$ denoting the
number of defaults in a given month, we use a log--PARX model to gain better
understanding of the dynamics of $Y_{n,t}$ as a function of its own past, $%
Y_{n,t-m}$, $m\geq 1$, but also in terms of additional covariates $%
X_{n,t}\in \mathbb{R}^{d_{x}}$, which include relevant macroeconomic and
financial factors as considered in \cite{agosto2016}. We model $Y_{n,t}$ as
a conditional Poisson distribution $Y_{n,t}|\mathcal{F}_{n,t-1}\sim \mathrm{%
Poisson}\left( \lambda _{n,t}\left( \theta \left( t/n\right) \right) \right) 
$, $t=1,2,\ldots ,n$, where the intensity $\lambda _{n,t}\left( \theta
\left( t/n\right) \right) $ depends on past counts, co--variates $X_{n,t-1}$
and a vector of time-varying parameters $\theta \left( t/n\right) $. Our
favoured specification of $\lambda _{n,t}\left( \theta \right) $ is the
log--PARX model of \cite{fokianos2011}, which is here augmented by the
chosen set of exogenous variables, $X_{n,t-1}$, 
\begin{equation}
\log \lambda _{n,t}\left( \theta \right) =\omega +\sum_{i=1}^{p}\alpha
_{i}\log \left( Y_{n,t-i}+1\right) +\gamma ^{\prime }X_{n,t-1},
\label{eq: LLPARX}
\end{equation}%
so that $\theta =\left( \omega ,\alpha _{1},...,\alpha _{p},\gamma ^{\prime
}\right) ^{\prime }$.

We here deviate from \cite{agosto2016} that specifies $\lambda _{n,t}$ to be
a linear function of past counts and factors. The reasons for us favouring
the above log--specification over the linear one are three--fold: First, (%
\ref{eq: LLPARX}) do not impose positivity constraints on the parameters
which facilitates the numerical computation of the estimators; second, it
allows us to include any predictors we wish without the need of first
transforming them to ensure that each component of the resulting $X_{n,t-1}$
is positive; third, when we estimate both models using the US default data,
we found that the log--specification delivers a better fit.

Similar to \cite{agosto2016}, we use $\sum_{i=1}^{p}\alpha _{i}\left(
t/n\right) $ as a measure of contagion in the financial markets: If $%
\sum_{i=1}^{p}\alpha _{i}\left( t/n\right) $ is large then firms defaulting
today will lead to a large increase in the risk of other firms defaulting
next period everything else equal.

As exogenous covariates, we consider the same financial, credit market, and
macroeconomic variables as in \cite{agosto2016}: Realized volatility ($RV$)
computed using daily squared return on the S\&P 500 index, the Leading Index
released by the Federal Reserve ($LI$), year-to-year change in Industrial
Production Index ($IP$), one-year return on the S\&P 500 index ($SPX$), the
three-month Treasury bill rate ($TB3$), and BAA Moody's rated to 10--year
Treasury spread ($SP$).\footnote{%
These covariates are tested for the existence of default covariates in \cite%
{das2007}, \cite{duffie2007}, and \cite{lando2010}.} \cite{agosto2016}
decompose $LI$ and $IP$ into their positive and negative parts to deal with
above--mentioned issue of $X_{n,t-1}$ having to be positive in the linear
specification. In contrast, no such transformations are needed for our
log-specification. Moreover, as well as $RV$, we also consider the logarithm
of $RV$, $\log \left( RV\right) $, to evaluate if the latter is a better
predictor; again, this would not be possible in the linear specification of 
\cite{agosto2016}.

The lag length $p$ is chosen using BIC where, for a given model, the
log--likelihood is evaluated at the estimated time-varying parameters.%
\footnote{%
As pointed out by a referee, we are unable to provide a formal justification
for using BIC to choose the lag-length in our time-varying parameter version
of the PARX model. So the application of this model selection criterion to
our setting is very ad hoc} According to this version of BIC, the preferred
specification is $p=2$. Importantly, by allowing for the parameters to be
time--varying, a much more parsimonious model is selected by BIC: If we do
model selection where for each model we restrict the estimated parameters to
be constant over time, the preferred model is $p=6$. Moreover, the
persistence, or contagion, of the estimated time--invariant version, as
measured by $\sum_{i=1}^{p}\hat{\alpha}_{i}$, is substantially higher than
for the time-varying version, as measured by $\sum_{i=1}^{p}\hat{\alpha}%
_{i}\left( u\right) $, $u\in \left[ 0,1\right] $. This is consistent with
the findings reported in \cite{Hillebrand2005} for GARCH\ models: Neglecting
structural changes in parameters causes the estimates of these to be
suffering from a strong upward bias which in turn leads BIC to selecting a
bigger model.

\begin{table}[tbp]
\caption{Estimation results of different LLPARX models. Standard errors are
in parentheses.}
\label{tab: USdef}\centering
{%
\begin{tabular}{ccccccc}
\hline\hline
& (1) & (2) & (3) & (4) &  &  \\ \hline
$\omega$ & 0.2186 & 1.0425 & 0.1003 & 1.3015 &  &  \\ 
& (0.1022) & (0.2491) & (0.2118) & (0.4188) &  &  \\ 
$\alpha_{1}$ & 0.3040 & 0.2777 & 0.3093 & 0.2836 &  &  \\ 
& (0.0492) & (0.0496) & (0.0497) & (0.0501) &  &  \\ 
$\alpha_{2}$ & 0.5033 & 0.4876 & 0.5082 & 0.4956 &  &  \\ 
& (0.0491) & (0.0486) & (0.0499) & (0.0492) &  &  \\ 
$RV$ & 5.8496 &  & 5.4864 &  &  &  \\ 
& (3.9738) &  & (4.4365) &  &  &  \\ 
$\log\left(RV\right)$ &  & 0.1230 &  & 0.1484 &  &  \\ 
&  & (0.0373) &  & (0.0437) &  &  \\ 
$LI$ & -0.1767 & -0.1482 & -0.1864 & -0.1860 &  &  \\ 
& (0.0310) & (0.0308) & (0.0490) & (0.0487) &  &  \\ 
$IP$ &  &  & 0.0029 & -0.0167 &  &  \\ 
&  &  & (0.0480) & (0.0481) &  &  \\ 
$SPX$ &  &  & 0.2074 & 0.2412 &  &  \\ 
&  &  & (0.2408) & (0.2413) &  &  \\ 
$TB3$ &  &  & 0.0065 & 0.0007 &  &  \\ 
&  &  & (0.0134) & (0.0136) &  &  \\ 
$SP$ &  &  & 0.0300 & -0.0426 &  &  \\ 
&  &  & (0.0585) & (0.0615) &  &  \\ \hline\hline
\end{tabular}%
}
\end{table}

As benchmark, we first estimate the time--invariant version of (\ref{eq:
LLPARX}) with $p=2$. Table \ref{tab: USdef} shows the estimation results for
four different specifications of the time--invariant LLPARX(2) model: Column
(1) contains the results when only $\left( RV,LI\right) $ are included;
column (2) when only $\left( \log RV,LI\right) $ are included; column (3)
when all covariates are included except for $\log \left( RV\right) $; and
column (4) when all covariates except $RV$ are included. As in \cite%
{agosto2016}, once we control for the information contained in $RV$ and $LI$%
, none of the other four covariates are found to be relevant in predicting
future defaults. We also observe that the two specifications using $\log
\left( RV\right) $ appear to perform better than the ones using $RV$. Based
on these results, our favoured specification of the time-varying version is
to use $LI$ and $\log \left( RV\right) $ as exogenous variables: 
\begin{equation}
\begin{split}
\log \lambda _{n,t}=\omega \left( t/n\right) & +\alpha _{1}\left( t/n\right)
\log \left( Y_{n,t-1}+1\right) +\alpha _{2}\left( t/n\right) \log \left(
Y_{n,t-2}+1\right) \\
& +\beta _{RV}\left( t/n\right) \log \left( RV_{n,t-1}\right) +\beta
_{LI}\left( t/n\right) LI_{n,t-1}.
\end{split}
\label{eq: tvLLPARX2}
\end{equation}

\begin{figure}[tbp]
\includegraphics[width=\textwidth]{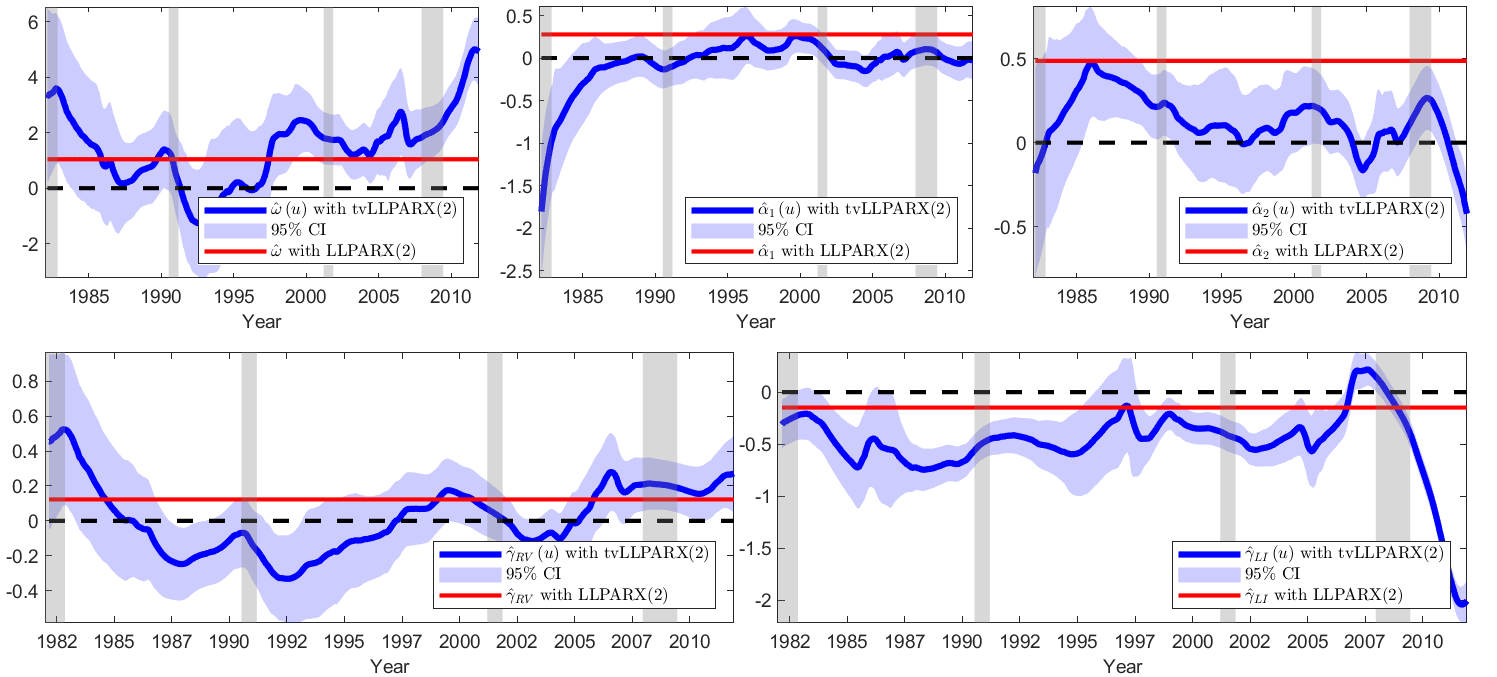}
\caption{Local linear estimate of time-varying parameter in eq.(\protect\ref%
{eq: tvLLPARX2}): Shaded areas are the 95\% confidence intervals.}
\label{fig: USdef_est}
\end{figure}

Figure \ref{fig: USdef_est} shows the time--series of the local linear
estimates of the time-varying parameters of (\ref{eq: tvLLPARX2}) together
with the time--invariant estimates reported in column (2) of Table \ref{tab:
USdef}. Pointwise confidence bands are computed based on the asymptotic
distribution derived in Theorem \ref{thm: Norm2}. The Leading Index is
pointwise significant for most of the sample period which highlights the
link between macroeconomic activity and corporate defaults also found in %
\citet{agosto2016}. At the same time, this link exhibits substantial
time--variation, in particular at the end of the sample during the Great
Recession. The link between realized volatility and defaults of industrial
firms is less significant but also appears to be changing over time. Similar
to \citet{agosto2016}, the realized volatility and the Leading Index are
strong explanatory variables during the Great Recession (2007--2011).
However, differently, both of them are still relevant in the late 1980s and
early 1990s. Especially, the effect of $\log \left( RV\right) $ tends to be
negative in this period which cannot be captured by the aforementioned
linear version of the PARX model.

Finally, judging from $\hat{\alpha}_{1}\left( t/n\right) +\hat{\alpha}%
_{2}\left( t/n\right) $, there is very little contagion in the financial
markets from 1990s and onwards, which is somewhat consistent with the
findings in \cite{agosto2016}. However, at the end of the sample, we
actually find a negative effect of today's log--defaults on tomorrow's
default risk. Also note that the time--varying estimates, $\hat{\alpha}%
_{1}\left( t/n\right) $ and $\hat{\alpha}_{2}\left( t/n\right) $, remain
below the corresponding time--invariant ones, $\hat{\alpha}_{1}$ and $\hat{%
\alpha}_{2}$ as marked by the horizontal red lines, throughout the sample
period. Again, this seems to indicate that ignoring time--variation in the
parameters of PARX models lead to over estimation of the level of
persistence/contagion.

\begin{figure}[tbp]
\includegraphics[width=\textwidth]{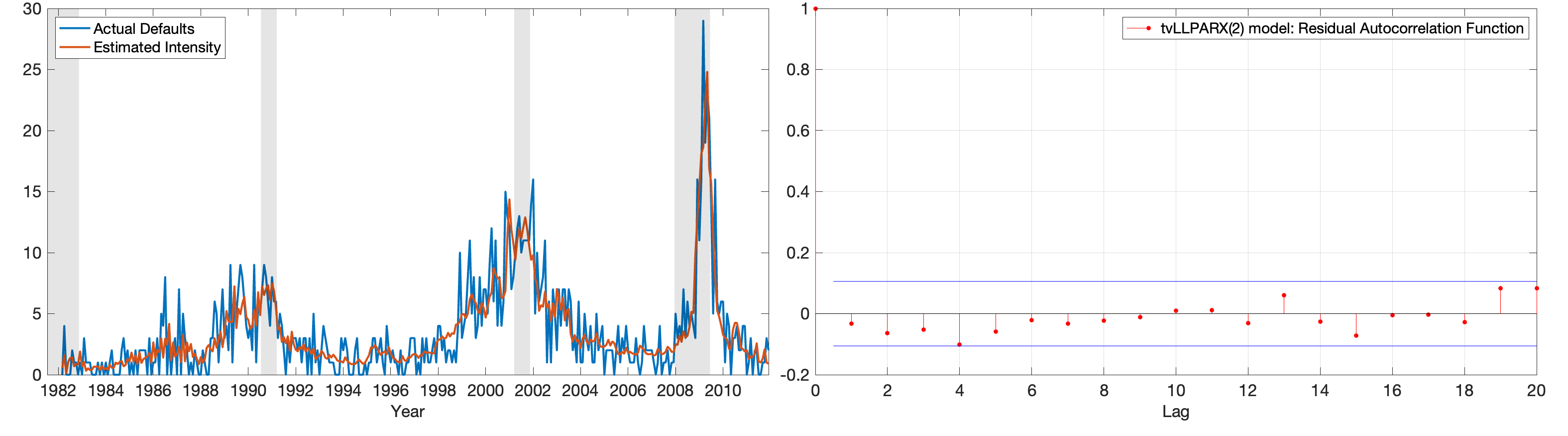}
\caption{Left: Actual number of defaults (blue) and estimated intensity
(red); Right: Sample autocorrelation function of Pearson residuals.}
\label{fig: tvLLPARX2-1}
\end{figure}

To assess in--sample fit and whether the reported time--variation in the
parameters is statistically significant, we carry out an array of graphical
and quantitative diagnostic tools for time series. First, we plot in the
left panel of Figure \ref{fig: tvLLPARX2-1} the actual default counts
together with the predicted defaults $\hat{Y}_{n,t}:=\hat{\lambda}%
_{n,t}=\lambda _{n,t}\left( \hat{\theta}\left( t/n\right) \right) $. As can
be seen from this plot, the time--varying LLPARX model captures the default
counts dynamics well. In the right panel of Figure \ref{fig: tvLLPARX2-1},
the sample autocorrelation function of the standardized Pearson residuals $%
\hat{e}_{n,t}=\hat{\lambda}_{n,t}^{-1/2}\left( Y_{n,t}-\hat{\lambda}%
_{n,t}\right) $ is plotted. Under correct specification, $e_{n,t}$ should be
white noise -- the plotted sample autocorrelation function supports this.

\begin{figure}[tbp]
\includegraphics[width=\textwidth]{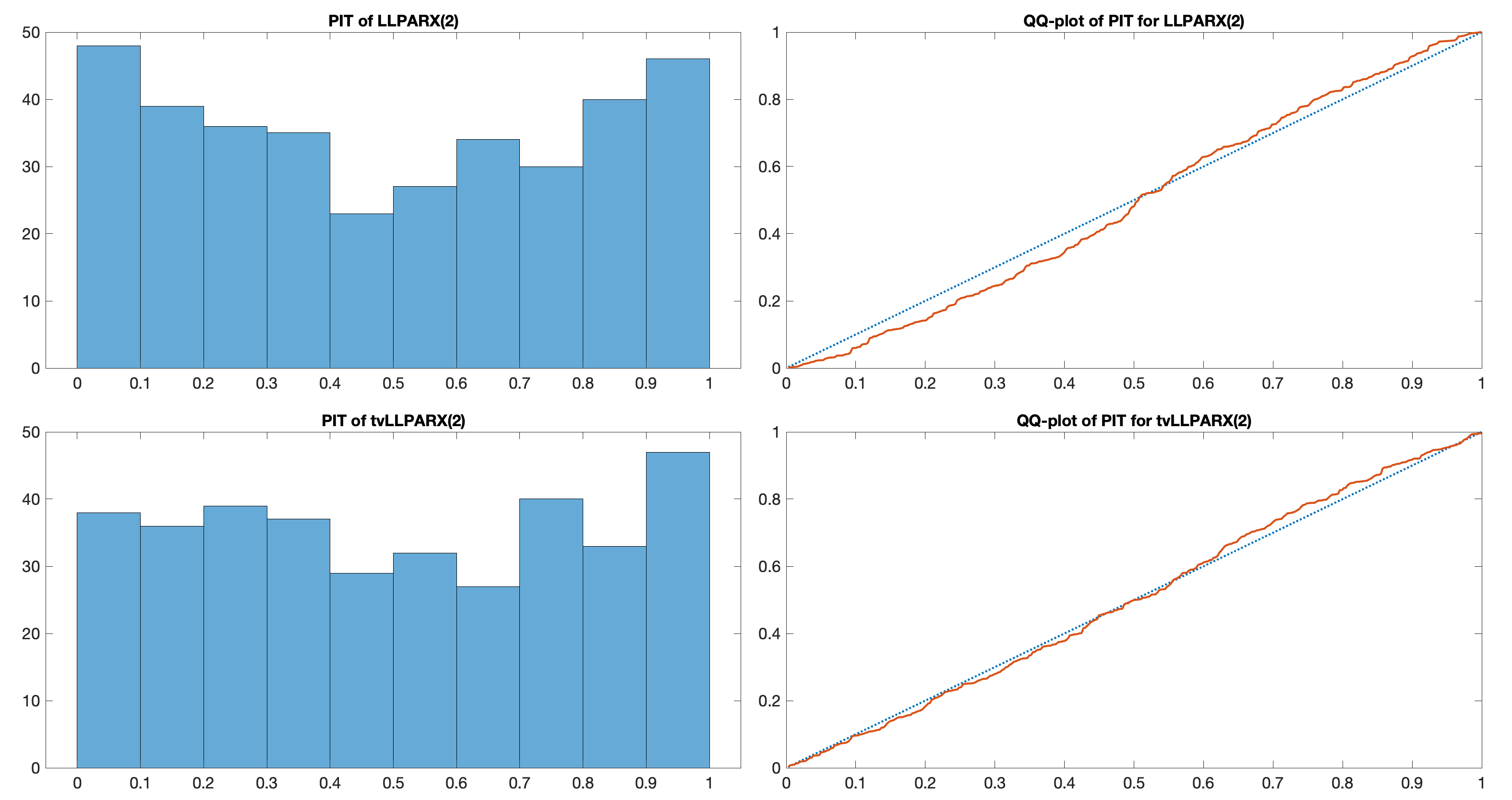}
\caption{Left: Histograms of randomized PIT's for log--linear PARX(2) and
time--varying log--linear PARX(2) models fitted to the US default data;
Right: QQ--plots of the randomized PIT against standard uniform distribution
for the corresponding models.}
\label{fig: tvLLPARX2-2}
\end{figure}

We also evaluate the adequacy of fit using the probability integral
transform (PIT). We follow \cite{Davis2016} and compute the PIT's by%
\begin{equation*}
\hat{u}_{n,t}:=F_{n,t}\left( Y_{n,t}-1\right) +\nu _{t}\left[ F_{n,t}\left(
Y_{n,t}\right) -F_{n,t}\left( Y_{n,t}-1\right) \right] ,
\end{equation*}%
where $\left\{ \nu _{t}\right\} $ is a sequence of i.i.d. random variables
from a standard uniform distribution, and $F_{n,t}$ is the CDF of a Poisson($%
\hat{\lambda}_{n,t})$ distribution. Under correct model specification, $\hat{%
u}_{n,t}$ is a sequence of i.i.d. random variables from the standard uniform
distribution. Figure \ref{fig: tvLLPARX2-2} depicts the histogram of PIT
which show that the tvLLPARX(2) model provides a better in--sample fit than
the corresponding time--invariant model.

Finally, in Table \ref{tab: USdef 2}, we report the log-likelihood, AIC and
BIC values and the $p$-value from a Kolmogorov-Smirnov test of the PIT's
being uniformly distributed of each of four specifications in columns
(1)--(4) of Table \ref{tab: USdef} together with the time-varying versions
of (2) and (4), labelled tv (2) and tv(4), respectively. From these, we see
that allowing for time--varying parameters increase the in--sample fit
dramatically. While this is not a formal statistical test of
time--variation, it provides strong informal evidence of such. As mentioned
earlier, we also see that specification (2) is favoured over (1), (3) and
(4) in the time-invariant case, and that (2) is favoured over (4) in the
time--varying case.

\begin{table}[tbp]
\caption{In--sample fit of time-invariant and time-varying LLPARX models}
\label{tab: USdef 2}\centering
{%
\begin{tabular}{ccccccc}
\hline\hline
& (1) & (2) & tv (2) & (3) & (4) & tv (4) \\ \hline
logL & 696.3 & 700.6 & 777.7 & 696.8 & 701.8 & 778.6 \\ 
AIC & -1382.5 & -1391.2 & -1526.6 & -1375.6 & -1385.5 & -1539.3 \\ 
BIC & -1363.1 & -1371.7 & -1507.2 & -1340.6 & -1350.5 & -1504.3 \\ 
$p$-value of PIT & 0.0726 & 0.1319 & 0.5432 & 0.0588 & 0.1544 & 0.6445 \\ 
\hline
\end{tabular}%
}
\end{table}

We complete the analysis by conducting a final sensitivity analysis of (\ref%
{eq: tvLLPARX2}). This is done by including the remaining exogenous
covariates in addition to realized volatility and Leading Index in the
time--varying version of the model: 
\begin{equation}
\begin{split}
\log \lambda _{n,t}=\omega \left( t/n\right) & +\alpha _{1}\left( t/n\right)
\log \left( Y_{n,t-1}+1\right) +\alpha _{2}\left( t/n\right) \log \left(
Y_{n,t-2}+1\right) \\
& +\gamma _{RV}\left( t/n\right) \log \left( RV_{n,t-1}\right) +\gamma
_{LI}\left( t/n\right) LI_{n,t-1}+\gamma _{IP}\left( t/n\right) IP_{n,t-1} \\
& +\gamma _{SPX}\left( t/n\right) SPX_{n,t-1}+\gamma_{TB3}\left( t/n\right)
TB3_{n,t-1}+\gamma_{SP}\left( t/n\right) SP_{n,t-1}.
\end{split}
\label{eq: tvLLPARX2all}
\end{equation}

\begin{figure}[tbp]
\includegraphics[width=\textwidth]{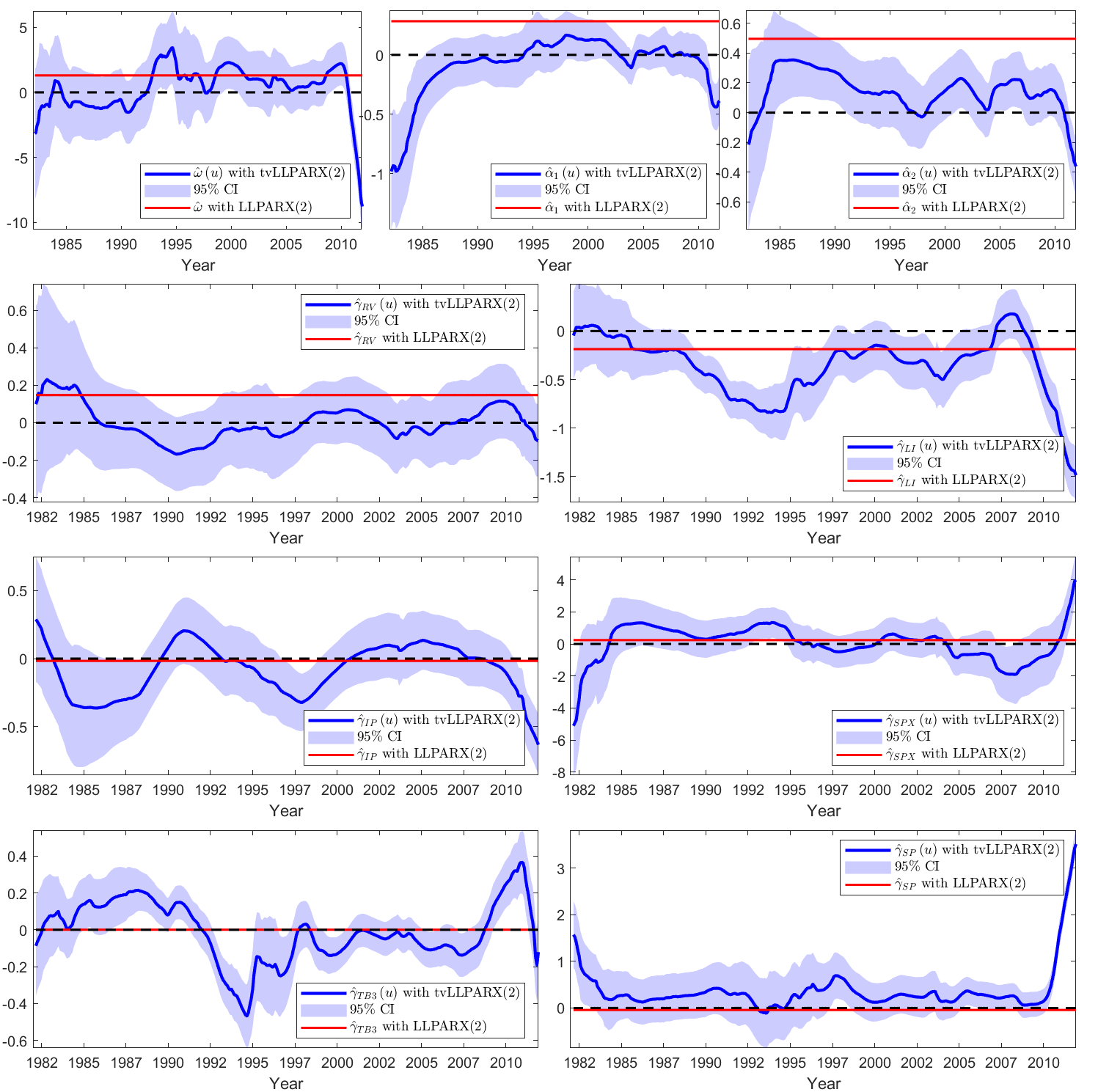}
\caption{Local linear estimate of time-varying parameter in eq.(\protect\ref%
{eq: tvLLPARX2all}): Shaded areas are the 95\% confidence intervals.}
\label{fig: tvLLPARX2all}
\end{figure}

The estimation results are provided in Figure \ref{fig: tvLLPARX2all}. The
estimates, except for the realized volatility, are consistent with our
baseline findings. The Leading Index remains highly significant, which shows
that macroeconomic factors are relevant in predicting future defaults. The
link between short--term interest rates and defaults of industrial firms
changes over time. During the late 1980s and the Great Recession
(2007--2011), interest rates play a role in determining the interest expense
of firms. In the 1990s, similar to the finding in \cite{duffie2007}, the
sign of the coefficient for the short--term rate is consistent with the fact
that the US Federal Reserve often increases the short--term rates to control
business expansions. Controlling for $LI$ and $TB3$, other covariates are
estimated to be insignificant for most of the sample period except for the
period of the Great Recession (2007--2011), in which financial, credit
market, and macroeconomic variables are significant explanators of the
default intensity. These are novel findings that the original analysis of 
\cite{agosto2016} did not reveal.

\newpage

\bibliographystyle{chicago}
\bibliography{Ref_Timevarying_MC}

\begin{thebibliography}{}

\bibitem[\protect\citeauthoryear{Agosto, Cavaliere, Kristensen, and
  Rahbek}{Agosto et~al.}{2016}]{agosto2016}
Agosto, A., G.~Cavaliere, D.~Kristensen, and A.~Rahbek (2016).
\newblock Modeling corporate defaults: Poisson autoregressions with exogenous
  covariates ({PARX}).
\newblock {\em Journal of Empirical Finance\/}~{\em 38, Part B}, 640 -- 663.

\bibitem[\protect\citeauthoryear{Bardet, Doukhan, and Wintenberger}{Bardet
  et~al.}{2022}]{bardet2022}
Bardet, J.-M., P.~Doukhan, and O.~Wintenberger (2022).
\newblock Contrast estimation of time-varying infinite memory processes.
\newblock {\em Stochastic Processes and their Applications\/}~{\em 152},
  32--85.

\bibitem[\protect\citeauthoryear{Brown}{Brown}{1971}]{Brown1971}
Brown, B. (1971).
\newblock Martingale central limit theorems.
\newblock {\em Annals of Mathematical Statistics\/}~{\em 42}, 59--66.

\bibitem[\protect\citeauthoryear{Caldara, Fern{\'{a}}ndez-Villaverde,
  Rubio-Ram{\'{\i}}rez, and Yao}{Caldara et~al.}{2012}]{Caldara2012}
Caldara, D., J.~Fern{\'{a}}ndez-Villaverde, J.~F. Rubio-Ram{\'{\i}}rez, and
  W.~Yao (2012).
\newblock Computing {DSGE} models with recursive preferences and stochastic
  volatility.
\newblock {\em Review of Economic Dynamics\/}~{\em 15\/}(2), 188--206.

\bibitem[\protect\citeauthoryear{Christoffersen, Jacobs, and
  Ornthanalai}{Christoffersen et~al.}{2012}]{Christoffersen2012}
Christoffersen, P., K.~Jacobs, and C.~Ornthanalai (2012).
\newblock Dynamic jump intensities and risk premiums: Evidence from {S}\&{P}
  500 returns and options.
\newblock {\em Journal of Financial Economics\/}~{\em 106\/}(3), 447--472.

\bibitem[\protect\citeauthoryear{Dahlhaus}{Dahlhaus}{1997}]{dahlhaus1997}
Dahlhaus, R. (1997).
\newblock Fitting time series models to nonstationary processes.
\newblock {\em Annals of Statistics\/}~{\em 25\/}(1), 1--37.

\bibitem[\protect\citeauthoryear{Dahlhaus, Richter, and Wu}{Dahlhaus
  et~al.}{2019}]{dahlhaus2017}
Dahlhaus, R., S.~Richter, and W.~B. Wu (2019).
\newblock Towards a general theory for nonlinear locally stationary processes.
\newblock {\em Bernoulli\/}~{\em 25\/}(2), 1013--1044.

\bibitem[\protect\citeauthoryear{Dahlhaus and Subba~Rao}{Dahlhaus and
  Subba~Rao}{2006}]{dahlhaus2006}
Dahlhaus, R. and S.~Subba~Rao (2006).
\newblock Statistical inference for time-varying {ARCH} processes.
\newblock {\em Ann. Statist.\/}~{\em 34\/}(3), 1075--1114.

\bibitem[\protect\citeauthoryear{Das, Duffie, Kapadia, and Saita}{Das
  et~al.}{2007}]{das2007}
Das, S.~R., D.~Duffie, N.~Kapadia, and L.~Saita (2007).
\newblock Common failings: How corporate defaults are correlated.
\newblock {\em The Journal of Finance\/}~{\em 62\/}(1), 93--117.

\bibitem[\protect\citeauthoryear{Davis and Liu}{Davis and
  Liu}{2016}]{Davis2016}
Davis, R.~A. and H.~Liu (2016).
\newblock Theory and inference for a class of nonlinear models with application
  to time series of counts.
\newblock {\em Statistica Sinica\/}~{\em 26\/}(4), 1673--1707.

\bibitem[\protect\citeauthoryear{Doukhan and Wintenberger}{Doukhan and
  Wintenberger}{2008}]{doukhan2008}
Doukhan, P. and O.~Wintenberger (2008).
\newblock Weakly dependent chains with infinite memory.
\newblock {\em Stochastic Processes and their Applications\/}~{\em 118\/}(11),
  1997--2013.

\bibitem[\protect\citeauthoryear{Duffie, Saita, and Wang}{Duffie
  et~al.}{2007}]{duffie2007}
Duffie, D., L.~Saita, and K.~Wang (2007).
\newblock Multi-period corporate default prediction with stochastic covariates.
\newblock {\em Journal of Financial Economics\/}~{\em 83\/}(3), 635--665.

\bibitem[\protect\citeauthoryear{Fan and Gijbels}{Fan and
  Gijbels}{2018}]{Fan2018}
Fan, J. and I.~Gijbels (2018).
\newblock {\em Local Polynomial Modelling and Its Applications}.
\newblock Routledge.

\bibitem[\protect\citeauthoryear{Fan, Heckman, and Wand}{Fan
  et~al.}{1995}]{fan1995JASA}
Fan, J., N.~E. Heckman, and M.~P. Wand (1995).
\newblock Local polynomial kernel regression for generalized linear models and
  quasi-likelihood functions.
\newblock {\em Journal of the American Statistical Association\/}~{\em
  90\/}(429), 141--150.

\bibitem[\protect\citeauthoryear{Fokianos and Tj\o{}stheim}{Fokianos and
  Tj\o{}stheim}{2011}]{fokianos2011}
Fokianos, K. and D.~Tj\o{}stheim (2011).
\newblock Log-linear poisson autoregression.
\newblock {\em Journal of Multivariate Analysis\/}~{\em 102\/}(3), 563--578.

\bibitem[\protect\citeauthoryear{Fryzlewicz, Sapatinas, and
  Subba~Rao}{Fryzlewicz et~al.}{2008}]{fryzlewicz2008}
Fryzlewicz, P., T.~Sapatinas, and S.~Subba~Rao (2008).
\newblock Normalized least-squares estimation in time-varying {ARCH} models.
\newblock {\em Annals of Statistics\/}~{\em 36\/}(2), 742--786.

\bibitem[\protect\citeauthoryear{Ghysels and Hall}{Ghysels and
  Hall}{1990}]{Ghysels1990}
Ghysels, E. and A.~Hall (1990).
\newblock Are consumption-based intertemporal capital asset pricing models
  structural?
\newblock {\em Journal of Econometrics\/}~{\em 45\/}(1-2), 121--139.

\bibitem[\protect\citeauthoryear{Giacomini and Rossi}{Giacomini and
  Rossi}{2016}]{giacomini2016}
Giacomini, R. and B.~Rossi (2016).
\newblock Model comparisons in unstable environments.
\newblock {\em International Economic Review\/}~{\em 57\/}(2), 369--392.

\bibitem[\protect\citeauthoryear{Han and Kristensen}{Han and
  Kristensen}{2014}]{han2014}
Han, H. and D.~Kristensen (2014).
\newblock Asymptotic theory for the {QMLE} in {GARCH-X} models with stationary
  and nonstationary covariates.
\newblock {\em Journal of Business \& Economic Statistics\/}~{\em 32\/}(3),
  416--429.

\bibitem[\protect\citeauthoryear{Hillebrand}{Hillebrand}{2005}]{Hillebrand2005}
Hillebrand, E. (2005).
\newblock Neglecting parameter changes in {GARCH} models.
\newblock {\em Journal of Econometrics\/}~{\em 129\/}(1-2), 121--138.

\bibitem[\protect\citeauthoryear{Inoue, Jin, and Pelletier}{Inoue
  et~al.}{2019}]{inoue2019}
Inoue, A., L.~Jin, and D.~Pelletier (2019).
\newblock Local-linear estimation of time-varying-parameter {GARCH} models and
  associated risk measures.
\newblock Working Paper.

\bibitem[\protect\citeauthoryear{Inoue and Rossi}{Inoue and
  Rossi}{2011}]{Inoue2011}
Inoue, A. and B.~Rossi (2011).
\newblock Identifying the sources of instabilities in macroeconomic
  fluctuations.
\newblock {\em Review of Economics and Statistics\/}~{\em 93\/}(4), 1186--1204.

\bibitem[\protect\citeauthoryear{Kristensen and Rahbek}{Kristensen and
  Rahbek}{2005}]{kristensen2005}
Kristensen, D. and A.~Rahbek (2005).
\newblock Asymptotics of the {QMLE} for a class of {ARCH}(q) models.
\newblock {\em Econometric Theory\/}~{\em 21\/}(5), 946--961.

\bibitem[\protect\citeauthoryear{Lando and Nielsen}{Lando and
  Nielsen}{2010}]{lando2010}
Lando, D. and M.~S. Nielsen (2010).
\newblock Correlation in corporate defaults: Contagion or conditional
  independence?
\newblock {\em Journal of Financial Intermediation\/}~{\em 19\/}(3), 355--372.

\bibitem[\protect\citeauthoryear{Loader}{Loader}{2006}]{loader2006}
Loader, C. (2006).
\newblock {\em Local regression and likelihood}.
\newblock Springer Science \& Business Media.

\bibitem[\protect\citeauthoryear{Meyn and Tweedie}{Meyn and
  Tweedie}{2009}]{meyn2009}
Meyn, S.~P. and R.~L. Tweedie (2009).
\newblock {\em Markov Chains and Stochastic Stability\/} (2nd ed.).
\newblock Springer London.

\bibitem[\protect\citeauthoryear{Newey and McFadden}{Newey and
  McFadden}{1994}]{newey1994}
Newey, W.~K. and D.~McFadden (1994).
\newblock Large sample estimation and hypothesis testing.
\newblock {\em Handbook of Econometrics\/}~{\em 4}, 2111--2245.

\bibitem[\protect\citeauthoryear{Richter and Dahlhaus}{Richter and
  Dahlhaus}{2019}]{richter}
Richter, S. and R.~Dahlhaus (2019).
\newblock Cross validation for locally stationary processes.
\newblock {\em Annals of Statistics\/}~{\em 47\/}(4), 2145--2173.

\bibitem[\protect\citeauthoryear{Truquet}{Truquet}{2019}]{Truquet2019}
Truquet, L. (2019).
\newblock Local stationarity and time-inhomogeneous markov chains.
\newblock {\em The Annals of Statistics\/}~{\em 47\/}(4), 2023--2049.

\bibitem[\protect\citeauthoryear{Truquet}{Truquet}{2020}]{Truquet2020}
Truquet, L. (2020).
\newblock A perturbation analysis of markov chains models with time-varying
  parameters.
\newblock {\em Bernoulli\/}~{\em 26\/}(4).

\bibitem[\protect\citeauthoryear{Vazquez-Abad and Kushner}{Vazquez-Abad and
  Kushner}{1992}]{VazquezAbad1992}
Vazquez-Abad, F.~J. and H.~J. Kushner (1992).
\newblock Estimation of the derivative of a stationary measure with respect to
  a control parameter.
\newblock {\em Journal of Applied Probability\/}~{\em 29\/}(2), 343--352.

\end{thebibliography}

\newpage

\appendix

\section{Appendix}

\subsection{Auxiliary Results\label{subsec: lemma}}

In the following, assume that $L$ satisfies: (i) $L\left( \cdot \right) $
has a compact support; (ii) for some $\Lambda <\infty $, $\left\vert
L(v)-L(v^{\prime })\right\vert \leq \Lambda \left\vert v-v^{\prime
}\right\vert $, $v,v^{\prime }\in \mathbb{R}$. We denote $L_{b}\left( \cdot
\right) :=L\left( \cdot /b\right) /b$.

\begin{lemma}
\label{thm: Aux result} The following hold as $b\rightarrow 0$ and $%
nb\rightarrow \infty $:

(i) Suppose $\left\{ W_{n,t}\left( \theta \right) \right\} $ is ULS$\left(
p,q,\Theta \right) $ with its stationary approximation $\left\{ W_{t}^{\ast
}\left( \theta |u\right) \right\} $ being $L_{p}$ continuous for some $p\geq
1,q>0$ and $\Theta $ is compact. Then, with $\mathcal{A}$ defined in
Assumption \ref{assu: compactness} and for any $u\in (0,1)$, 
\begin{equation*}
\sup_{\alpha \in \mathcal{A}}\left\Vert \frac{1}{n}\sum_{t=1}^{n}L_{b}\left(
t/n-u\right) W_{n,t}\left( D_{b}\left( t/n-u\right) \alpha \right) -\int
L\left( v\right) \mathbb{E}\left[ W_{t}^{\ast }\left( D\left( v\right)
\alpha |u\right) \right] dv\right\Vert =o_{p}\left( 1\right) .
\end{equation*}%
(ii) Suppose $\left\{ W_{n,t}\left( \theta \left( t/n\right) \right) ,%
\mathcal{F}_{n,t}\right\} $ is a MGD array; $V_{n,t}\left( \theta \right)
=W_{n,t}\left( \theta \right) W_{n,t}^{\prime }\left( \theta \right) $ is ULS%
$\left( p,q,\left\{ \theta :\left\Vert \theta -\theta \left( u\right)
\right\Vert <\epsilon \right\} \right) $ for some $p\geq 1$ and $q,\epsilon
>0$ with its stationary approximation $V_{t}^{\ast }\left( \theta |u\right) $
being $L_{p}$ continuous; and $v\mapsto \theta \left( v\right) $ is
continuous at $v=u$. Then, for any $u\in (0,1)$, 
\begin{align*}
\sqrt{\frac{b}{n}}\sum_{t=1}^{n}L_{b}\left( t/n-u\right) W_{n,t}\left(
\theta \left( t/n\right) \right) & \rightarrow ^{d}N\left( 0,\int
L^{2}\left( v\right) dv\times \mathbb{E}\left[ V_{t}^{\ast }\left( \theta
\left( u\right) |u\right) \right] \right) ; \\
\sqrt{\frac{b}{n}}\sum_{t=1}^{n}L_{b}\left( t/n-cb\right) W_{n,t}\left(
\theta \left( t/n\right) \right) & \rightarrow ^{d}N\left(
0,\int_{-c}^{+\infty }L^{2}\left( v\right) dv\times \mathbb{E}\left[
V_{t}^{\ast }\left( \theta \left( u\right) |u\right) \right] \right) .
\end{align*}%
(iii) Suppose $W_{t}^{\ast }$ is stationary and ergodic with $%
\sum_{s=0}^{\infty }\left\vert cov\left( W_{t}^{\ast },W_{t+s}^{\ast
}\right) \right\vert <\infty $. Then, for any $u\in (0,1)$, 
\begin{equation*}
\left\vert \frac{1}{n}\sum_{t=1}^{n}L_{b}\left( t/n-u\right) W_{t}^{\ast
}-\int L\left( v\right) dv\times \mathbb{E}\left[ W_{t}^{\ast }\right]
\right\vert =o_{p}\left( 1/\sqrt{nb}\right) .
\end{equation*}
\end{lemma}

\begin{proof}[Proof of (i)]
We first show that for all $\theta \in \Theta $, 
\begin{equation*}
\frac{1}{n}\sum_{t=1}^{n}L_{b}\left( t/n-u\right) W_{n,t}\left( \theta
\right) \rightarrow ^{p}\int L\left( v\right) dv\times \mathbb{E}\left[
W_{t}^{\ast }\left( \theta |u\right) \right] .
\end{equation*}%
Note that $L(v)=0$ for $\left\vert v\right\vert \geq \bar{v}$ for some $\bar{%
v}>0$. Then, Minkowski's inequality implies that 
\begin{align*}
& \mathbb{E}\left[ \left\Vert \frac{1}{n}\sum_{t=1}^{n}L_{b}\left(
t/n-u\right) \left\{ W_{n,t}\left( \theta \right) -W_{t}^{\ast }\left(
\theta |u\right) \right\} \right\Vert ^{p}\right] ^{1/p} \\
& \leq \frac{1}{n}\sum_{t=1}^{n}\left\vert L_{b}\left( t/n-u\right)
\right\vert \mathbb{E}\left[ \left\Vert W_{n,t}\left( \theta \right)
-W_{t}^{\ast }\left( \theta |u\right) \right\Vert ^{p}\right] ^{1/p} \\
& \leq \frac{C}{n}\sum_{t=1}^{n}\left\vert L_{b}\left( t/n-u\right)
\right\vert \left( b^{q}\left\vert \frac{t/n-u}{b}\right\vert
^{q}+1/n^{q}+\rho ^{t}\right) \\
& \leq \frac{C}{n}\sum_{t=1}^{n}\left\vert L_{b}\left( t/n-u\right)
\right\vert \times \left( b^{q}\bar{v}^{q}+1/n^{q}+\rho ^{t}\right) =O\left(
b^{q}\right) +O\left( n^{-q}\right) +O\left( \frac{1}{\sqrt{nb}}\right) ,
\end{align*}%
where we have used that 
\begin{equation*}
\frac{1}{n}\sum_{t=1}^{n}\left\vert L_{b}\left( t/n-u\right) \right\vert
\rho ^{qt}\leq \frac{1}{\sqrt{nb}}\sqrt{\frac{1}{n}\sum_{t=1}^{n}\left(
L^{2}\right) _{b}\left( t/n-u\right) }\sqrt{\sum_{t=1}^{n}\rho ^{2qt}}%
=O\left( \frac{1}{\sqrt{nb}}\right) .
\end{equation*}%
Next, with $\bar{W}_{t}=W_{t}^{\ast }\left( \theta |u\right) -\mathbb{E}%
\left[ W_{t}^{\ast }\left( \theta |u\right) \right] $, $\frac{1}{n}%
\sum_{t=1}^{n}L_{b}\left( t/n-u\right) \bar{W}_{t}=\frac{1}{nb}\sum_{t=%
\underline{t}}^{\bar{t}}L_{b}\left( t/n-u\right) \bar{W}_{t}$ for
sufficiently large $n$, where $\bar{t}=\left[ n\left( u+\bar{v}b\right) %
\right] $ and $\underline{t}=\left[ n\left( u-\bar{v}b\right) \right] $.
Here, $\left[ x\right] $ denotes the integer part of any real number $x$. By
summation by parts, we have, with $S_{n,t}=\sum_{j=\underline{t}}^{t}\bar{%
W_{j}}$, 
\begin{align*}
\frac{1}{n}\sum_{t=\underline{t}}^{\bar{t}}L_{b}\left( t/n-u\right) \bar{W}%
_{t}& =\frac{1}{n}\sum_{t=\underline{t}}^{\bar{t}}L_{b}\left( t/n-u\right)
\left( S_{n,t}-S_{n,t-1}\right) \\
& =\frac{1}{n}\sum_{t=\underline{t}}^{\bar{t}-1}\left[ L_{b}\left(
t/n-u\right) -L_{b}\left( \left( t+1\right) /n-u\right) \right] S_{n,t}+%
\frac{1}{n}L_{b}\left( \bar{t}/n-u\right) S_{n,\bar{t}}.
\end{align*}%
Since $\left\{ \bar{W}_{t}\right\} $ is stationary, $S_{n,t}$ has the same
distribution as $\tilde{S}_{n,t}=\sum_{j=1}^{t-\underline{t}+1}\bar{W}_{j}$.
Thus, for some constant $M$, $\left\vert \frac{1}{n}\sum_{t=1}^{n}L_{b}%
\left( t/n-u\right) \bar{W}_{t}\right\vert \leq \frac{M}{nb}\sup_{t\leq \bar{%
t}-\underline{t}+1}\left\vert \tilde{S}_{n,t}\right\vert $. The ergodic
theorem yields $\tilde{S}_{n,t}/t\rightarrow 0$ which in turn implies that $%
\frac{1}{n}\sum_{t=1}^{n}L_{b}\left( t/n-u\right) \bar{W}_{t}$ tends to zero
almost surely. Finally, using the mean value theorem, there exists $%
v_{n,t}\in \left[ \frac{t-1}{n},\frac{t}{n}\right] $ so that with $\bar{L}%
=\sup_{v}L\left( v\right) $, 
\begin{align*}
\left\vert \frac{1}{n}\sum_{t=1}^{n}L_{b}\left( t/n-u\right) -\int
L_{b}\left( x-u\right) dx\right\vert & =\left\vert \frac{1}{nb}%
\sum_{t=1}^{n}L_{b}\left( t/n-u\right)
-\sum_{t=1}^{n}\int_{(t-1)/n}^{t/n}L_{b}\left( x-u\right) dx\right\vert \\
& \leq \frac{1}{nb}\sum_{t=1}^{n}\left\vert L_{b}\left( t/n-u\right)
-L{}_{b}\left( v_{n,t}-u\right) \right\vert \\
& \leq \frac{1}{nb}\sum_{t=1}^{n}\Lambda \left\vert \frac{t/n-v_{n,t}}{b}%
\right\vert =O\left( \frac{1}{nb}\right) ,
\end{align*}%
which shows that $\frac{1}{n}\sum_{t=1}^{n}L_{b}\left( t/n-u\right) \mathbb{E%
}\left[ W_{t}^{\ast }\left( \theta |u\right) \right] =\int L_{b}\left(
x-u\right) dx\mathbb{E}\left[ W_{t}^{\ast }\left( \theta |u\right) \right]
+O\left( 1/\left( nb\right) \right) $.

For the uniform convergence, we note that by definition of $\mathcal{A}$, $%
D_{b}\left(v-u\right)\alpha\in\Theta$ for all $v\in\mathrm{supp}%
\left(L\right)$ and $\alpha\in\mathcal{A}$. Thus, $\frac{1}{n}%
\sum_{t=1}^{n}K_{b}\left(t/n-u\right)W_{n,t}\left(D_{n,t}\left(u\right)%
\alpha\right)$, where $D_{n,t}\left(u\right)=D_{b}\left(t/n-u\right)$, is
well-defined for $\alpha\in\mathcal{A}$, and 
\begin{align*}
\mathbb{E}\left[\sup_{\alpha\in\mathcal{A}}\left\Vert
W_{n,t}\left(D_{n,t}\left(u\right)\alpha\right)-W_{t}^{*}\left(D_{n,t}%
\left(u\right)\alpha|u\right)\right\Vert ^{p}\right] & \leq\mathbb{E}\left[%
\sup_{\theta\in\Theta}\left\Vert
W_{n,t}\left(\theta\right)-W_{t}^{*}\left(\theta|u\right)\right\Vert ^{p}%
\right] \\
& \leq C\left(\left|t/n-u\right|^{q}+1/n^{q}+\rho^{t}\right)^{p}.
\end{align*}
Using H{\"o}lder's and Minkowski's inequality, 
\begin{align*}
& \mathbb{E}\left[\sup_{\alpha\in\mathcal{A}}\left\Vert \frac{1}{n}%
\sum_{t=1}^{n}L_{b}\left(t/n-u\right)\left\{
W_{n,t}\left(D_{n,t}\left(u\right)\alpha\right)-W_{t}^{*}\left(D_{n,t}%
\left(u\right)\alpha|u\right)\right\} \right\Vert \right] \\
& \leq\frac{1}{n}\sum_{t=1}^{n}\left|L_{b}\left(t/n-u\right)\right|\mathbb{E}%
\left[\sup_{\alpha\in\mathcal{A}}\left\Vert
W_{n,t}\left(D_{n,t}\left(u\right)\alpha\right)-W_{t}^{*}\left(D_{n,t}%
\left(u\right)\alpha|u\right)\right\Vert ^{p}\right]^{1/p} \\
& \leq C\frac{b^{q}}{n}\sum_{t=1}^{n}\left|L_{b}\left(t/n-u\right)\right|%
\left(\left|\frac{t/n-u}{b}\right|^{q}+1/n^{q}+\rho^{t}\right)=O\left(b^{q}%
\right).
\end{align*}
Next, 
\begin{align*}
& \sup_{\alpha\in\mathcal{A}}\left\Vert \frac{1}{n}\sum_{t=1}^{n}L_{b}%
\left(t/n-u\right)\left\{
W_{t}^{*}\left(D_{n,t}\left(u\right)\alpha|u\right)-\mathbb{E}\left[%
W_{t}^{*}\left(D_{n,t}\left(u\right)\alpha|u\right)\right]\right\}
\right\Vert \\
& \leq\sup_{\theta\in\Theta}\left\Vert \frac{1}{n}\sum_{t=1}^{n}L_{b}%
\left(t/n-u\right)\left\{ W_{t}^{*}\left(\theta|u\right)-\mathbb{E}\left[%
W_{t}^{*}\left(\theta|u\right)\right]\right\} \right\Vert
+o_{p}\left(1\right)
\end{align*}
where $\frac{1}{n}\sum_{t=1}^{n}L_{b}\left(t/n-u\right)\left\{
W_{t}^{*}\left(\theta|u\right)-\mathbb{E}\left[W_{t}^{*}\left(\theta|u\right)%
\right]\right\} =o_{P}\left(1\right)$ for all $\theta\in\Theta$. Thus, the
result will follow if we can show stochastic equicontinuity of $\theta\mapsto%
\frac{1}{n}\sum_{t=1}^{n}L_{b}\left(t/n-u\right)W_{t}^{*}\left(\theta|u%
\right)$ but this follows from the assumption of $\theta\mapsto
W_{t}^{*}\left(\theta|u\right)$ being $L_{p}$ continuous: For a given $%
\theta\in\Theta$ and $\epsilon>0$ there exists $\delta>0$ so that 
\begin{align*}
& \mathbb{E}\left[\sup_{\theta^{\prime}:\left\Vert
\theta-\theta^{\prime}\right\Vert <\delta}\left\Vert \frac{1}{n}%
\sum_{t=1}^{n}L_{b}\left(t/n-u\right)W_{t}^{*}\left(\theta|u\right)-\frac{1}{%
n}\sum_{t=1}^{n}L_{b}\left(t/n-u\right)W_{t}^{*}\left(\theta^{\prime}|u%
\right)\right\Vert \right]\, \\
\leq & \frac{1}{n}\sum_{t=1}^{n}\left|L_{b}\left(t/n-u\right)\right|\mathbb{E%
}\left[\sup_{\theta^{\prime}:\left\Vert \theta-\theta^{\prime}\right\Vert
<\delta}\left\Vert
W_{t}^{*}\left(\theta|u\right)-W_{t}^{*}\left(\theta^{\prime}|u\right)\right%
\Vert \right] \\
= & \frac{\epsilon}{n}\sum_{t=1}^{n}\left|L_{b}\left(t/n-u\right)\right|=O%
\left(\epsilon\right).
\end{align*}

\textbf{Proof of (ii).} Observe that $\sqrt{b/n}\sum_{t=1}^{n}L_{b}\left(
t/n-u\right) W_{n,t}\left( \theta \left( t/n\right) \right) $ is a
martingale with quadratic variation $Q_{n}=\frac{b}{n}%
\sum_{t=1}^{n}L_{b}^{2}\left( t/n-u\right) V_{n,t}\left( \theta \left(
t/n\right) \right) $. To derive the limit of $Q_{n}$, write 
\begin{align*}
Q_{n}=& \frac{b}{n}\sum_{t=1}^{n}L_{b}^{2}\left( t/n-u\right) \mathbb{E}%
\left[ V_{t}^{\ast }\left( \theta \left( t/n\right) |u\right) \right] +\frac{%
b}{n}\sum_{t=1}^{n}L_{b}^{2}\left( t/n-u\right) \left\{ V_{n,t}\left( \theta
\left( t/n\right) \right) -V_{t}^{\ast }\left( \theta \left( t/n\right)
|u\right) \right\} \\
& +\frac{b}{n}\sum_{t=1}^{n}L_{b}^{2}\left( t/n-u\right) \left\{ V_{t}^{\ast
}\left( \theta \left( t/n\right) |u\right) -\mathbb{E}\left[ V_{t}^{\ast
}\left( \theta \left( t/n\right) |u\right) \right] \right\} .
\end{align*}%
For the first term, employing standard results for kernel averages together
with the fact that $\theta \mapsto \mathbb{E}\left[ V_{t}^{\ast }\left(
\theta |u\right) \right] $ is continuous (because $V_{t}^{\ast }\left(
\theta |u\right) $ is $L_{1}$-continuous), 
\begin{equation*}
\frac{b}{n}\sum_{t=1}^{n}L_{b}^{2}\left( t/n-u\right) \mathbb{E}\left[
V_{t}^{\ast }\left( \theta \left( t/n\right) |u\right) \right] \rightarrow
\int L^{2}\left( x\right) dx\mathbb{E}\left[ V_{t}^{\ast }\left( \theta
\left( u\right) |u\right) \right] .
\end{equation*}%
Applying arguments similar to those in the proof of Theorem \ref{thm: Aux
result}(i) together with continuity of $v\mapsto \theta \left( v\right) $, $%
L_{1}$-continuity of $\theta \mapsto V_{t}^{\ast }\left( \theta |u\right) $
and $L$ having compact support, we have for all $n$ large enough, 
\begin{align*}
& \frac{b}{n}\sum_{t=1}^{n}L_{b}^{2}\left( t/n-u\right) \mathbb{E}\left[
\left\Vert V_{n,t}\left( \theta \left( t/n\right) \right) -V_{t}^{\ast
}\left( \theta \left( t/n\right) |u\right) \right\Vert \right] \\
& \leq \frac{b}{n}\sum_{t=1}^{n}L_{b}^{2}\left( t/n-u\right)
\sup_{\left\Vert \theta -\theta \left( u\right) \right\Vert <\epsilon }%
\mathbb{E}\left[ \left\Vert V_{n,t}\left( \theta \right) -V_{t}^{\ast
}\left( \theta |u\right) \right\Vert \right] =o\left( 1\right) ,
\end{align*}%
and 
\begin{align*}
& \frac{b}{n}\sum_{t=1}^{n}L_{b}^{2}\left( t/n-u\right) \left\{ V_{t}^{\ast
}\left( \theta \left( t/n\right) |u\right) -\mathbb{E}\left[ V_{t}^{\ast
}\left( \theta \left( t/n\right) |u\right) \right] \right\} \\
& \leq \frac{b}{n}\sum_{t=1}^{n}L_{b}^{2}\left( t/n-u\right)
\sup_{\left\Vert \theta -\theta \left( u\right) \right\Vert <\epsilon }%
\mathbb{E}\left[ \left\Vert V_{t}^{\ast }\left( \theta |u\right) -\mathbb{E}%
\left[ V_{t}^{\ast }\left( \theta |u\right) \right] \right\Vert \right]
=o\left( 1\right) .
\end{align*}%
The result now follows if the Lindeberg condition is satisfied, c.f. %
\citet{Brown1971}. But, as $nb\rightarrow \infty $, with $m_{n,t}\left(
\theta \right) =\sqrt{b/n}L_{b}\left( t/n-u\right) W_{t}^{\ast }\left(
\theta |u\right) $, 
\begin{align*}
& \sum_{t=1}^{n}\left\Vert m_{n,t}\left( \theta \left( t/n\right) \right)
\right\Vert ^{2}1\left( \left\Vert m_{n,t}\left( \theta \left( t/n\right)
\right) \right\Vert >\varepsilon \right) \\
\leq & \sum_{t=1}^{n}\left( \left\Vert m_{n,t}\left( \theta \left(
t/n\right) \right) \right\Vert ^{2}-\left\Vert m_{t}^{\ast }\left( \theta
\left( u\right) |u\right) \right\Vert ^{2}\right) 1\left( \left\Vert
m_{n,t}\left( \theta \left( t/n\right) \right) \right\Vert >\varepsilon
\right) \\
& +\sum_{t=1}^{n}\left\Vert m_{t}^{\ast }\left( \theta \left( u\right)
|u\right) \right\Vert ^{2}1\left( \left\Vert m_{n,t}\left( \theta \left(
t/n\right) \right) \right\Vert >\varepsilon ,\left\Vert m_{t}^{\ast }\left(
\theta \left( u\right) |u\right) \right\Vert \leq \varepsilon /\sqrt{2}%
\right) \\
& +\sum_{t=1}^{n}\left\Vert m_{t}^{\ast }\left( \theta \left( u\right)
|u\right) \right\Vert ^{2}1\left( \left\Vert m_{t}^{\ast }\left( \theta
\left( u\right) |u\right) \right\Vert >\varepsilon /\sqrt{2}\right) .
\end{align*}%
Recycling the arguments used in the analysis of $Q_{n}$, it follows that the
first and third terms are $o_{p}\left( 1\right) $. Similarly, the
convergence of the second term is obtained with the following inequality and
Markov's inequality: 
\begin{align*}
& \sum_{t=1}^{n}\left\Vert m_{t}^{\ast }\left( \theta \left( u\right)
|u\right) \right\Vert ^{2}1\left( \left\Vert m_{n,t}\left( \theta \left(
t/n\right) \right) \right\Vert >\varepsilon ,\left\Vert m_{t}^{\ast }\left(
\theta \left( u\right) |u\right) \right\Vert \leq \varepsilon /\sqrt{2}%
\right) \\
& \leq \frac{\varepsilon ^{2}}{2}\sum_{t=1}^{n}1\left( \left\Vert
m_{n,t}\left( \theta \left( t/n\right) \right) \right\Vert ^{2}-\left\Vert
m_{t}^{\ast }\left( \theta \left( u\right) |u\right) \right\Vert
^{2}>\varepsilon ^{2}/2\right) .
\end{align*}

\textbf{Proof of (iii). }Assume w.l.o.g. that $\mathbb{E}\left[ W_{t}^{\ast }%
\right] =0$ and then use 
\begin{align*}
Var\left( A_{n}\right) & \leq \frac{1}{n}\sum_{t_{1},t_{2}=1}^{n}\left\vert
L_{b}\left( t_{1}/n-u\right) \right\vert \left\vert L_{b}\left(
t_{2}/n-u\right) \right\vert \left\vert cov\left( W_{t_{1}}^{\ast
},W_{t_{2}}^{\ast }\right) \right\vert \\
& \leq \frac{\bar{L}}{\left( nb\right) ^{2}}\sum_{t_{1},t_{2}=1}^{n}\left%
\vert L\left( \frac{t_{1}/n-u}{b}\right) \right\vert \left\vert cov\left(
W_{t_{1}}^{\ast },W_{t_{2}}^{\ast }\right) \right\vert =O\left( \frac{1}{nb}%
\right) .
\end{align*}
\end{proof}

\subsection{\label{subsec: proof}Proofs of results in Section \protect\ref%
{sec: asymptotics}}

\medskip

\begin{proof}[Proof of Theorem \protect\ref{thm: Con}]
By Lemma \ref{thm: Aux result}(i), $\sup_{\alpha \in \mathcal{A}}\left\vert
Q_{n}\left( \alpha |u\right) -Q^{\ast }\left( \alpha |u\right) \right\vert
=o_{P}\left( 1\right) $, where $Q^{\ast }\left( \alpha |u\right) =\int_{%
\mathbb{R}}K\left( v\right) \mathbb{E}\left[ \ell _{t}^{\ast }\left(
D_{m}\left( v\right) \alpha |u\right) \right] dv$. Now, observe that for any 
$\alpha =\left( \alpha _{1},...,\alpha _{m+1}\right) $ with $\alpha _{i}\neq
0$ for some $i\geq 2$, the polynomial $v\mapsto D_{m}\left( v\right) \alpha $
is non-constant almost everywhere. Thus, for any $\alpha \neq \alpha ^{\ast
}=\left( \theta \left( u\right) ,0,...,0\right) $, $D_{m}\left( v\right)
\alpha \neq \theta \left( u\right) =D_{m}\left( v\right) \alpha ^{\ast }$
for almost all $v\in \left[ 0,1\right] $ and so by Assumption \ref{assu:
compactness}(iii) $\mathbb{E}\left[ \ell _{t}^{\ast }\left( D\left( v\right)
\alpha |u\right) \right] <\mathbb{E}\left[ \ell _{t}^{\ast }\left( \theta
\left( u\right) |u\right) \right] =\mathbb{E}\left[ \ell _{t}^{\ast }\left(
D_{m}\left( v\right) \alpha ^{\ast }|u\right) \right] $ for almost every $v$%
. Since $K\left( \cdot \right) \geq 0$ this in turn implies that $Q^{\ast
}\left( \alpha |u\right) \leq Q^{\ast }\left( \alpha ^{\ast }|u\right) $.
Finally, by the dominated convergence theorem together with Assumption \ref%
{assu: compactness}(ii) $\alpha \mapsto Q^{\ast }\left( \alpha |u\right) $
is continuous. This proves $\hat{\alpha}\rightarrow ^{p}\alpha ^{\ast }$,
c.f. Theorem 2.1 in \citet{newey1994}.
\end{proof}

\medskip

\begin{proof}[Proof of Theorem \protect\ref{thm: Norm1}]
From Theorem \ref{thm: Con} we know that $\hat{\alpha}\rightarrow ^{p}\alpha
^{\ast }:=\left( \theta \left( u\right) ,0,....,0\right) $. It is easily
checked that the limit is situated in the interior of $\mathcal{A}$ and so
w.p.a.1. so will $\hat{\alpha}$. As a consequence, $\hat{\alpha}$ will
satisfy (\ref{eq: FOC}) w.p.a.1. Adding and subtracting $S_{n}\left(
u\right) $ and then rearranging yields 
\begin{equation*}
0=\sqrt{nb}S_{n}\left( u\right) +H_{n}\left( \bar{\alpha}|u\right) \sqrt{nb}%
\left( \hat{\alpha}-\alpha _{0}-H_{n}^{-1}\left( \bar{\alpha}|u\right)
\left\{ S_{n}\left( \alpha _{0}|u\right) -S_{n}\left( u\right) \right\}
\right) .
\end{equation*}%
Here, $H_{n}^{-1}\left( \bar{\alpha}|u\right) $ is well-defined w.p.a.1
since, as shown below, it converges towards an invertible matrix. The
claimed asymptotic result now follows if we can verify the claims of eqs. (%
\ref{eq: as norm 1})-(\ref{eq: as norm 2}):

\textbf{Proof of eq. (\ref{eq: as norm 1}).} First note that, with $%
L(u)=K(u)D_{m}\left( u\right) $, $\sqrt{nb}S_{n}\left( u\right) =\sqrt{\frac{%
b}{n}}\sum_{t=1}^{n}L_{b}(t/n-u)\otimes s_{n,t}\left( \theta \left(
t/n\right) \right) $. The result now follows from Lemma \ref{thm: Aux result}%
(ii) under Assumptions \ref{assu: smoothness}(iii) and \ref{assu: score}(i).

\textbf{Proof of first claim of eq. (\ref{eq: as norm 2}).} With $%
L(u)=K(u)D_{m}\left( u\right) D_{m}\left( u\right) ^{\prime }$, we can write 
$H_{n}\left( \beta |u\right) =\frac{1}{n}\sum_{t=1}^{n}L_{b}(t/n-u)\otimes
h_{n,t}\left( D_{n,t}\left( u\right) \beta \right) $. It follows from Lemma %
\ref{thm: Aux result}(i) and Assumption \ref{assu: score}(ii) that $%
\sup_{\alpha \in \mathcal{B}(\epsilon )}\lVert H_{n}\left( \alpha |u\right) -%
\mathbb{K}_{1}\otimes H\left( D_{m}\left( u\right) \alpha |u\right) \rVert
=o_{p}(1)$, where $H\left( \theta |u\right) =\mathbb{E}\left[ h_{t}^{\ast
}\left( \theta |u\right) \right] $ is continuous w.r.t. $\theta $ and $%
\mathcal{B}(\epsilon )=\left\{ \alpha :\lVert \alpha -\alpha ^{\ast }\rVert
<\epsilon \right\} $ for some small $\epsilon >0$. Thus, given that $\bar{%
\alpha}\rightarrow ^{p}\alpha ^{\ast }$, $H_{n}\left( \bar{\alpha}|u\right)
\rightarrow ^{p}\mathbb{K}_{1}\otimes H\left( \theta \left( u\right)
|u\right) $. Finally, note here that since $K$ is a probability density
function, $\mathbb{K}_{1}$ is invertible, while $H\left( \theta \left(
u\right) |u\right) =H\left( u\right) $ is invertible by assumption.

\textbf{Proof of second claim of eq. (\ref{eq: as norm 2}).} First observe
that $D_{n,t}\left( u\right) \alpha ^{\ast }=\theta _{u}^{\ast }\left(
t/n\right) $ where $\theta _{u}^{\ast }\left( t/n\right) $ was defined in
eq. (\ref{eq: theta* def}). Now, employ the mean-value theorem to obtain
that, for some $\bar{\theta}_{n,t}$ lying between $\theta _{u}^{\ast }\left(
t/n\right) $ and $\theta \left( t/n\right) $ and some $u_{n,t}\in \left[
t/n,u\right] $, 
\begin{align*}
b_{n,t}& =s_{n,t}\left( \theta _{u}^{\ast }\left( t/n\right) \right)
-s_{n,t}\left( \theta \left( t/n\right) \right) \\
& =h_{n,t}\left( \bar{\theta}_{n,t}\right) \left\{ \theta _{u}^{\ast }\left(
t/n\right) -\theta \left( t/n\right) \right\} =-h_{n,t}\left( \bar{\theta}%
_{n,t}\right) \frac{\theta ^{\left( m+1\right) }\left( u_{n,t}\right) }{%
\left( m+1\right) !}\left( t/n-u\right) ^{m+1} \\
& =-h_{n,t}\left( \theta \left( t/n\right) \right) \frac{\theta ^{\left(
m+1\right) }\left( t/n\right) }{\left( m+1\right) !}\left( t/n-u\right)
^{m+1} \\
& +\left\{ h_{n,t}\left( \theta \left( t/n\right) \right) \theta ^{\left(
m+1\right) }\left( t/n\right) -h_{n,t}\left( \bar{\theta}_{n,t}\right)
\theta ^{\left( m+1\right) }\left( u_{n,t}\right) \right\} \frac{\left(
t/n-u\right) ^{m+1}}{\left( m+1\right) !} \\
& =:b_{n,t}^{\left( 1\right) }+b_{n,t}^{\left( 2\right) }.
\end{align*}%
The first term is locally stationary and so by the same arguments as in the
proof of Lemma \ref{thm: Aux result}(ii),%
\begin{eqnarray*}
\frac{1}{n}\sum_{t=1}^{n}K_{n,t}\left( u\right) D_{n,t}\left( u\right)
^{\prime }b_{n,t}^{\left( 1\right) } &=&-\frac{b^{m+1}}{n}%
\sum_{t=1}^{n}K_{n,t}\left( u\right) D_{n,t}\left( u\right) ^{\prime }\left( 
\frac{t/n-u}{b}\right) ^{m+1}h_{n,t}\left( \theta \left( t/n\right) \right) 
\frac{\theta ^{\left( m+1\right) }\left( t/n\right) }{\left( m+1\right) !} \\
&=&-b^{m+1}\left\{ \mu _{1}\otimes H\left( u\right) \frac{\theta ^{\left(
m+1\right) }\left( u\right) }{\left( m+1\right) !}+o_{p}\left( 1\right)
\right\} .
\end{eqnarray*}%
For the second term, observe that for $\left\vert t/n-u\right\vert \leq Cb$, 
$\lVert \bar{\theta}_{n,t}-\theta \left( t/n\right) \rVert \leq \lVert
\theta _{u}^{\ast }\left( t/n\right) -\theta \left( t/n\right) \rVert \leq 
\tilde{C}b^{m+1}$ and so, using the ULS property of $h_{n,t}\left( \theta
\right) $,%
\begin{eqnarray*}
&&\sup_{n,t}\mathbb{E}\left[ \lVert h_{n,t}\left( \theta \left( t/n\right)
\right) -h_{n,t}\left( \bar{\theta}_{n,t}\right) \rVert \right] \\
&\leq &C\left( b^{q}\left\vert \frac{t/n-u}{b}\right\vert
^{q}+1/n^{q}\right) +\sup_{\lVert \theta -\theta ^{\ast }\rVert \leq \tilde{C%
}b^{m+1}}\mathbb{E}\left[ \lVert h_{t}^{\ast }\left( \theta |u\right)
-h_{t}^{\ast }\left( \theta ^{\prime }|u\right) \rVert \right] \rightarrow 0,
\end{eqnarray*}%
as $n\rightarrow \infty $ and $b\rightarrow 0$. Similarly, $\sup_{n,t}\lVert
\theta ^{\left( m+1\right) }\left( t/n\right) -\theta ^{\left( m+1\right)
}\left( u_{n,t}\right) \rVert \rightarrow 0$ as $n\rightarrow \infty $.
These two results combined show that the $\frac{b^{m+1}}{n}%
\sum_{t=1}^{n}K_{n,t}\left( u\right) D_{n,t}\left( u\right) ^{\prime }\left( 
\frac{t/n-u}{b}\right) ^{m+1}b_{n,t}^{\left( 2\right) }=o_{p}\left( 1\right) 
$.
\end{proof}

\medskip

\begin{proof}[Proof of Theorem \protect\ref{thm: Norm2}]
Write%
\begin{equation*}
b_{n,t}=s_{n,t}\left( \theta _{u}^{\ast }\left( t/n\right) \right)
-s_{n,t}\left( \theta \left( t/n\right) \right) =s_{t}^{\ast }\left( \theta
_{u}^{\ast }\left( t/n\right) ,t/n\right) -s_{t}^{\ast }\left( \theta \left(
t/n\right) ,t/n\right) +O_{P}\left( n^{-q}\right) .
\end{equation*}

The proof proceeds exactly as the one of Theorem \ref{thm: Norm1} except
that the first-order expansion of $b_{n,t}$ is replaced by a second-order
expansion of $s_{n,t}\left( \theta \right) $ w.r.t. $\theta $ combined with 
\begin{equation*}
\theta \left( t/n\right) =\theta _{u}^{\ast }\left( t/n\right) +\frac{\theta
^{\left( m+1\right) }\left( u\right) }{\left( m+1\right) !}\left\{
t/n-u\right\} ^{m+1}+\frac{\theta ^{\left( m+2\right) }\left( \bar{u}%
_{n,t}\right) }{\left( m+2\right) !}\left\{ t/n-u\right\} ^{m+2},
\end{equation*}%
where $u_{n,t}$ lies between $u$ and $t/n$. This yields%
\begin{eqnarray*}
b_{n,t} &=&h_{n,t}\left( \theta \left( t/n\right) \right) \left\{ \theta
_{u}^{\ast }\left( t/n\right) -\theta \left( t/n\right) \right\} +\frac{1}{2}%
\sum_{i=1}^{d_{\theta }}\left\{ \theta _{u,i}^{\ast }\left( t/n\right)
-\theta _{i}\left( t/n\right) \right\} \frac{\partial h_{n,t}\left( \bar{%
\theta}_{n,t}\right) }{\partial \theta _{i}}\left\{ \theta _{u}^{\ast
}\left( t/n\right) -\theta \left( t/n\right) \right\} \\
&=&-h_{n,t}\left( \theta \left( t/n\right) \right) \left[ \frac{\theta
^{\left( m+1\right) }\left( u\right) }{\left( m+1\right) !}\left\{
t/n-u\right\} ^{m+1}+\frac{\theta ^{\left( m+2\right) }\left( u_{n,t}\right) 
}{\left( m+2\right) !}\left\{ t/n-u\right\} ^{m+2}\right] \\
&&+\frac{1}{2}\sum_{i=1}^{d_{\theta }}\frac{\theta _{i}^{\left( m+1\right)
}\left( u_{n,t}\right) }{\left( m+1\right) !}\frac{\partial h_{n,t}\left( 
\bar{\theta}_{n,t}\right) }{\partial \theta _{i}}\frac{\theta ^{\left(
m+1\right) }\left( u_{n,t}\right) }{\left( m+1\right) !}\left\{
t/n-u\right\} ^{2m+2}.
\end{eqnarray*}%
Thus, with $L_{n,t}^{(m)}(u)=K_{n,t}\left( u\right) D_{n,t}\left( u\right)
^{\prime }\left( \frac{t/n-u}{b}\right) ^{m}$,%
\begin{eqnarray*}
B_{n}\left( u\right) &=&\frac{1}{n}\sum_{t=1}^{n}K_{n,t}\left( u\right)
D_{n,t}\left( u\right) ^{\prime }b_{n,t} \\
&=&-b^{m+1}\left\{ \frac{1}{n}\sum_{t=1}^{n}L_{n,t}^{(m+1)}(u)h_{n,t}\left(
\theta \left( t/n\right) \right) \right\} \frac{\theta ^{\left( m+1\right)
}\left( u\right) }{\left( m+1\right) !} \\
&&-b^{m+2}\left\{ \frac{1}{n}\sum_{t=1}^{n}L_{n,t}^{(m+2)}(u)h_{n,t}\left(
\theta \left( t/n\right) \right) \theta ^{\left( m+2\right) }\left(
u_{n,t}\right) \right\} \frac{1}{\left( m+2\right) !} \\
&&+\frac{b^{2m+2}}{2\left\{ \left( m+1\right) !\right\} ^{2}}%
\sum_{i=1}^{d_{\theta }}\left\{ \frac{1}{n}\sum_{t=1}^{n}\theta _{i}^{\left(
m+1\right) }\left( u_{n,t}\right) L_{n,t}^{(2m+2)}(u)\frac{\partial
h_{n,t}\left( \bar{\theta}_{n,t}\right) }{\partial \theta _{i}}\theta
^{\left( m+1\right) }\left( u_{n,t}\right) \right\} .
\end{eqnarray*}%
We now analyse the three bracketed terms. First,%
\begin{align*}
& \frac{1}{n}\sum_{t=1}^{n}L_{n,t}^{(m+1)}(u)h_{n,t}\left( \theta \left(
t/n\right) \right) \\
& =\frac{1}{n}\sum_{t=1}^{n}L_{n,t}^{(m+1)}(u)\left[ h_{n,t}\left( \theta
\left( t/n\right) \right) -h_{n,t}\left( \theta \left( u\right) \right) %
\right] \\
& +\frac{1}{n}\sum_{t=1}^{n}L_{n,t}^{(m+1)}(u)h_{t}^{\ast }\left( \theta
\left( u\right) |u\right) +\frac{b}{n}\sum_{t=1}^{n}L_{n,t}^{(m+2)}(u)%
\partial _{u}h_{t}^{\ast }\left( \theta \left( u\right) |u\right) \\
& +\frac{1}{n}\sum_{t=1}^{n}L_{n,t}^{(m+1)}(u)\left[ h_{n,t}\left( \theta
\left( u\right) \right) -h_{t}^{\ast }\left( \theta \left( u\right)
|u\right) -\partial _{u}h_{t}^{\ast }\left( \theta \left( u\right) |u\right)
\left\{ t/n-u\right\} \right] .
\end{align*}%
By Lemma \ref{thm: Aux result}(iii) together with Assumption \ref{assu:
hessian shortdep} and Lemma \ref{thm: Aux result}(i), 
\begin{align}
\frac{1}{n}\sum_{t=1}^{n}L_{n,t}^{(m+1)}(u)\left[ h_{n,t}\left( \theta
\left( t/n\right) \right) -h_{n,t}\left( \theta \left( u\right) \right) %
\right] & =b\mu _{m+2}\sum_{i=1}^{d_{\theta }}\theta _{i}^{\left( 1\right)
}\left( u\right) \partial _{\theta _{i}}H\left( u\right) +o_{P}\left(
b\right) ,  \label{eq: term 1} \\
\frac{1}{n}\sum_{t=1}^{n}L_{n,t}^{(m+1)}(u)h_{t}^{\ast }\left( \theta \left(
u\right) |u\right) & =\mu _{m+1}H\left( u\right) +o_{P}\left( 1/\sqrt{nb}%
\right) ,  \notag \\
\frac{1}{n}\sum_{t=1}^{n}L_{n,t}^{(m+2)}(u)\partial _{u}h_{t}^{\ast }\left(
\theta \left( u\right) |u\right) & =\mu _{m+2}\partial _{u}H\left( u\right)
+o_{P}\left( 1\right) ,  \notag
\end{align}%
while, using Assumption \ref{assu: derivprocess}, with $h_{n,t}\left(
u\right) =h_{n,t}\left( \theta \left( u\right) \right) $ and $h_{t}^{\ast
}\left( u\right) =h_{t}^{\ast }\left( \theta \left( u\right) |u\right) $,%
\begin{align*}
& \frac{1}{n}\sum_{t=1}^{n}\lVert L_{n,t}^{(m+1)}(u)\rVert \mathbb{E}\left[
\lVert h_{n,t}\left( u\right) -h_{t}^{\ast }\left( u\right) -\partial
_{u}h_{t}^{\ast }\left( u\right) \left\{ t/n-u\right\} \rVert \right] \\
& \leq \frac{1}{n}\sum_{t=1}^{n}\lVert L_{n,t}^{(m+1)}(u)\rVert C\left(
1/n^{q}+\rho ^{t}\right) \\
& +\frac{1}{n}\sum_{t=1}^{n}\lVert L_{n,t}^{(m+1)}(u)\left( u\right) \rVert 
\mathbb{E}\left[ \lVert h_{t}^{\ast }\left( t/n\right) -h_{t}^{\ast }\left(
u\right) -\partial _{u}h_{t}^{\ast }\left( u\right) \left\{ t/n-u\right\}
\rVert \right] \\
& =O_{P}\left( n^{-q}\right) +O_{P}\left( 1/\sqrt{nb}\right) +o_{P}(b).
\end{align*}%
For the second and the third terms, similar to the proof of eq. (\ref{eq: as
norm 2}) and using Assumption \ref{assu: 3rdderiv}, 
\begin{equation*}
\sum_{t=1}^{n}L_{n,t}^{(m+2)}\left( u\right) h_{n,t}\left( \theta \left(
t/n\right) \right) \theta ^{\left( m+2\right) }\left( u_{n,t}\right)
\rightarrow ^{p}\mu _{m+2}H\left( u\right) \theta ^{\left( m+2\right)
}\left( u\right) ,
\end{equation*}%
\begin{equation*}
\sum_{t=1}^{n}L_{n,t}^{(2m+2)}\left( u\right) \theta _{i}^{\left( m+1\right)
}\left( u_{n,t}\right) \frac{\partial h_{n,t}\left( \bar{\theta}%
_{n,t}\right) }{\partial \theta _{i}}\theta ^{\left( m+1\right) }\left(
u_{n,t}\right) \rightarrow ^{p}\mu _{2m+2}\theta _{i}^{\left( m+1\right)
}\left( u\right) \partial _{\theta _{i}}H\left( u\right) \theta ^{\left(
m+1\right) }\left( u\right) .
\end{equation*}%
Collecting terms now yield the claimed result.
\end{proof}

\medskip

\begin{proof}[Proof of Corollary \protect\ref{cor: LC norm}]
The first part is a direct consequence of Theorem \ref{thm: Norm2}. To show
the second part, first observe that%
\begin{equation*}
B_{1}\left( u\right) +B_{2}\left( u\right) =H\left( u\right) \frac{\theta
^{\left( 2\right) }\left( u\right) }{2}+\left( \partial _{u}H\left( u\right)
+\frac{1}{2}\sum_{i=1}^{d_{\theta }}\theta _{i}^{\left( 1\right) }\left(
u\right) \partial _{\theta _{i}}H\left( u\right) \right) \theta ^{\left(
1\right) }\left( u\right) .
\end{equation*}%
Next, using that $E\left[ s_{t}^{\ast }\left( \theta \left( v\right)
|v\right) \right] =0$ for all $v$,%
\begin{eqnarray*}
0 &=&\frac{\partial ^{2}}{\partial v^{2}}E\left[ s_{t}^{\ast }\left( \theta
\left( v\right) |v\right) \right] _{v=u} \\
&=&\frac{\partial ^{2}}{\partial u^{2}}E\left[ s_{t}^{\ast }\left( \theta
\left( v\right) |u\right) \right] _{v=u}+2\frac{\partial ^{2}}{\partial
u\partial v}E\left[ s_{t}^{\ast }\left( \theta \left( v\right) |u\right) %
\right] _{v=u}+\frac{\partial ^{2}}{\partial v^{2}}E\left[ s_{t}^{\ast
}\left( \theta \left( v\right) |u\right) \right] _{v=u} \\
&=&E\left[ \partial _{u}^{2}s_{t}^{\ast }\left( \theta \left( v\right)
|u\right) \right] _{v=u}+2\frac{\partial ^{2}}{\partial u\partial v}E\left[
s_{t}^{\ast }\left( \theta \left( v\right) |u\right) \right] _{v=u}+\frac{%
\partial ^{2}}{\partial v^{2}}E\left[ s_{t}^{\ast }\left( \theta \left(
v\right) |u\right) \right] _{v=u}
\end{eqnarray*}%
where it is easily checked that%
\begin{equation*}
2\frac{\partial ^{2}}{\partial u\partial v}E\left[ s_{t}^{\ast }\left(
\theta \left( v\right) |u\right) \right] _{v=u}+\frac{\partial ^{2}}{%
\partial v^{2}}E\left[ s_{t}^{\ast }\left( \theta \left( v\right) |u\right) %
\right] _{v=u}=2\left\{ B_{1}\left( u\right) +B_{2}\left( u\right) \right\} .
\end{equation*}
\end{proof}

\medskip

\begin{proof}[Proof of Theorem \protect\ref{thm: Norm3}]
Inspecting the proof of Theorem \ref{thm: Norm2}, we observe that
Assumptions \ref{assu: derivprocess}--\ref{assu: hessian shortdep} are only
invoked in the analysis of $\frac{1}{n}%
\sum_{t=1}^{n}L_{n,t}^{(m+1)}(u)h_{n,t}\left( \theta \left( t/n\right)
\right) $. We here re--analyse this term under Assumptions \ref{assu:
derivprocess alt}--\ref{assu: hessian shortdep alt}: Write%
\begin{eqnarray*}
\frac{1}{n}\sum_{t=1}^{n}L_{n,t}^{(m+1)}(u)h_{n,t}\left( \theta \left(
t/n\right) \right) &=&\frac{1}{n}\sum_{t=1}^{n}L_{n,t}^{(m+1)}(u)\left[
h_{n,t}\left( \theta \left( t/n\right) \right) -h_{n,t}\left( \theta \left(
u\right) \right) \right] \\
&&+\frac{1}{n}\sum_{t=1}^{n}L_{n,t}^{(m+1)}(u)h_{n,t}\left( \theta \left(
u\right) \right) ,
\end{eqnarray*}%
where the first term satisfies eq. (\ref{eq: term 1}). Now, with $%
h_{n,t}:=h_{n,t}\left( \theta \left( u\right) \right) $ and $h_{t}^{\ast
}\left( v\right) :=h_{t}^{\ast }\left( \theta \left( u\right) |v\right) $, 
\begin{eqnarray*}
\frac{1}{n}\sum_{t=1}^{n}L_{b}\left( t/n-u\right) h_{n,t} &=&\frac{1}{n}%
\sum_{t=1}^{n}L_{b}\left( t/n-u\right) \left\{ h_{n,t}-h_{t}^{\ast }\left(
t/n\right) \right\} +\frac{1}{n}\sum_{t=1}^{n}L_{b}\left( t/n-u\right)
\left\{ h_{t}^{\ast }\left( t/n\right) -E\left[ h_{t}^{\ast }\left(
t/n\right) \right] \right\} \\
&&+\frac{1}{n}\sum_{t=1}^{n}L_{b}\left( t/n-u\right) E\left[ h_{t}^{\ast
}\left( t/n\right) \right] \\
&=&:A_{1}+A_{2}+A_{3}.
\end{eqnarray*}

By Assumption \ref{assu: score}(ii), 
\begin{equation*}
E\left[ \left\Vert A_{1}\right\Vert \right] \leq \frac{1}{n}%
\sum_{t=1}^{n}\left\vert L_{n,t}^{(m+1)}(u)\right\vert E\left[ \left\Vert
h_{n,t}-h_{t}^{\ast }\left( t/n\right) \right\Vert \right] \leq \frac{C}{%
n^{1+q}}\sum_{t=1}^{n}\left\vert L_{n,t}^{(m+1)}(u)\right\vert =O\left(
n^{-q}\right) ;
\end{equation*}%
for any element $A_{2,i,j}$ of $A_{2}$, by Assumption \ref{assu: hessian
shortdep alt} and with $V_{\left\vert t_{1}-t_{2}\right\vert }^{\left(
i,j\right) }\left( v_{1},v_{2}\right) =$Cov$\left( h_{t_{1},i,j}^{\ast
}\left( v_{1}\right) ,h_{t_{2},i,j}^{\ast }\left( \theta \left( v_{2}\right)
|u\right) \right) $,%
\begin{align*}
E\left[ A_{2,i,j}^{2}\right] & =\frac{1}{n^{2}}%
\sum_{t_{1},t_{2}=1}^{n}L_{n,t_{1}}^{(m+1)}(u)L_{n,t_{2}}^{(m+1)}(u)V_{\left%
\vert t_{1}-t_{2}\right\vert }^{\left( i,j\right) }\left(
t_{1}/n,t_{2}/n\right) \\
& =\frac{1}{n^{2}}%
\sum_{t_{1},t_{2}=1}^{n}L_{n,t_{1}}^{(m+1)}(u)L_{n,t_{2}}^{(m+1)}(u)\left\{
V_{\left\vert t_{1}-t_{2}\right\vert }^{\left( i,j\right) }\left(
t_{1}/n,t_{2}/n\right) -V_{\left\vert t_{1}-t_{2}\right\vert }^{\left(
i,j\right) }\left( u,u\right) \right\} \\
& +\frac{1}{n^{2}}%
\sum_{t_{1},t_{2}=1}^{n}L_{n,t_{1}}^{(m+1)}(u)L_{n,t_{2}}^{(m+1)}(u)V_{\left%
\vert t_{1}-t_{2}\right\vert }^{\left( i,j\right) }\left( u,u\right) \\
& =\frac{1}{n^{2}}%
\sum_{t_{1},t_{2}=1}^{n}L_{n,t_{1}}^{(m+1)}(u)L_{n,t_{2}}^{(m+1)}(u)\left\{
V_{\left\vert t_{1}-t_{2}\right\vert }^{\left( i,j\right) }\left(
t_{1}/n,t_{2}/n\right) -V_{\left\vert t_{1}-t_{2}\right\vert }^{\left(
i,j\right) }\left( u,u\right) \right\} \\
& +O\left( \frac{1}{nb}\right)
\end{align*}%
where, with 
\begin{eqnarray*}
\Delta V_{t_{1},t_{2}}^{\left( i,j\right) } &:&=V_{\left\vert
t_{1}-t_{2}\right\vert }^{\left( i,j\right) }\left( t_{1}/n,t_{2}/n\right)
-V_{\left\vert t_{1}-t_{2}\right\vert }^{\left( i,j\right) }\left(
u,u\right) -\frac{\partial V_{\left\vert t_{1}-t_{2}\right\vert }^{\left(
i,j\right) }\left( u,u\right) }{\partial u}\left[ \left( t_{1}/n-u\right)
+\left( t_{2}/n-u\right) \right] \\
&=&O\left( \left( t_{1}/n-u\right) ^{2}+\left( t_{2}/n-u\right) ^{2}\right) ,
\end{eqnarray*}%
\begin{eqnarray*}
&&\left\vert \frac{1}{n^{2}}%
\sum_{t_{1},t_{2}=1}^{n}L_{n,t_{1}}^{(m+1)}(u)L_{n,t_{2}}^{(m+1)}(u)\left\{
V_{\left\vert t_{1}-t_{2}\right\vert }^{\left( i,j\right) }\left(
t_{1}/n,t_{2}/n\right) -V_{\left\vert t_{1}-t_{2}\right\vert }^{\left(
i,j\right) }\left( u,u\right) \right\} \right\vert \\
&\leq &\frac{1}{n^{2}}\sum_{t_{1},t_{2}=1}^{n}\left\vert
L_{n,t_{1}}^{(m+1)}(u)\right\vert \left\vert
L_{n,t_{2}}^{(m+1)}(u)\right\vert \left\vert \Delta V_{t_{1},t_{2}}^{\left(
i,j\right) }\right\vert \\
&&+\frac{b}{n^{2}}\sum_{t_{1},t_{2}=1}^{n}\left\vert
L_{n,t_{1}}^{(m+1)}(u)\right\vert \left\vert
L_{n,t_{2}}^{(m+1)}(u)\right\vert \left\vert \frac{\partial V_{\left\vert
t_{1}-t_{2}\right\vert }\left( u,u\right) }{\partial u}\right\vert \left[
\left\vert t_{1}/n-u\right\vert +\left\vert t_{1}/n-u\right\vert \right] \\
&\leq &\frac{C}{n^{2}}\sum_{t_{1},t_{2}=1}^{n}\left\vert
L_{n,t_{1}}^{(m+1)}(u)\right\vert \left\vert
L_{n,t_{2}}^{(m+1)}(u)\right\vert \left\{ \left( t_{1}/n-u\right)
^{2}+\left( t_{1}/n-u\right) ^{2}\right\} \\
&&+\frac{\bar{L}}{\left( nb\right) ^{2}}\sum_{t_{1},t_{2}=1}^{n}\left\vert
L\left( \frac{t_{1}/n-u}{b}\right) \right\vert \left\vert \frac{\partial
V_{\left\vert t_{1}-t_{2}\right\vert }\left( u,u\right) }{\partial u}\left[
\left\vert t_{1}/n-u\right\vert +\left\vert t_{1}/n-u\right\vert \right]
\right\vert \\
&=&O\left( \frac{1}{nb}\right)
\end{eqnarray*}%
Finally, by Assumption \ref{assu: derivprocess alt}, 
\begin{eqnarray*}
A_{3} &=&\frac{1}{n}\sum_{t=1}^{n}L_{n,t}^{(m+1)}(u)E\left[ h_{t}^{\ast
}\left( u\right) \right] -b\times \frac{1}{n}%
\sum_{t=1}^{n}L_{n,t}^{(m+2)}(u)\partial _{u}E\left[ h_{t}^{\ast }\left(
u\right) \right] \\
&&+\frac{1}{n}\sum_{t=1}^{n}L_{n,t}^{(m+1)}(u)\left\{ E\left[ h_{t}^{\ast
}\left( t/n\right) \right] -E\left[ h_{t}^{\ast }\left( u\right) \right]
-\partial _{u}E\left[ h_{t}^{\ast }\left( u\right) \right] \left(
t/n-u\right) \right\} \\
&=&\int_{\mathbb{R}}L^{\left( m+1\right) }\left( v\right) dv\times E\left[
h_{t}^{\ast }\left( u\right) \right] +b\int_{\mathbb{R}}L^{\left( m+1\right)
}\left( v\right) vdv\times \partial _{u}E\left[ h_{t}^{\ast }\left( u\right) %
\right] +o\left( b\right)
\end{eqnarray*}
\end{proof}

\subsection{Proofs of results in Section \protect\ref{sec: ex}}

\begin{proof}[Proof of Lemma \protect\ref{lemma: LS transform}]
First note that $f\left( \mathcal{W}_{t}^{\ast }\left( \theta |u\right)
;\theta \right) $ is well-defined in the $L_{1}$--sense under (\ref{eq: f
conds}) together with $\mathbb{E}\left[ \left\Vert W_{t}^{\ast }\left(
\theta |u\right) \right\Vert \right] <\infty $. Next, with $p_{W}=p/\left(
r+1\right) $ and $\left\Vert \mathcal{W}_{n,t}\left( \theta \right)
\right\Vert _{a,p}=\left( \sum_{i=1}^{\infty }a_{i}\left\Vert W_{t-i}^{\ast
}\left( \theta |u\right) \right\Vert ^{p}\right) ^{1/p}$,%
\begin{align*}
& \mathbb{E}\left[ \sup_{\theta \in \Theta }\lVert f\left( \mathcal{W}%
_{n,t}\left( \theta \right) ;\theta \right) -f\left( \mathcal{W}_{t}^{\ast
}\left( \theta |t/n\right) ;\theta \right) \rVert ^{p_{W}}\right] ^{1/p_{W}}
\\
& \leq C\mathbb{E}\left[ \sup_{\theta \in \Theta }\left( 1+\left\Vert 
\mathcal{W}_{n,t}\left( \theta \right) \right\Vert
_{a_{1},r}^{p_{Z}r}+\left\Vert \mathcal{W}_{t}^{\ast }\left( \theta
|t/n\right) \right\Vert _{a_{1},r}^{p_{Z}r}\right) \left\Vert \mathcal{W}%
_{n,t}\left( \theta \right) -\mathcal{W}_{t}^{\ast }\left( \theta
|t/n\right) \right\Vert _{a_{2},1}^{p_{Z}}\right] ^{1/p_{W}} \\
& \leq C\sum_{i=1}^{\infty }a_{2,i}\mathbb{E}\left[ \sup_{\theta \in \Theta
}\left\Vert W_{n,t}\left( \theta \right) -W_{t}^{\ast }\left( \theta
|t/n\right) \right\Vert ^{p}\right] ^{1/p}\leq \left( C\sum_{i=1}^{\infty
}a_{2,i}\right) n^{-q}
\end{align*}%
where we have employed H{oe}lder's inequality. Similarly, $\mathbb{E}\left[
\sup_{\theta \in \Theta }\lVert f\left( \mathcal{W}_{t}^{\ast }\left( \theta
|u\right) ;\theta \right) -f\left( \mathcal{W}_{t}^{\ast }\left( \theta
|v\right) ;\theta \right) \rVert ^{p_{Z}}\right] ^{1/p_{Z}}\leq C\left\vert
u-v\right\vert $.

To show the second part, let $\tilde{W}_{t}^{\ast }\left( \theta |u\right) $
be an independent copy of $W_{t}^{\ast }\left( \theta |u\right) $. By
Hoelder's inequality and then Loeve's $c_{r}$ inequality,%
\begin{eqnarray*}
&&\mathbb{E}\left[ \lVert f\left( \mathcal{\tilde{W}}_{t}^{\ast }\left(
\theta |u\right) ;\theta \right) -f\left( \mathcal{W}_{t}^{\ast }\left(
\theta |u\right) ;\theta \right) \rVert ^{\alpha /\beta \left( p+1\right) }%
\right] \\
&\leq &C\mathbb{E}\left[ \lVert \mathcal{\tilde{W}}_{t}^{\ast }\left( \theta
|u\right) -\mathcal{W}_{t}^{\ast }\left( \theta |u\right) \rVert ^{\alpha }%
\right] ^{1/\left( p+1\right) }\left( 1+\mathbb{E}\left[ \lVert \mathcal{%
\tilde{W}}_{t}^{\ast }\left( \theta |u\right) \rVert ^{\alpha /\beta }\right]
+\mathbb{E}\left[ \lVert \mathcal{W}_{t}^{\ast }\left( \theta |u\right)
\rVert ^{\alpha /\beta }\right] \right) ^{p/\left( p+1\right) }
\end{eqnarray*}

The second part now follows from the definition of $\tau $-weak dependence,
c.f. p. 1999 of \cite{doukhan2008}. This in turn implies that the
short-memory condition is satisfied, c.f. \cite{doukhan2008}.
\end{proof}

\bigskip

\begin{proof}[Proof of Lemma \protect\ref{thm: Markov LS}]
The first part of the result follows from Corollary 3.1 of \cite{doukhan2008}%
. For the second part, 
\begin{align*}
\mathbb{E}\left[ \lVert Y_{t}^{\ast }\left( u\right) -Y_{t}^{\ast }\left(
v\right) \rVert ^{\tilde{p}}\right] ^{1/\tilde{p}}& =\mathbb{E}\left[ \lVert
F\left( Y_{t-1}^{\ast }\left( u\right) ,\varepsilon _{t},\theta \left(
u\right) \right) -F\left( Y_{t-1}^{\ast }\left( v\right) ,\varepsilon
_{t},\theta \left( v\right) \right) \rVert ^{\tilde{p}}\right] ^{1/\tilde{p}}
\\
& \leq \mathbb{E}\left[ \lVert F\left( Y_{t-1}^{\ast }\left( u\right)
,\varepsilon _{t},\theta \left( u\right) \right) -F\left( Y_{t-1}^{\ast
}\left( u\right) ,\varepsilon _{t},\theta \left( v\right) \right) \rVert ^{%
\tilde{p}}\right] ^{1/\tilde{p}} \\
& +\mathbb{E}\left[ \lVert F\left( Y_{t-1}^{\ast }\left( u\right)
,\varepsilon _{t},\theta \left( v\right) \right) -F\left( Y_{t-1}^{\ast
}\left( v\right) ,\varepsilon _{t},\theta \left( v\right) \right) \rVert ^{%
\tilde{p}}\right] ^{1/\tilde{p}} \\
& \leq C\left( 1+\mathbb{E}\left[ \lVert Y_{t-1}^{\ast }\left( u\right)
\rVert ^{r\tilde{p}}\right] ^{1/\tilde{p}}\right) \left\vert u-v\right\vert
^{q}+\rho \mathbb{E}\left[ \lVert Y_{t-1}^{\ast }\left( u\right)
-Y_{t-1}^{\ast }\left( v\right) \rVert ^{\tilde{p}}\right] ^{1/\tilde{p}} \\
& \vdots \\
& \leq C_{1}\left\vert u-v\right\vert ^{q},
\end{align*}%
where $C_{1}=C\left( 1+\sup_{u\in \left[ 0,1\right] }\mathbb{E}\left[ \lVert
Y_{t-1}^{\ast }\left( u\right) \rVert ^{r\tilde{p}}\right] ^{1/\tilde{p}%
}\right) /\left( 1-\rho \right) $. By setting $v=t/n$, we have that 
\begin{equation*}
\mathbb{E}\left[ \lVert Y_{t}^{\ast }\left( t/n\right) -Y_{t}^{\ast }\left(
u\right) \rVert ^{p}\right] ^{1/\tilde{p}}\leq C_{1}\left\vert
t/n-u\right\vert ^{q}.
\end{equation*}%
In addition, 
\begin{align*}
& \mathbb{E}\left[ \lVert Y_{n,t}-Y_{t}^{\ast }\left( t/n\right) \rVert ^{%
\tilde{p}}\right] ^{1/\tilde{p}} \\
& =\mathbb{E}\left[ \lVert F\left( Y_{n,t-1},\varepsilon _{t},\theta \left(
t/n\right) \right) -F\left( Y_{t-1}^{\ast }\left( t/n\right) ,\varepsilon
_{t},\theta \left( t/n\right) \right) \rVert ^{\tilde{p}}\right] ^{1/\tilde{p%
}} \\
& \leq \rho \mathbb{E}\left[ \lVert Y_{n,t-1}-Y_{t-1}^{\ast }\left(
t/n\right) \rVert ^{\tilde{p}}\right] ^{1/\tilde{p}} \\
& \leq \rho \left( \mathbb{E}\left[ \lVert Y_{n,t-1}-Y_{t-1}^{\ast }\left(
\left( t-1\right) /n\right) \rVert ^{\tilde{p}}\right] ^{1/\tilde{p}}+%
\mathbb{E}\left[ \lVert Y_{t-1}^{\ast }\left( t/n\right) -Y_{t-1}^{\ast
}\left( \left( t-1\right) /n\right) \rVert ^{\tilde{p}}\right] ^{1/\tilde{p}%
}\right) \\
& \leq \rho \left( \mathbb{E}\left[ \lVert Y_{n,t-1}-Y_{t-1}^{\ast }\left(
\left( t-1\right) /n\right) \rVert ^{\tilde{p}}\right] ^{1/\tilde{p}%
}+C_{1}/n^{q}\right) \\
& \vdots \\
& \leq \frac{C_{1}/n^{q}}{\left( 1-\rho \right) }.
\end{align*}%
Continuing the above two recursions yields the desired results.
\end{proof}

\bigskip

\begin{proof}[Proof of Lemma \protect\ref{lem: MarkovX LS}]
Define $\xi _{t}=\left( \varepsilon _{t},\eta _{t}\right) $ and $F\left(
y,x,\xi _{t},t/n\right) =\left[ G\left( y,x,\varepsilon _{t};\theta \left(
t/n\right) \right) ,H\left( x,\eta _{t};t/n\right) \right] $. Then the
results follow by applying Lemma \ref{thm: Markov LS} to the Markov process $%
Z_{n,t}=\left( Y_{n,t},X_{n,t}\right) $ which solves $Z_{n,t}=F\left(
Z_{n,t-1},\xi _{t},t/n\right) $.
\end{proof}

\bigskip

\begin{proof}[Proof of Corollary \protect\ref{cor: tvTAR1}]
We can here apply our theory with $\Theta =\mathbb{R}^{1+2q+d_{X}}$ since
the least-squares criterion used for estimation is concave in $\theta $,
c.f. the comments following Assumptions \ref{assu: kernel}-\ref{assu: LS}.
We first show that $Y_{n,t}$ is locally stationary with $p\geq 2$ moments
when $\mathbb{E}\left[ \left\Vert \varepsilon _{t}\right\Vert ^{2}\right]
<\infty $ by verifying the conditions of Lemma \ref{lem: MarkovX LS} for $%
G\left( y,x,e,\theta \right) :=\omega +\sum_{i=1}^{q}\alpha
_{1,i}y_{i}^{+}+\sum_{i=1}^{q}\alpha _{2,i}y_{i}^{-}+\gamma ^{\prime }x+e$:
First, $\mathbb{E}\left[ G\left( 0,0,\varepsilon _{t};\theta \right) ^{2}%
\right] =\mathbb{E}\left[ \varepsilon _{t}^{2}\right] <\infty $; second, for
all $y,y\in \mathcal{\mathbb{R}}^{q}$ and $x,x^{\prime }\in \mathcal{\mathbb{%
R}}^{d_{X}}$, 
\begin{equation*}
\mathbb{E}\left[ \left( G\left( y,x,\varepsilon _{t};\theta \right) -G\left(
y^{\prime },x^{\prime },\varepsilon _{t};\theta \right) \right) ^{2}\right]
^{1/2}\leq \sum_{i=1}^{q}\max \left\{ \left\vert \alpha _{1,i}\right\vert
,\left\vert \alpha _{2,i}\right\vert \right\} \left\Vert y-y^{\prime
}\right\Vert +\left\Vert \gamma \right\Vert \left\Vert x-x^{\prime
}\right\Vert ;
\end{equation*}%
third, for all $\theta ,\theta ^{\prime }\in \Theta $, 
\begin{eqnarray*}
\mathbb{E}\left[ \left\Vert G\left( y,x,\varepsilon _{t};\theta \right)
-G\left( y,x,\varepsilon _{t};\theta ^{\prime }\right) \right\Vert ^{p}%
\right] ^{1/p} &=&\left\vert \omega -\omega ^{\prime }\right\vert +\left\{
\left\Vert \alpha _{1}-\alpha _{1}^{\prime }\right\Vert +\left\Vert \alpha
_{2}-\alpha _{2}^{\prime }\right\Vert \right\} \left\Vert y\right\Vert
+\left\Vert \gamma -\gamma ^{\prime }\right\Vert \left\Vert x\right\Vert \\
&\leq &C\left( 1+\left\Vert x\right\Vert +\left\Vert y\right\Vert \right)
\left\Vert \theta -\theta ^{\prime }\right\Vert .
\end{eqnarray*}%
Thus, under (\ref{eq: TAR stat}), $\left( Y_{n,t},X_{n,t}\right) $ is
locally stationary with $\left( Y_{t}^{\ast }\left( u\right) ,X_{t}^{\ast
}\left( u\right) \right) $ being $\tau $--weakly dependent, $\mathbb{E}\left[
Y_{t}^{\ast }\left( u\right) ^{2}\right] <\infty $ and $\mathbb{E}\left[
\left\Vert X_{t}^{\ast }\left( u\right) \right\Vert ^{2}\right] <\infty $%
Next, write $\ell _{n,t}\left( \theta \right) =-\left( Y_{n,t}-\theta
^{\prime }\tilde{X}_{n,t}\right) ^{2}$, where $\tilde{X}_{n,t}=\left(
1,Y_{n,t-1}^{+},...,Y_{n,t-q}^{+},Y_{n,t-1}^{-},...,Y_{n,t-q}^{-},X_{n,t-1}^{\prime }\right) ^{\prime } 
$, so that $s_{n,t}\left( \theta \right) =\left( Y_{n,t}-\theta ^{\prime }%
\tilde{X}_{n,t}\right) \tilde{X}_{n,t}$ and $h_{n,t}\left( \theta \right) =%
\tilde{X}_{n,t}\tilde{X}_{n,t}^{\prime }$. All three functions satisfy (\ref%
{eq: f conds}) with $r=1$. It now follows from the first part of Corollary %
\ref{cor: Master} that Theorem \ref{thm: Norm1} applies to the local linear
estimator.

Next, we verify the conditions of the second part of Corollary \ref{cor:
Master}. First, note that Lemma \ref{lem: MarkovX LS} implies that if $%
\mathbb{E}\left[ \varepsilon _{t}^{4}\right] <\infty $ then $\mathbb{E[}%
Y_{t}^{\ast }\left( u\right) ^{4}]<\infty $ and so $h_{t}^{\ast }\left(
\theta \left( u\right) |u\right) =\tilde{X}_{t}^{\ast }\left( u\right) 
\tilde{X}_{t}^{\ast }\left( u\right) ^{\prime }$ is $\tau $--weakly
dependent with $\mathbb{E}\left[ \left\Vert h_{t}^{\ast }\left( \theta
\left( u\right) |u\right) \right\Vert ^{2}\right] <\infty $, and hence
Assumption \ref{assu: hessian shortdep alt} is satisfied. Finally, to verify
Assumption \ref{assu: derivprocess}, we show that $\partial _{u}h_{t}^{\ast
}\left( \theta \left( u\right) |u\right) =2\tilde{X}_{t}^{\ast }\left(
u\right) \partial _{u}\tilde{X}_{t}^{\ast }\left( u\right) ^{\prime }$ is
well--defined almost surely and has first moments. This will hold if $%
\partial _{u}\left( Y_{t}^{\ast }\left( u\right) ,X_{t}^{\ast }\left(
u\right) \right) $ exists almost surely and has second moments. We show this
by applying Theorem 4.8 of \cite{dahlhaus2017}. This theorem requires $%
G\left( y,x,\varepsilon _{t};\theta \right) $ to be differentiable w.r.t. $%
\left( y,x,\theta \right) $ for all values of $\left( y,x,\theta \right) $
which fails to hold at $y=0$. However, since $\varepsilon _{t}$ is assumed
to have a continuous distribution then $Y_{t}^{\ast }\left( u\right) |%
\mathcal{F}_{t-1}^{\ast }\left( u\right) $ will also have a continuous
distribution and so $\Pr \left( Y_{t}^{\ast }\left( u\right) =0|\mathcal{F}%
_{t-1}^{\ast }\left( u\right) \right) =0$. Thus, $G\left( Y_{t-1}^{\ast
}\left( u\right) ,....,Y_{t-q}^{\ast }\left( u\right) ,X_{t-1}^{\ast }\left(
u\right) .\varepsilon _{t};\theta \right) $ is differentiable w.r.t. $\left(
Y_{t-1}^{\ast }\left( u\right) ,....,Y_{t-q}^{\ast }\left( u\right)
,X_{t-1}^{\ast }\left( u\right) ,\theta \right) $ almost surely. By
inspection of the proof of Theorem 4.8, this suffices for the result to hold.
\end{proof}

\bigskip

\begin{proof}[Proof of Corollary \protect\ref{cor: ARCH}]
We verify the conditions of Corollary \ref{cor: Master}. Verification of the
conditions for the stationary version, including identification and
existence of relevant moments, follow from \citet{kristensen2005}. For the
analysis of the local linear estimator, what remains to be shown is local
stationarity of the log-likelihood function, the conditional variance of the
score function and the hessian. First, by Lemma \ref{lem: MarkovX LS} with $%
G\left( y,x,\varepsilon ,\theta \right) =\left( \omega +\sum_{i=1}^{q}\alpha
_{i}y_{i}+\gamma ^{\prime }x\right) \varepsilon ^{2}$, where $%
\sum_{i=1}^{q}\alpha _{i}<1$, it follows that $\left( Y_{n,t},X_{n,t}\right) 
$ is locally stationary with $\sup_{n,t}\mathbb{E}\left[ Y_{n,t}\right]
<\infty $ and $\mathbb{E}\left[ Y_{t}^{\ast }\left( u\right) \right] <\infty 
$. Next, we verify that $\ell _{n,t}\left( \theta \right) $ and its first
two derivatives satisfy the conditions of Lemma \ref{lemma: LS transform}:

Recall that $\ell _{n,t}\left( \theta \right) =\log \left( \lambda
_{n,t}\left( \theta \right) \right) +Y_{n,t}/\lambda _{n,t}\left( \theta
\right) $ with $\tilde{X}_{n,t}=\left(
1,Y_{n,t-1},...,Y_{n,t-q},X_{n,t-1}^{\prime }\right) ^{\prime }$ and $%
\lambda _{n,t}\left( \theta \right) =\theta ^{\prime }\tilde{X}_{n,t}$.
Here, $\lambda _{n,t}\left( \theta \right) $ is trivially ULS$\left(
1,1,\Theta \right) $ while 
\begin{equation*}
\left\vert \frac{\partial \ell _{n,t}\left( \theta \right) }{\partial Y_{n,t}%
}\right\vert =\frac{1}{\lambda _{n,t}\left( \theta \right) }\leq \frac{1}{%
\omega }\leq \frac{1}{\delta _{L}},\;\left\Vert \frac{\partial \ell
_{n,t}\left( \theta \right) }{\partial \lambda _{n,t}\left( \theta \right) }%
\right\Vert \leq \frac{1}{\lambda _{n,t}\left( \theta \right) }+\frac{Y_{n,t}%
}{\lambda _{n,t}^{2}\left( \theta \right) }\leq \frac{1}{\delta _{L}}+\frac{%
Y_{n,t}}{\delta _{L}\lambda _{n,t}\left( \theta \right) },
\end{equation*}%
where 
\begin{equation}
\frac{Y_{n,t}}{\lambda _{n,t}\left( \theta \right) }=\frac{\theta
(t/n)^{\prime }\tilde{X}_{n,t}}{\theta ^{\prime }\tilde{X}_{n,t}}\varepsilon
_{t}^{2}\leq \frac{\sup_{u}\omega \left( u\right)
+\sum_{i=1}^{q}\sup_{u}\alpha _{i}\left( u\right)
+\sum_{i=1}^{d_{X}}\sup_{u}\beta _{i}\left( u\right) }{\delta _{L}}%
\varepsilon _{t}^{2}.  \label{eq: ARCH ineq}
\end{equation}%
Thus, $\ell _{n,t}\left( \theta \right) $ satisfies the conditions of Lemma %
\ref{lemma: LS transform} with $r=0$ and $q=1$.

Next, we analyse the score function $s_{n,t}\left( \theta \right) =\left(
1-Y_{n,t}/\lambda _{n,t}\left( \theta \right) \right) \partial _{\theta
}\lambda _{n,t}\left( \theta \right) /\lambda _{n,t}\left( \theta \right) $.
Since $\mathbb{E}\left[ Y_{n,t}|\mathcal{F}_{n,t-1}\right] =\lambda
_{n,t}\left( \theta \left( t/n\right) \right) \mathbb{E}\left[ \varepsilon
_{t}^{2}|\mathcal{F}_{n,t-1}\right] =\lambda _{n,t}\left( \theta \left(
t/n\right) \right) $, $s_{n,t}\left( \theta \left( t/n\right) \right) $ is a
MGD with 
\begin{equation*}
\omega _{n,t}\left( \theta \right) :=s_{n,t}\left( \theta \right)
s_{n,t}\left( \theta \right) ^{\prime }=\frac{\partial _{\theta }\lambda
_{n,t}\left( \theta \right) \left( \partial _{\theta }\lambda _{n,t}\left(
\theta \right) \right) ^{\prime }}{\lambda _{n,t}^{2}\left( \theta \right) }%
\left( 1-Y_{n,t}/\lambda _{n,t}\left( \theta \right) \right) ^{2}.
\end{equation*}
Here, $\partial _{\theta }\lambda _{n,t}\left( \theta \right) =\tilde{X}%
_{n,t}$ is trivially ULS$\left( 1,1,\Theta \right) $. Next, $\sup_{\left\{
\theta :\left\Vert \theta -\theta \left( u\right) \right\Vert <\epsilon
\right\} }\left\Vert \partial _{\theta }\lambda _{n,t}\left( \theta \right)
/\lambda _{n,t}\left( \theta \right) \right\Vert \leq C$ by the same
arguments as in (\ref{eq: ARCH ineq}), Thus, suppressing dependence on $%
\theta \left( t/n\right) $, 
\begin{align*}
\left\vert \frac{\partial \omega _{n,t}}{\partial Y_{n,t}}\right\vert & =%
\frac{2\partial _{\theta }\lambda _{n,t}^{2}}{\lambda _{n,t}^{3}}\left\vert
1-Y_{n,t}/\lambda _{n,t}\right\vert \leq C\left( 1+\varepsilon
_{t}^{2}\right) , \\
\left\vert \frac{\partial \omega _{n,t}}{\partial \lambda _{n,t}}\right\vert
& =\frac{2\partial _{\theta }\lambda _{n,t}^{2}}{\lambda _{n,t}^{3}}%
\left\vert 1-\frac{3Y_{n,t}}{\lambda _{n,t}}+\frac{4Y_{n,t}^{2}}{\lambda
_{n,t}^{2}}\right\vert \leq C\left( 1+\varepsilon _{t}^{2}+\varepsilon
_{t}^{4}\right) , \\
\left\Vert \frac{\partial \omega _{n,t}}{\partial \left( \partial _{\theta
}\lambda _{n,t}\right) }\right\Vert & =\frac{2\partial _{\theta }\lambda
_{n,t}}{\lambda _{n,t}^{2}}\left( 1-Y_{n,t}/\lambda _{n,t}\right) ^{2}\leq
C\left( 1+\varepsilon _{t}^{2}+\varepsilon _{t}^{4}\right) ,
\end{align*}%
and so $\omega _{n,t}\left( \theta \left( t/n\right) \right) $ satisfies the
conditions of Lemma \ref{lemma: LS transform} with $r=0$ and $q=1$.

The hessian takes the form 
\begin{equation*}
h_{n,t}\left( \theta \right) =\frac{\partial _{\theta }\lambda _{n,t}\left(
\theta \right) \partial _{\theta }\lambda _{n,t}\left( \theta \right)
^{\prime }}{\lambda _{n,t}^{2}\left( \theta \right) }\left[ \frac{2Y_{n,t}}{%
\lambda _{n,t}\left( \theta \right) }-1\right] ,
\end{equation*}%
and recycling the inequalities established above it follows that the hessian
is also ULS(1,1,$\Theta $). This verifies the conditions for Theorem \ref%
{thm: Norm1}.

For the analysis of the local constant estimator, observe that $%
h_{ij,t}^{\ast }\left( \theta \left( u\right) |u\right) $ is $\tau $--weakly
dependent by Lemma \ref{lemma: LS transform}; moreover, if $\mathbb{E}\left[
\lVert \varepsilon _{t}\rVert ^{4+\gamma }\right] <\infty $ then $\mathbb{E}%
\left[ \lVert h_{ij,t}^{\ast }\left( \theta \left( u\right) |u\right) \rVert
|^{2+\gamma /2}\right] <\infty $. Thus, Assumption \ref{assu: hessian
shortdep alt} is satisfied holds. To verify Assumption \ref{assu:
derivprocess alt}, first note that $\partial _{u}X_{t}^{\ast }\left(
u\right) $ exists under the assumptions of the theorem. We then follow the
same arguments as in Example 5.5 of \cite{dahlhaus2017} to show that $%
\partial _{u}h_{ij,t}^{\ast }\left( \theta |u\right) $ exists.
\end{proof}

\medskip

\begin{proof}[Proof of Corollary \protect\ref{cor: PARX}]
We first show that the PARX process $\left( Y_{n,t},X_{n,t}\right) $ is
locally stationary by verifying the conditions of Theorem \ref{lem: MarkovX
LS} with 
\begin{equation*}
G\left( y,x,\varepsilon _{t};\theta \right) :=N_{t}\left( \omega
+\sum_{i=1}^{q}\alpha _{i}y_{i}+\gamma ^{\prime }x\right) ,
\end{equation*}%
where $\varepsilon _{t}:=N_{t}\left( \cdot \right) $, $t=1,2,...$, are
i.i.d. copies of a Poisson process (see \citet{agosto2016} for details):%
\begin{equation*}
\mathbb{E}\left[ \left\vert G\left( 0,0,\varepsilon _{t};\theta \right)
\right\vert \right] \leq \mathbb{E}\left[ N_{t}\left( \omega \right) \right]
=\omega <\infty ;
\end{equation*}%
and for all $x,y,x^{\prime },y^{\prime }$, 
\begin{eqnarray*}
\mathbb{E}\left[ \left\Vert G\left( y,x,\varepsilon _{t};\theta \right)
-G\left( y,x^{\prime },\varepsilon _{t};\theta \right) \right\Vert \right]
&\leq &\mathbb{E}\left[ \left\Vert N_{t}\left( \sum_{i=1}^{q}\alpha
_{i}\left\vert y_{i}-y_{i}^{\prime }\right\vert +\left\Vert \gamma
\right\Vert \left\Vert x-x^{\prime }\right\Vert \right) \right\Vert \right]
\\
&=&\sum_{i=1}^{q}\alpha _{i}\left\vert y_{i}-y_{i}^{\prime }\right\vert
+\left\Vert \gamma \right\Vert \left\Vert x-x^{\prime }\right\Vert ,
\end{eqnarray*}%
where $\sum_{i=1}^{q}\alpha _{i}\left( u\right) <1$. Finally, 
\begin{eqnarray*}
\mathbb{E}\left[ \left\Vert G\left( y,x,\varepsilon _{t};\theta \right)
-G\left( y,x,\varepsilon _{t};\theta ^{\prime }\right) \right\Vert \right]
&=&\left\vert \omega -\omega ^{\prime }\right\vert +\sum_{i=1}^{q}\left\vert
\alpha _{i}-\alpha _{i}^{\prime }\right\vert y_{i}+\left\Vert \gamma -\gamma
^{\prime }\right\Vert \left\Vert x\right\Vert \\
&\leq &C\left( 1+\left\Vert x\right\Vert +\left\Vert y\right\Vert \right)
\left\Vert \theta -\theta ^{\prime }\right\Vert .
\end{eqnarray*}%
We will also need the existence of higher-order moments, and so we
demonstrate by induction that $E\left[ \lambda _{t}^{\ast }\left( u\right)
^{r}\right] <\infty $ for all $r<\infty $: First, $E[\lambda _{t}^{\ast
}\left( u\right) ]=\omega \left( u\right) +\sum_{i=1}^{q}\alpha _{i}\left(
u\right) E\left[ \lambda _{t}^{\ast }\left( u\right) \right] $ which has a
well-defined solution while 
\begin{equation*}
\lambda _{t}^{\ast }\left( u\right) ^{r}=\sum_{j=0}^{r}\binom{r}{j}\left(
\sum_{i=1}^{q}\alpha _{i}\left( u\right) Y_{t-i}^{\ast }\left( u\right)
\right) ^{j}\omega ^{r-j}\left( u\right) ,
\end{equation*}%
and so 
\begin{align*}
E[\lambda _{t}^{\ast }\left( u\right) ^{r}]& =\sum_{j=0}^{r}\binom{r}{j}E%
\left[ \left( \sum_{i=1}^{q}\alpha _{i}\left( u\right) Y_{t-i}^{\ast }\left(
u\right) \right) ^{j}\right] \left( \omega \left( u\right) \right) ^{r-j} \\
& =\omega \left( u\right) ^{r}+E\left[ \left( \sum_{i=1}^{q}\alpha
_{i}\left( u\right) Y_{t-i}^{\ast }\left( u\right) \right) ^{r}\right] +E%
\left[ p_{r-1}\left( X_{t}^{\ast }\left( u\right) \right) \right] ,
\end{align*}%
with $p_{r-1}\left( x\right) $ being an $\left( r-1\right) $th order
polynomial. By induction, $E\left[ p_{r-1}\left( X_{t}^{\ast }\left(
u\right) \right) \right] <\infty $, and we are left with considering terms
of the form, for some constants $c_{ij}$, 
\begin{align*}
E\left[ \left( \sum_{i=1}^{q}\alpha _{i}\left( u\right) Y_{t-i}^{\ast
}\left( u\right) \right) ^{r}\right] &
=\sum_{i=1}^{q}\sum_{j=0}^{r}c_{ij}\alpha _{i}^{j}\left( u\right) E\left[
Y_{t-i}^{\ast }\left( u\right) ^{j}\right] \\
& =\sum_{i=1}^{q}\alpha _{i}^{r}\left( u\right) E\left[ Y_{t-i}^{\ast
}\left( u\right) ^{r}\right] +C_{r} \\
& =\sum_{i=1}^{q}\alpha _{i}^{r}\left( u\right) E\left[ \lambda _{t}^{\ast
}\left( u\right) ^{r}\right] +C_{r}
\end{align*}%
where, again by induction, $C_{r}<\infty $. Collecting terms, $E\left[
\left( \lambda _{t}^{\ast }\left( u\right) \right) ^{r}\right]
=\sum_{i=1}^{q}\alpha _{i}^{r}\left( u\right) E\left[ \lambda _{t}^{\ast
}\left( u\right) ^{r}\right] +\tilde{C}_{r}$ which has a well-defined
solution since $\sum_{i=1}^{q}\alpha _{i}^{r}\left( u\right) <1$. This in
turn implies that $E\left[ Y_{t}^{\ast }\left( u\right) ^{r}\right] <\infty $
for all $r<\infty $. Thus, $\lambda _{n,t}$ and $Y_{n,t}$ are LS$\left( r,1\right) $ with $%
E[\lambda _{t}^{\ast }\left( u\right) ^{r}]<\infty $ and $E[Y_{t}^{\ast
}\left( u\right) ^{r}]<\infty $ for any $r\geq 1$.

Next, we observe that $\lambda _{n,t}\left( \theta \right) $, $\partial
_{\theta }\lambda _{n,t}\left( \theta \right) $ and $\partial _{\theta
\theta }^{2}\lambda _{n,t}\left( \theta \right) $ are on the same form as in
the GARCH model. In particular, it is easily checked that $\lambda
_{n,t}\left( \theta \right) ,$ $\partial _{\theta }\lambda _{n,t}\left(
\theta \right) $ and $\partial _{\theta \theta }^{2}\lambda _{n,t}\left(
\theta \right) $ are ULS$\left( 1,1,\Theta \right) $ and with their
stationary versions having all polynomial moments. This in turn implies that
the log-likelihood and its first two derivatives w.r.t. $\theta $ satisfy
the conditions of Lemma \ref{lemma: LS transform}. First, 
\begin{equation*}
\left\vert \frac{\partial \ell _{n,t}\left( \theta \right) }{\partial Y_{n,t}%
}\right\vert =\left\vert \log \left\{ \lambda _{n,t}\left( \theta \right)
\right\} \right\vert \leq \max \left\{ \left\vert \log \delta
_{L}\right\vert ,\lambda _{n,t}\left( \theta \right) \right\} ,
\end{equation*}%
where $\lambda _{n,t}\left( \theta \right) $ has all relevant moments.
Second, 
\begin{equation*}
\left\vert \frac{\partial \ell _{n,t}\left( \theta \right) }{\partial
\lambda _{n,t}\left( \theta \right) }\right\vert =\frac{Y_{n,t}}{\lambda
_{n,t}\left( \theta \right) }+1\leq \frac{Y_{n,t}}{\delta _{L}}+1,
\end{equation*}%
where again the right-hand side has all relevant moments. The score function
takes the form $s_{n,t}\left( \theta \right) =\left( Y_{n,t}/\lambda
_{n,t}\left( \theta \right) -1\right) \partial {}_{\theta }\lambda
_{n,t}\left( \theta \right) $ which satisfies the Martingale difference
condition with conditional variance $\omega _{n,t}\left( \theta \right)
=\partial _{\theta }\lambda _{n,t}\left( \theta \right) \partial _{\theta
}\lambda _{n,t}\left( \theta \right) ^{\prime }/\lambda _{n,t}\left( \theta
\right) $. As before, due to all polynomial moments existing, it is easily
checked that the $\omega _{n,t}\left( \theta \right) $ satisfies the
conditions of Lemma \ref{lemma: LS transform} and similarly for the hessian which
is on the form 
\begin{equation*}
h_{n,t}\left( \theta \right) =\frac{Y_{n,t}}{\lambda _{n,t}^{2}\left( \theta
\right) }\partial _{\theta }\lambda _{n,t}\left( \theta \right) \partial
_{\theta }\lambda _{n,t}\left( \theta \right) ^{\prime }-\left( \frac{Y_{n,t}%
}{\lambda _{n,t}\left( \theta \right) }-1\right) \partial _{\theta \theta
}^{2}\lambda _{n,t}\left( \theta \right) .
\end{equation*}

Finally, for the local constant estimator, we need to show that $\partial
_{u}H\left( u\right) $ exists. We do this for the case of $q=1$ to keep
notation simple; the general case follows by the same arguments. Under the
restrictions of the corollary, $h_{t}^{\ast }\left( \theta |u\right) $ has
all relevant moments and so the result will hold if the distribution of $%
\left( Y_{t}^{\ast }\left( u\right) ,Y_{t-1},X_{t-1}^{\ast }\left( u\right)
\right) $ is differentiable, c.f. Lemma \ref{lemma: time deriv conds}. To
show this property holds, first observe that the transition kernel of $%
\left( Y_{t}^{\ast },X_{t}^{\ast }\right) \left( u\right) |\left(
Y_{t-1}^{\ast },X_{t-1}^{\ast }\right) \left( u\right) $ takes the form%
\begin{equation*}
p\left( y,x|y_{0},x_{0};u\right) =\frac{\lambda \left( y_{0},x_{0};u\right)
^{y}e^{-\lambda \left( y_{0},x_{0};u\right) }}{y!}p_{X}\left(
x|x_{0};u\right) ,
\end{equation*}%
where $\lambda \left( y_{0},x_{0};u\right) =\omega +\alpha \left( u\right)
y_{0}+\gamma \left( u\right) ^{\prime }x_{0}$. This is clearly
differentiable w.r.t. $u$ since $\theta \left( u\right) $ is. What remains
to be shown is that the stationary density $\pi $ solving $\pi \left(
y,x;u\right) =\sum_{y_{0}=0}^{\infty }\int_{\mathbb{R}}p\left(
y,x|y_{0},x_{0};u\right) \pi \left( y_{0},x_{0};u\right) dx_{0}$ is also
differentiable w.r.t $u$. We do this by verifying the assumptions A1--A.5 in 
\cite{VazquezAbad1992} so that their Theorem 4 applies. A1 and A2 clearly
holds. To show A3, observe that $p\left( y,x|y_{0},x_{0};u\right) >0$ for
all $\left( y,x|y_{0},x_{0};u\right) $ since $p_{X}\left( x|x_{0};u\right)
>0 $ for all $\left( x,x_{0}\right) $ by assumption. Thus, the Markov chain
is irreducible and aperiodic which, together with stationarity, implies it
is recurrent, c.f. Proposition 10.1.1 in \cite{meyn2009}. Finally, A4 and A5
follows by the Markov chain being $\tau $-weakly dependent, c.f. \cite%
{doukhan2008}
\end{proof}

\end{document}